\documentclass[a4paper,11pt]{article}
\pdfoutput=1
\usepackage{jheppub}
\usepackage{subcaption}
\usepackage{xcolor}
\usepackage{enumitem}
\newcommand{\be}{\begin{equation}}
\newcommand{\ee}{\end{equation}}
\newcommand{\reply}[1]{#1}

\title{Multi-component Dark Matter in a Novel Three-Loop \\ Inverse Scotogenic Seesaw Model}

\author[a,b]{Asmaa Abada,}
\affiliation[a]{Pôle Théorie, Laboratoire de Physique des 2 Infinis Irène Joliot-Curie (UMR 9012), \\
CNRS/IN2P3, 15 Rue Georges Clemenceau, 91400 Orsay, France}
\affiliation[b]{Institut universitaire de France (IUF)}
\emailAdd{asmaa.abada@ijclab.in2p3.fr}

\author[c]{Nicolás Bernal,}
\affiliation[c]{New York University Abu Dhabi, \\
PO Box 129188, Saadiyat Island, Abu Dhabi, United Arab Emirates}
\emailAdd{nicolas.bernal@nyu.edu}

\author[d,e,f]{A.~E.~Cárcamo Hernández,}
\affiliation[d]{Universidad Técnica Federico Santa María, Casilla 110-V, Valparaíso, Chile}
\affiliation[e]{Centro Científico-Tecnológico de Valparaíso, Casilla 110-V, Valparaíso, Chile}
\affiliation[f]{Millennium Institute for Subatomic Physics at the High-Energy Frontier, SAPHIR, Chile}
\emailAdd{antonio.carcamo@usm.cl}

\author[g]{Vishnudath K.~N,}
\emailAdd{vishnudath.kn@vit.ac.in}
\affiliation[g]{School of Advanced Sciences, Vellore Institute of Technology, Vellore 632014, Tamil Nadu, India}

\author[f,h]{Sergey Kovalenko,}
\affiliation[h]{Center for Theoretical and Experimental Particle Physics, Facultad de Ciencias Exactas, \\
Universidad Andrés Bello, Fernández Concha 700, Santiago, Chile}
\emailAdd{sergey.kovalenko@unab.cl}

\author[i]{Téssio B.~de Melo,}
\affiliation[i]{Universidad Viña del Mar, Formación en Ciencias Exactas,\\
Agua Santa 7055, Rodelillo, Viña del Mar, Chile}
\emailAdd{tessio.melo@uvm.cl}

\author[a]{Salvador Urrea}
\emailAdd{salvador.urrea@ijclab.in2p3.fr}

\abstract{We consider a model which provides an explanation of the origin of light neutrino masses, the baryon asymmetry of the Universe -- via leptogenesis -- and explains the observed dark matter relic abundance with several components. In this scenario,  Majorana masses of the active neutrinos are produced by the inverse seesaw (ISS) mechanism with the lepton-number-violating mass parameter being dynamically generated at three loops with a novel topology. This model is based on an extension of the Standard Model gauge symmetry by a global $U(1)'$ and a discrete $\mathbb{Z}_2\otimes \mathbb{Z}_3$. The latter, which is responsible for the radiative origin of the ISS lepton-number-violating parameter, survives the spontaneous breaking of the global $U(1)'$ and, at the same time, ensures the stabilization of the dark sector. The lightest particles carrying non-trivial residual charges are stable, becoming potentially viable dark matter candidates. The model complies with bounds and constraints from neutrino data, collider and high-intensity charged lepton flavor-violating observables, as well as dark matter relic density and direct detection. To efficiently explore the model's high-dimensional parameter space and identify phenomenologically viable regions, we perform a global numerical scan using the MultiNest algorithm.
}


\begin{document}
\begin{flushright}
\end{flushright}
\maketitle

\section{Introduction} \label{sec:Introduction}
The absence of a viable dark matter (DM) candidate, together with the problem of generating the tiny masses of the active neutrinos and their mixing, as observed in neutrino oscillation phenomena, constitute compelling indications of physics Beyond the Standard Model (BSM). In addition, the measured baryon asymmetry of the Universe (BAU) cannot be accommodated within the Standard Model (SM). In this work, we propose a model that can successfully address these three major observational problems.

A minimal extension of the SM consisting of the addition of right-handed Majorana neutrinos can generate active neutrino masses through the type-I seesaw mechanism~\cite{Minkowski:1977sc, Yanagida:1979as, Glashow:1979nm, Mohapatra:1979ia, Gell-Mann:1979vob, Schechter:1980gr, Schechter:1981cv}. However, in the conventional type-I seesaw, successfully accommodating neutrino data typically requires the right-handed Majorana neutrinos to be either extremely heavy or very weakly coupled to the SM leptons through tiny Dirac neutrino Yukawa couplings. In both cases, the mixing between active and heavy neutrinos is strongly suppressed, leading to charged lepton flavor violating (CLFV) rates several orders of magnitude below the current experimental sensitivity. This makes conventional type-I seesaw realizations very difficult to probe through CLFV observables. Furthermore, extending the SM only by right-handed Majorana neutrinos does not provide a viable explanation for the observed DM relic abundance.

A possible way to realize a low-scale seesaw scenario with sizable Yukawa couplings is to introduce additional sterile fermions together with an approximate lepton number symmetry, as in the inverse seesaw (ISS) mechanism~\cite{Wyler:1982dd, Mohapatra:1986bd, Gonzalez-Garcia:1988okv, GonzalezGarcia:1988rw, Akhmedov:1995ip, Akhmedov:1995vm, Malinsky:2005bi, Malinsky:2009df, Abada:2014vea}. In the ISS, active neutrino masses are generated at tree level and are controlled by a small lepton-number-violating (LNV) Majorana mass parameter. The smallness of this parameter suppresses the active neutrino masses while still allowing for comparatively large Yukawa couplings and sizable active-heavy neutrino mixing. Although this small LNV parameter is technically natural in the sense of ’t Hooft~\cite{tHooft:1979rat}, since lepton number symmetry is restored in the limit in which it vanishes, its origin remains unexplained in the minimal ISS setup. In this sense, the ISS lacks a more fundamental motivation for the smallness of the LNV scale.
 
Many extensions of the SM propose a radiative origin for the Majorana masses of new neutral fermions, or more generally for the LNV sources responsible for neutrino mass generation. These constructions typically rely on preserved discrete symmetries that forbid the corresponding tree-level mass terms but allow them to arise at loop level~\cite{Balakrishna:1988ks, Ma:1988fp, Ma:1989ys, Ma:1990ce, Ma:1998dn, Tao:1996vb, Ma:2006km, Gu:2007ug, Ma:2008cu, Hirsch:2013ola, Aranda:2015xoa, Restrepo:2015ura, Longas:2015sxk, Fraser:2015zed, Fraser:2015mhb, Wang:2015saa, Arbelaez:2016mhg, vonderPahlen:2016cbw, Nomura:2016emz, Kownacki:2016hpm,Abada:2025ozu, Nomura:2017emk, Nomura:2017vzp, Bernal:2017xat, Wang:2017mcy, Bonilla:2018ynb, Calle:2018ovc, Avila:2019hhv, CarcamoHernandez:2018aon, Alvarado:2021fbw, Mandal:2021yph, Arbelaez:2022ejo, Cepedello:2022xgb, CarcamoHernandez:2022vjk, Leite:2023gzl, Kumar:2025cte, Kumar:2025zvv}. Such models naturally connect DM with neutrino mass generation and are commonly referred to as ``scotogenic'' models. \reply{Some examples of scotogenic models where active neutrino masses arise at three loop level are provided in Refs.~\cite{Krauss:2002px, Aoki:2008av, Kajiyama:2013lja, Ahriche:2014cda, Ahriche:2014oda, Hatanaka:2014tba, Chen:2014ska, Jin:2014glp, Okada:2015hia, Nishiwaki:2015iqa, Ahriche:2015wha, CarcamoHernandez:2015hjp, Gu:2016xno, Cheung:2017efc, Dutta:2018qei, CarcamoHernandez:2019cbd, Cepedello:2020lul, Hernandez:2021iss, Abada:2022dvm, Bonilla:2023wok, Abada:2023znk, Abada:2023zbb, Loualidi:2025jqn}}

This radiative approach can be naturally combined with the ISS mechanism. In such scotogenic ISS constructions, the small LNV mass term that controls the active neutrino masses is not introduced by hand but is instead generated at loop level. Extensions of the SM embedding this idea~\cite{Wyler:1982dd, Mohapatra:1986bd, GonzalezGarcia:1988rw, Akhmedov:1995ip, Akhmedov:1995vm, Malinsky:2005bi, Ma:2009gu, Malinsky:2009df, Bazzocchi:2010dt, Law:2012mj, Das:2012ze, Okada:2012np, Abada:2014vea, Fraser:2014yha, Ahriche:2016acx, CarcamoHernandez:2013krw, Das:2017ski, CarcamoHernandez:2017owh, CarcamoHernandez:2018hst, CarcamoHernandez:2018iel, Bertuzzo:2018ftf, Mandal:2019oth, Das:2019pua, CarcamoHernandez:2019eme, CarcamoHernandez:2019pmy, CarcamoHernandez:2019vih, CarcamoHernandez:2019lhv, Hernandez:2021uxx, Hernandez:2021xet, Hernandez:2021kju, Nomura:2021adf, Hernandez:2021mxo, Abada:2021yot, Abada:2023zbb, Bonilla:2023egs, Bonilla:2023wok, Binh:2024lez, Abada:2024xpf, deMelo:2024lux, Pathak:2024sei, Wang:2024qhe, CarcamoHernandez:2024hll, Huong:2025uwx} provide interesting and testable explanations for the smallness of active neutrino masses, with the LNV parameter typically arising at the one-, two- or three-loop level. This radiative origin offers a dynamical explanation for the small LNV scale while preserving the main phenomenological advantages of the ISS. In particular, sizable active-heavy neutrino mixing can be achieved, allowing CLFV processes to reach rates within the sensitivity of next-generation experiments. Moreover, these models establish a natural connection between DM and neutrino mass generation, since some of the neutral seesaw messengers can serve as viable DM candidates. Their stability is guaranteed by preserved discrete symmetries, which are also crucial for implementing the radiative seesaw mechanism. The annihilation of the DM candidates into SM particles, as well as into other BSM states, can successfully reproduce the observed DM relic abundance in suitable regions of parameter space. Furthermore, ISS realizations can also provide a viable framework for resonant leptogenesis owing to the small mass splitting between the heavy pseudo-Dirac neutral leptons induced by the ISS structure.

In this work, we propose a new extension of the SM in which active neutrino masses are generated through an ISS mechanism, which is radiatively induced at the three-loop level with  a novel topology, thereby providing a new scotogenic origin for the ISS mechanism. The SM particle content is enlarged by several gauge-singlet scalars and neutral leptons, and the model is endowed with a global $U(1)'$ symmetry, which is spontaneously broken, together with a preserved discrete $\mathbb{Z}_2\otimes \mathbb{Z}_3$ symmetry. This extension allows the LNV Majorana mass term to be generated radiatively at the three-loop level and, at the same time, ensures the stability of the dark sector: the lightest particles carrying non-trivial residual charges cannot decay into SM states and can therefore serve as viable DM candidates. The model thus accommodates scalar and/or fermionic DM candidates, giving rise to several multi-component DM scenarios with up to three stable DM components. We perform a detailed phenomenological and cosmological analysis, imposing the relevant constraints from neutrino oscillation data, charged lepton flavor violation searches, DM relic abundance, direct detection limits, and the present observed baryon asymmetry of the Universe, assuming it to be generated via leptogenesis. We identify viable regions of parameter space in which multi-component DM scenarios remain compatible with current experimental bounds while successfully reproducing the observed relic abundance, and we assess their potential for resonant leptogenesis.

The main component of the numerical analysis relies on exploring the possibilities allowed within our model by performing a scan over the fundamental parameters of the theory and identifying the regions of parameter space that can successfully reproduce the measured DM relic abundance, $\Omega h^2 \simeq 0.12$~\cite{Planck:2018vyg}, using micrOMEGAs~\cite{Belanger:2001fz,Alguero:2023zol,Belanger:2026asz}. The novelty of our work lies in the use of the MultiNest algorithm~\cite{Feroz:2008xx,Feroz:2013hea}, which allowed us to efficiently sample the multidimensional parameter space of the model instead of performing a fully fine-grained scan over all the model parameters, which would have been computationally inefficient and prohibitively time-consuming.

The remainder of the paper is organized as follows. Section~\ref{sec:model} presents the field content, symmetries, the scalar sector, and the neutrino-mass mechanism. Section~\ref{constraints} summarizes the constraints used in the phenomenological analysis, including CLFV expressions and experimental sensitivities. Section~\ref{cosmo} discusses multi-component DM regimes and the implications for resonant leptogenesis. Section~\ref{sec:interplay} presents the interplay between CLFV and cosmological observables. We conclude in Section~\ref{Sec:conclusions}. Technical details of the numerical analysis, together with the loop functions entering the CLFV observables and neutrino masses, are collected in Appendices~\ref{Appendix:mu-loop}, \ref{app:formfactors}, and~\ref{app:multinest_and_importance}.

\section{The model} \label{sec:model}
In this section, we construct a model providing a new scotogenic origin of the ISS mechanism. It is radiatively induced at the three-loop level with a novel topology not used before for the ISS mechanism. The particle content is enlarged by the gauge-singlet scalar fields $\varphi_1$, $\varphi_2$, $\eta$, and $\sigma$, as well as by two generations of right-handed neutrinos $\nu_{1R},\nu_{2R}$, sterile fermions $N_{1R},N_{2R}$, and by two pairs of vector-like neutral leptons $\Psi_{kR},\Psi_{kL}$, with $k = 1, 2$. The SM gauge symmetry is supplemented by a global $U(1)'$ symmetry and a discrete $\mathbb{Z}_2 \otimes \mathbb{Z}_3$ symmetry. The model scalar potential develops a tree-level instability, generating vacuum expectation values (VEVs) $\langle \phi \rangle =v_\phi $ and $\langle \sigma \rangle =v_\sigma $, which break the symmetry in the following way:
\begin{align}
    & SU(3)_C\otimes SU(2)_L\otimes U(1)_Y\otimes U(1)'\otimes \mathbb{Z}_2\otimes \mathbb{Z}_3 \nonumber \\
    & \hspace{30mm}\Downarrow v_{\sigma }  \notag \\
    & SU(3)_C\otimes SU(2)_L\otimes U(1)_Y\otimes \mathbb{Z}_2\otimes \mathbb{Z}_3 \nonumber\\
    & \hspace{30mm}\Downarrow v_{\phi }  \notag \\
    & SU(3)_C\otimes U(1)_{\text{em}}\ \otimes \mathbb{Z}_2\otimes \mathbb{Z}_3\,, \nonumber
\end{align}
where $\phi $ corresponds to the SM Higgs boson. The global $U(1)'$ symmetry is spontaneously broken at the TeV scale, the scalar field $\sigma$ acquiring its classical solution $\langle \sigma \rangle =v_\sigma$, while the discrete $\mathbb{Z}_2\otimes \mathbb{Z}_3$ symmetry remains preserved. The other new scalars $\varphi_1$, $\varphi_2$ and $\eta$, which carry non-trivial $\mathbb{Z}_2\otimes \mathbb{Z}_3$ charges, are inert and therefore do not acquire VEVs. 


\reply{
The three inert scalars play complementary roles in realizing the specific
three-loop topology (see Feynman diagram of Fig.~\ref{Neutrinodiagram}) responsible for the radiative generation of the ISS
lepton-number-violating parameter $\mu$. The field $\varphi_1$ connects
$N_R$ to the vector-like fermions $\Psi$ through the Yukawa interaction
$y_N\overline{\Psi}_L\varphi_1N_R$, whereas $\eta$ couples to the $\Psi$
sector through $y_\Psi\overline{\Psi_R^C}\eta\Psi_R$, with the
$A_\eta\eta^3$ interaction being required to close the scalar part of the
three-loop diagram. The field $\varphi_2$ connects the $\varphi_1$ and
$\eta$ lines through the $A_1\varphi_1\varphi_2\eta^\ast$ interaction.
Furthermore, the coupling
$\lambda_{16}\varphi_2^2\sigma^{\ast2}+{\rm H.c.}$ communicates the
spontaneous $U(1)'$ breaking to the loop by generating the required
$\varphi_{2R}$--$\varphi_{2I}$ mass splitting. 
Their inert nature is a consequence of the preserved $\mathbb Z_2\otimes\mathbb Z_3$ symmetry, which guarantees the stability of the dark-sector candidates.
}


The charge assignments of the model fields under the extended symmetry of our model are shown in Table~\ref{model}. Given this particle content and symmetries, the following neutrino Yukawa terms are allowed:
\begin{align}\label{Yukawa-neutrino}
    -\mathcal{L}_Y^{\left( \nu \right) } &= \sum_{i=1}^{3}\sum_{k=1}^2\left(y_{\nu }\right)_{ik}\overline{l}_{iL}\widetilde{\phi }\nu_{kR}+\sum_{n=1}^2\sum_{k=1}^2M_{nk}\overline{\nu^C _{nR}}N_{kR}+ \sum_{n=1}^2\sum_{k=1}^2\left( y_N \right) _{nk}\overline{\Psi }_{kL}\varphi _1N_{nR}  \notag \\
    %
&\qquad +\reply{
\sum_{n=1}^2\sum_{k=1}^2
\left(y_{\nu\Psi}\right)_{nk}
\overline{\nu_{nR}^{C}}\,
\varphi_1^\ast\Psi_{kR}
+   
\sum_{n=1}^2\sum_{k=1}^2
\left(x_{\Psi}\right)_{nk}
\overline{\Psi}_{nL}\Psi_{kL}^{C}\eta^\ast
}\notag \\
&\qquad+\sum_{n=1}^2\sum_{k=1}^2\left( y_{\Psi }\right)_{nk}\overline{\Psi _{kR}^{C}}\eta~ \Psi _{nR} +\sum_{n=1}^2\sum_{k=1}^2\left( m_{\Psi}\right) _{nk}\overline{\Psi }_{nL}\Psi _{kR} + {\rm H.c.}
\end{align}
\begin{table}[t]
\renewcommand{\arraystretch}{1.3} \centering%
\begin{tabular}{|c||c|c|c|c|c|c|c|c|c|c|c|}
\hline
Field & $l_{iL}$ & $l_{iR}$ & $\nu _{kR}$ & $N_{kR}$ & $\Psi _{kR}$ & $\Psi
_{kL}$ & $\phi $ & $\varphi _1$ & $\varphi _2$ & $\eta $ & $\sigma $ \\ 
\hline\hline
$SU(3)_C$ & $\mathbf{1}$ & $\mathbf{1}$ & $\mathbf{1}$ & $\mathbf{1}$ & $%
\mathbf{1}$ & $\mathbf{1}$ & $\mathbf{1}$ & $\mathbf{1}$ & $\mathbf{1}$ & $\mathbf{1}$
& $\mathbf{1}$ \\ \hline
$SU(2)_L$ & $\mathbf{2}$ & $\mathbf{1}$ & $\mathbf{1}$ & $\mathbf{1}$ & $%
\mathbf{1}$ & $\mathbf{1}$ & $\mathbf{2}$ & $\mathbf{1}$ & $\mathbf{1}$ & $\mathbf{1}$
& $\mathbf{1}$ \\ \hline
$U(1)_Y$ & $-\frac{1}{2}$ & $-1$ & $0$ & $0$ & $0$ & $0$ & $\frac{1}{2}$ & 
$0$ & $0$ & $0$ & $0$ \\ \hline
$U(1)'$ & $0$ & $1$ & $-1$ & $1$ & $0$ & $0$ & $-1$ & $-1$ & $1$ & $0$ & $1$ \\ \hline
$\mathbb{Z}_2$ & $0$ & $0$ & $0$ & $0$ & $1$ & $1$ & $0$ & $1$ & $1$ & $0$ & $0$ \\ 
\hline
$\mathbb{Z}_3$ & $0$ & $0$ & $0$ & $0$ & $2$ & $2$ & $0$ & $2$ & $0$ & $2$ & $0$ \\ 
\hline
\end{tabular}%
\caption{Particle charge assignments under the $SU(3)_C \otimes SU(2)_L \otimes U(1)_Y \otimes U(1)' \otimes \mathbb{Z}_2 \otimes \mathbb{Z}_3$ symmetry. Here $i=1,2,3$ and $k=1,2$.}
\label{model}
\end{table}

\subsection{The scalar potential}
We now discuss the scalar sector of the model. As described above, only the scalar fields $\phi$ and $\sigma$ acquire VEVs, while the remaining scalar singlets are inert in order to preserve the residual symmetry $\mathbb{Z}_2\otimes \mathbb{Z}_3$. The most general renormalizable scalar potential invariant under the charge assignments shown in Table~\ref{model} is given by
\begin{align} \label{potential}
    V & = -\mu_\phi ^2 \phi^\dag \phi - \mu_\sigma^2 \sigma^* \sigma + \mu_{\varphi 1}^2  \varphi_1^* \varphi_1  + \mu_{\varphi 2}^2  \varphi_2^* \varphi_2 + \mu_\eta^2 \eta^* \eta \nonumber \\
    &\quad + \lambda_1 \left(\phi^\dag \phi\right)^2 + \lambda_2 \left(\sigma^* \sigma\right)^2 + \lambda_3 \left(\varphi_1^* \varphi_1\right)^2  +  \lambda_4 \left(\varphi_2^* \varphi_2\right)^2   +  \lambda_5 \left(\eta^* \eta\right)^2  +  \lambda_6 \left(\phi^\dag \phi\right) \left(\sigma^* \sigma\right) \nonumber \\
    &\quad +  \lambda_7 \left(\phi^\dag \phi\right) \left(\varphi_1^* \varphi_1\right)  +  \lambda_8 \left(\phi^\dag \phi\right) \left(\varphi_2^* \varphi_2\right)  +  \lambda_9 \left(\phi^\dag \phi\right) \left(\eta^* \eta\right) +  \lambda_{10} \left(\sigma^* \sigma\right) \left(\varphi_1^* \varphi_1\right)  \nonumber \\
    &\quad  +  \lambda_{11} \left(\sigma^* \sigma\right) \left(\varphi_2^* \varphi_2\right)  +   \lambda_{12} \left(\sigma^* \sigma\right) \left(\eta^* \eta\right)  +  \lambda_{13} \left(\varphi_1^* \varphi_1\right) \left(\varphi_2^* \varphi_2\right) +   \lambda_{14} \left(\varphi_1^* \varphi_1\right) \left(\eta^* \eta\right)  \nonumber \\
    &\quad + \lambda_{15} \left(\varphi_2^* \varphi_2\right) \left(\eta^* \eta\right) + \lambda_{16} \left(\varphi_2^2 {\sigma^*}^2 + {\varphi_2^*}^2 {\sigma}^2\right)  +  \lambda_{17} \left(\varphi_1 \varphi_2 \eta^2 + \varphi_1^* \varphi_2^* {\eta^*}^2\right) \nonumber \\
    &\quad + A_1 \left(\varphi_1 \varphi_2 \eta^* + \varphi_1^* \varphi_2^* \eta\right) + A_\eta  \left(\eta^3 + {\eta^*}^3\right).
\end{align}
Here, the trilinear couplings $A_1$ and $A_\eta $ are dimensionful parameters, while the remaining couplings are dimensionless. For simplicity, all parameters in the scalar potential are taken to be real.

The scalar fields that acquire VEVs are expanded around the electroweak and $U(1)'$-breaking vacua as
\be 
    {\phi} = \frac{1}{\sqrt{2}} \left(
    \begin{array}{c}
        \phi^{+} \\ 
        \phi^0_R + v_\phi +  i~\phi^0_I 
    \end{array}
    \right) \,\, , \,\, \sigma  = \frac{1}{\sqrt{2}} (\sigma_R + v_\sigma + i~\sigma_I ), 
\ee
whereas the other three singlets that do not acquire VEVs are expanded in components as
\be
    \varphi_1 = \frac{1}{\sqrt{2}} ({\varphi_1}_R + i~ {\varphi_1}_I) \,\, , \,\,  \varphi_2 = \frac{1}{\sqrt{2}} ({\varphi_2}_R + i~ {\varphi_2}_I) \,\, , \,\,  \eta = \frac{1}{\sqrt{2}} (\eta_R + i~\eta_I)\,.
\ee

Once the $U(1)'$ and the electroweak symmetries are broken by the VEVs of $\sigma$ and $\phi$, respectively, the scalar spectrum consists of:
\begin{itemize}
    \item Five massive CP-even scalars. Two of them arise from the CP-even fields $\phi^0_R $ and $\sigma_R $, which mix through the Higgs-portal interaction proportional to $\lambda_6$; see Eq.~\eqref{potential}. In the basis $(\phi^0_R ,\sigma_R )$, the corresponding mass-squared matrix is
    \be
        \begin{pmatrix}
            2\lambda_1v_\phi ^2 & \lambda_6v_\phi v_\sigma   \\ 
            \lambda_6v_\phi v_\sigma  & 2\lambda_2v_\sigma ^2 
        \end{pmatrix}.
    \ee
    In this work, we take the scalar alignment limit, motivated by the stringent LHC constraints on deviations from the SM Higgs properties~\cite{ATLAS:2022vkf, CMS:2022dwd, ParticleDataGroup:2024cfk}. In practice, this amounts to suppressing the Higgs--singlet mixing controlled by the portal coupling $\lambda_6$, which we set to $\lambda_6 = 0$, while fixing $\lambda_1=m_h^2/(2v_\phi ^2)\simeq 0.13$ for $m_h\simeq 125~\mathrm{GeV}$ and $v_\phi \simeq 246~\mathrm{GeV}$. Therefore, the observed Higgs boson remains SM-like, $h\simeq \phi^0_R $. The remaining three CP-even scalars are $\varphi_{1R}$, $\varphi_{2R}$, and $\eta_R $, whose tree-level masses are given by
    \begin{align}
        m_{\varphi_{1R}}^2 &= \mu_{\varphi_1}^2 +\frac{\lambda_7}{2}v_\phi ^2 +\frac{\lambda_{10}}{2}v_\sigma ^2\,,\\
        \label{eq:varphi_2R}
        m_{\varphi_{2R}}^2 &= \mu_{\varphi_2}^2 +\frac{\lambda_8}{2}v_\phi ^2 +\frac{\lambda_{11}}{2}v_\sigma ^2 +\lambda_{16}v_\sigma ^2\,,\\
        m_{\eta_R }^2 &= \mu_\eta ^2 +\frac{\lambda_9}{2}v_\phi ^2 +\frac{\lambda_{12}}{2}v_\sigma ^2\,.
    \end{align}
    \item Three massive CP-odd scalars, corresponding to $\varphi_{1I}$, $\varphi_{2I}$ and $\eta_I $, whose tree-level masses are given by
    \begin{align}
        m_{\varphi_{1I}}^2 &= \mu_{\varphi_1}^2 +\frac{\lambda_7}{2}v_\phi ^2 +\frac{\lambda_{10}}{2}v_\sigma ^2\,,\\
        \label{eq:varphi_2I}
        m_{\varphi_{2I}}^2 &= \mu_{\varphi_2}^2 +\frac{\lambda_8}{2}v_\phi ^2 +\frac{\lambda_{11}}{2}v_\sigma ^2 -\lambda_{16}v_\sigma ^2\,,\\
        m_{\eta_I }^2 &= \mu_\eta ^2 +\frac{\lambda_9}{2}v_\phi ^2 +\frac{\lambda_{12}}{2}v_\sigma ^2\,.
    \end{align}
    Notice that at tree level, the CP-even and CP-odd components of both  $\varphi_1$ and $\eta$ are degenerate, while the degeneracy in the $\varphi_2$ sector is lifted by the term $\lambda_{16}$, which gives opposite contributions to $m_{\varphi_{2R}}^2$ and $m_{\varphi_{2I}}^2$ after $U(1)'$ breaking.
    \end{itemize}
    
    Before concluding this section, we briefly discuss the nature and possible phenomenological implications of the Goldstone boson present in our model. There are two massless states in the CP-odd sector: the would-be-Goldstone $G_Z\sim \phi^0_I$ absorbed by the $Z$-boson and one physical  Goldstone boson $J\sim \sigma_I$. The latter - despite being a sterile state from the SM gauge symmetry - may have various phenomenological and cosmological implications depending on the quark sector details of the model, which is not specified in our present work, since we focused on the neutrino mass generation and the DM candidates. Nevertheless, we can cleanly split the possible realizations of the $J$-boson into two cases: 
    \begin{enumerate}[label=\Alph*.]
    \item Exact Nambu--Goldstone boson (true NGB): symmetry is exact and anomaly-free. In this case the NGB is massless $m_J = 0$ and the dominant constraints typically arise from~\cite{ParticleDataGroup:2024cfk}: SN1987A and stellar cooling (quark/nucleon and possible $\nu\nu J$ couplings)~\cite{Raffelt:1996wa, Kachelriess:2000qc, Farzan:2002wx, Keung:2013mfa}; dark radiation through $\Delta N_{\rm eff}$ if $J$ thermalizes~\cite{Weinberg:2013kea}; collider bounds from $h\to JJ$ via $\phi$--$\sigma$ portal mixing~\cite{Bonilla:2015uwa}.
    \item Pseudo-Nambu--Goldstone boson (pNGB): symmetry is explicitly broken by mixed SM--$U(1)'$ anomalies, by soft ad hoc Lagrangian terms, or by higher-dimensional operators. This case differs in its phenomenology from the previous true NGB one. For example, a QCD-anomalous $U(1)'$ gives rise to an axion-like pNGB whose mass is generated by non-perturbative QCD effects~\cite{Peccei:1977hh, Peccei:1977ur, Weinberg:1977ma, Wilczek:1977pj, DiLuzio:2020wdo}. More generally, soft-breaking terms or Planck-suppressed operators can modify the pNGB potential, with implications for axion quality, neutrino physics, cosmology, and DM~\cite{Barr:1992qq, Holman:1992us, Kamionkowski:1992mf, Akhmedov:1992hi, Rothstein:1992rh, Gu:2010ys, deGiorgi:2023tvn, Georis:2025kzv, Sheng:2025sou}.
\end{enumerate}

In our model, the mixed $[SU(3)_c]^2-U(1)'$ and $[U(1)_{\rm em}]^2-U(1)'$ anomalies cancel automatically, with the latter cancellation including the charged-lepton contribution. The $[SU(2)_L]^2-U(1)'$ and $[U(1)_Y]^2-U(1)'$ anomalies can also be canceled by assigning vanishing $U(1)'$ charges to the left-handed quark doublets. Therefore, within the renormalizable theory, $J$ is an exact Nambu--Goldstone boson and remains massless. 

%

%
\reply{
For completeness, we briefly discuss whether the phenomenological and
cosmological constraints on the massless Nambu--Goldstone boson $J$ can
affect the parameter regions considered below.

For the anomaly-free quark charge assignment $X'(Q_L)=0$, invariance of
the SM Yukawa interactions implies
\begin{equation}
    X'(u_R) = -1 \qquad\text{and} \qquad X'(d_R) = 1\,.
\end{equation}
For the purpose of identifying the physical Goldstone direction, the
global generator can be redefined by adding a multiple of the gauged
hypercharge generator. It is convenient to introduce
\begin{equation}
    \widetilde X=X'+2Y\,.
\end{equation}
With this choice,
\begin{equation}
    \widetilde X(\phi)=0 \qquad\text{and}\qquad \widetilde X(\sigma)=1\,,
\end{equation}
so that the SM Higgs vacuum expectation value does not break
$\widetilde U(1)$, while $\langle\sigma\rangle$ does. Therefore, after
removing the electroweak Goldstone mode, the physical Nambu--Goldstone
boson is identified at tree level with
\begin{equation}
    J\simeq\sigma_I\,.
\end{equation}

An important consequence of this charge assignment is that the charged
SM fermions transform vectorially under $\widetilde U(1)$. Indeed,
\begin{equation}
    \widetilde X(e_L)=\widetilde X(e_R)=-1\,,
\end{equation}
whereas in the quark sector
\begin{equation}
    \widetilde X(Q_L) = \widetilde X(u_R) = \widetilde X(d_R) = \frac13\,.
\end{equation}
The derivative Goldstone interaction with a Dirac fermion $f$ can be
written schematically as
\begin{equation}
    {\cal L}_{Jff} = \frac{\partial_\mu J}{2v_\sigma} \left[ (\widetilde X_L+\widetilde X_R)\, \overline f\gamma^\mu f + (\widetilde X_R-\widetilde X_L)\, \overline f\gamma^\mu\gamma_5 f \right].
\end{equation}
Since $\widetilde X_L=\widetilde X_R$ for the charged SM fermions, the
axial part of the Goldstone current vanishes. Consequently, the usual
tree-level pseudoscalar interactions
$J\overline f i\gamma_5 f$ are absent. The standard stellar-cooling and
supernova constraints associated with single-Goldstone emission through
tree-level electron or nucleon pseudoscalar couplings are therefore
strongly suppressed~\cite{Raffelt:1996wa, Keung:2013mfa}.
Possible loop-induced effects are not considered in the present analysis.

Another potentially relevant constraint arises from invisible Higgs
decays, which can provide an important probe of Majoron-like models~\cite{Bonilla:2015uwa}. In the present analysis we work in the scalar
alignment limit $\lambda_6=0$, which is consistent with current experimental measurements~\cite{Fischer:2026bka}. Since the tree-level $hJJ$ interaction is
generated by the Higgs--singlet portal term
$\lambda_6(\phi^\dagger\phi)(\sigma^\ast\sigma)$, one has
\begin{equation}
    g_{hJJ}^{\rm tree}=0 \qquad (\lambda_6=0),
\end{equation}
and therefore the decay $h\to JJ$ does not constrain the parameter space
considered here at tree level.

In the neutrino sector, the coupling of $J$ to the light neutrinos follows
the usual Majoron scaling~\cite{Chikashige:1980ui, Schechter:1981cv},
\begin{equation}
    g_{J\nu_i\nu_j} \sim \frac{(m_\nu)_{ij}}{v_\sigma}.
    \label{eq:Jnunucoupling}
\end{equation}
For the parameter range considered in our scan,
$v_\sigma\gtrsim100~{\rm GeV}$, this gives parametrically
\begin{equation}
    g_{J\nu\nu} \lesssim {\cal O}(10^{-12})
\end{equation}
for sub-eV neutrino masses. This value lies well below the region probed
by supernova constraints on neutrino--Majoron interactions~\cite{Farzan:2002wx, Ivanez-Ballesteros:2024nwsz}.
Hence, the neutrino coupling of $J$ does not provide a relevant additional
constraint on the parameter space explored here.

The remaining potentially relevant cosmological effect is the
contribution of a thermalized massless $J$ to dark radiation. The
Goldstone boson interacts with the inert scalar sector through the
quartic interactions proportional to $\lambda_{10}$, $\lambda_{11}$, $\lambda_{12}$ and $\lambda_{16}$, which generate annihilation processes such as
$\Phi\Phi^\ast\to JJ$, already included in our dark-matter analysis;
see Fig.~\ref{fig:annihilation_scalar}. If $J$ reaches thermal
equilibrium and subsequently decouples while still relativistic, entropy
conservation gives~\cite{Weinberg:2013kea}
\begin{equation}
    \Delta N_{\rm eff} = \frac47 \left[ \frac{g_{*s}(T_\nu^{\rm dec})}{g_{*s}(T_J^{\rm dec})} \right]^{4/3},
    \label{eq:DeltaNeffJ}
\end{equation}
where $T_J^{\rm dec}$ denotes the Goldstone decoupling temperature and
$g_{*s}$ is the effective number of entropy degrees of freedom.

The QCD crossover provides a useful reference scale for this estimate,
because $g_{*s}$ decreases rapidly below this epoch. If $J$ decouples
above the QCD crossover, one typically has
\begin{equation}
    g_{*s}(T_J^{\rm dec})\gtrsim60\,,
\end{equation}
and therefore
\begin{equation}
    \Delta N_{\rm eff} \lesssim \frac47 \left(\frac{10.75}{60}\right)^{4/3} \simeq0.06.
\end{equation}
The subsequent entropy release into the SM plasma thus dilutes the
temperature of the already decoupled Goldstone relative to the neutrinos.
Such a contribution is compatible with present cosmological constraints
on additional relativistic degrees of freedom~\cite{Planck:2018vyg}.

Conversely, if $J$ remained in thermal equilibrium significantly below the
QCD crossover, the smaller value of $g_{*s}(T_J^{\rm dec})$ would imply a
larger $\Delta N_{\rm eff}$, and the dark-radiation constraint could become
relevant. In the parameter region studied here, an early decoupling is
plausible because the particles communicating the Goldstone sector with
the thermal bath have electroweak- or TeV-scale masses and become
Boltzmann suppressed as the temperature decreases. Nevertheless, a
precise determination of $T_J^{\rm dec}$ over the full parameter scan
would require a dedicated Boltzmann analysis, which is beyond the scope
of the present work. If the relevant portal interactions are too weak to
bring $J$ into thermal equilibrium in the first place, its contribution
to $\Delta N_{\rm eff}$ is smaller, assuming a negligible primordial
Goldstone abundance.

We therefore conclude that the most immediate constraints from invisible
Higgs decays, charged-fermion couplings, and neutrino--Majoron interactions
do not significantly affect the parameter space considered in this work.
The dark-radiation constraint is also mild for Goldstone decoupling above
the QCD crossover, while a dedicated study of the late-decoupling region
would be required for a fully quantitative determination of
$\Delta N_{\rm eff}$.
}


\subsection{Neutrino mass matrix}
\begin{figure}[t!]
    \centering
    \includegraphics[width=0.55\textwidth]{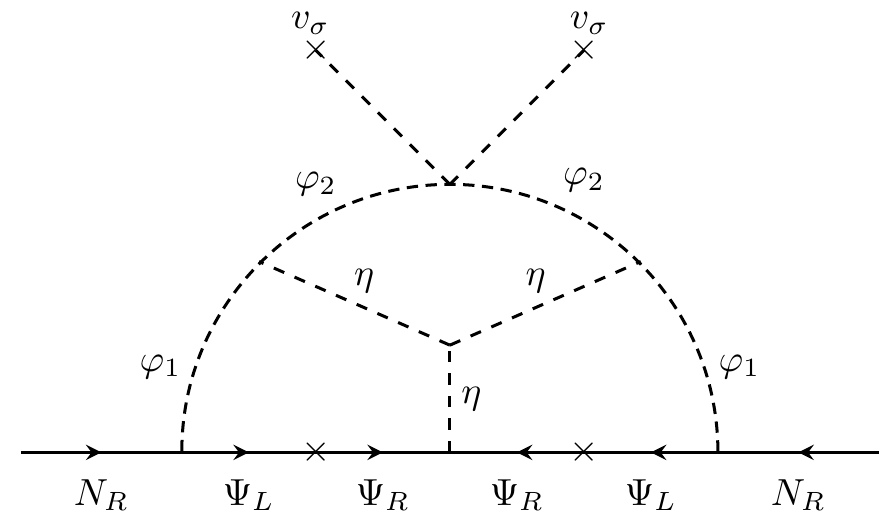} 
    \caption{Three-loop diagram for $2\times 2$  LNV mass matrix $\mu$.}
    \label{Neutrinodiagram}
\end{figure}
We now discuss the generation of active neutrino masses. Based on the Yukawa Lagrangian of Eq.~\eqref{Yukawa-neutrino}, and after electroweak symmetry breaking, the neutral-fermion mass matrix relevant for the light active neutrino mass generation can be written as
\begin{equation}
    -\mathcal{L}_{\mathrm{mass}}=\frac{1}{2}
    \begin{pmatrix}
        \bar{\nu_L} &  & \bar{\nu_R^C} &  & \bar{N_R^C}%
    \end{pmatrix}
    \begin{pmatrix}
        0 &  & M_D &  & 0 \\ 
        {M_D^T} &  & \mu^{\prime} &  & M \\ 
        0 &  & M^T &  & \mu%
    \end{pmatrix}%
    \begin{pmatrix}
        \nu_L \\ 
        \nu_R \\ 
        N_R%
    \end{pmatrix}
    +\text{H.c.}\, ,
    \label{nuLang} 
\end{equation}
%
%
where generation indices are suppressed.  Here, $M_D$ is a $3\times 2$ matrix given by $M_D = y_\nu v_\phi/\sqrt{2}$, while $M$ is a $2\times 2$ matrix. 
The entries $\mu$ and $\mu'$ are the LNV Majorana mass $2\times 2$ matrices of $N_R$ and $\nu_R$, respectively. 
In the ISS limit, these mass matrices satisfy the hierarchy $M\gg M_D\gg \mu,\mu^{\prime}$. The matrix $M$ corresponds to a bare mass term, while the small LNV terms $\mu$ and $\mu'$ are generated in our model at the three-loop level 
through the same topology
proportional to the products of the Yukawa couplings $y_N y_\Psi$ and $y_{\nu \Psi} x_\Psi$, respectively.  
An important property of the generalized inverse-seesaw structure in Eq.~\eqref{nuLang} is that
$\mu'$ does not contribute to the leading-order light-neutrino mass matrix, as shown for example in Refs.~\cite{Law:2013gma, Hernandez:2021uxx}. Its main contribution will be to the heavy-neutrino spectrum, where it will enter in the same way as $\mu$ and it will not affect the phenomenology considered here. 
In the present phenomenological analysis we restrict ourselves
to the minimal-ISS parameter subspace
$y_{\nu\Psi}=x_\Psi=0$, and consequently set $\mu'=0$.
The generalized case with $\mu'\neq0$ is left for future work.
    %
%

The leading diagram contributing to the LNV term  $\mu$ is shown in Fig.~\ref{Neutrinodiagram}.
    The corresponding expression  is~\cite{Hatanaka:2014tba}
\begin{equation}
\label{eq:mu-loop}
    \mu _{nk} = \frac{4\, A_1^2\, A_\eta }{\left( 4\pi \right) ^{6}M^{4}}\sum_{r=1}^2\sum_{s=1}^2\left(y_N \right) _{nr}\left( m_{\Psi }\right) _r \left( y_{\Psi }\right)_{rs}\left( m_{\Psi }\right) _{s}\left( y_N ^T\right) _{sk}\, G^{rs}_{\rm loop}\,,
\end{equation}
where $G^{rs}_{\rm loop}$ is a loop function given by
\begin{equation}
    G^{rs}_{\rm loop} \equiv F\Big(\frac{m_{\varphi_1}^2}{M^2},\frac{m_\eta ^2}{M^2}, \frac{m_{\Psi_r}^2}{M^2},\frac{m_{\Psi_s}^2}{M^2}, \frac{m_{{\varphi_2}_R}^2}{M^2}\Big) - F\Big(\frac{m_{\varphi_1}^2}{M^2},\frac{m_\eta ^2}{M^2}, \frac{m_{\Psi_r}^2}{M^2},\frac{m_{\Psi_s}^2}{M^2}, \frac{m_{{\varphi_2}_I}^2}{M^2}\Big)\,,
\end{equation}
with the explicit form of loop integrals $F$ shown in Appendix~\ref{Appendix:mu-loop}. The mass scale is defined as $M = \max[m_{\varphi_1},m_\eta, m_{\Psi_{r}}, m_{\varphi_{2R,I}}]$. The loop function $G^{rs}_{\rm loop}$ vanishes in the limit of the degenerate $\varphi_2$ sector. As seen from Eqs.~\eqref{eq:varphi_2R} and~\eqref{eq:varphi_2I}, the mass squared difference is non-zero in our model $\Delta m^2_{\varphi_2} = 2\lambda_{16}v^2_\sigma$ after the $U(1)'$ symmetry breaking. This is just the contribution of the quartic vertex in the diagram of Fig.~\ref{Neutrinodiagram}. 
%
%
%
%
The mass matrix in Eq.~\eqref{nuLang} can be conveniently rewritten as
\begin{equation}
    \begin{pmatrix}
        0 &   {\hat{M}}_D  \\ 
        {\hat{M}}_D^T &   \hat{M} 
    \end{pmatrix} ,
    \label{mseesaw}
\end{equation}
where ${\hat{M}}_D = (M_D \,\,\, 0_{3 \times 2})$ is a $3 \times 4$ matrix, and the heavy-neutrino mass matrix $\hat{M}$ is defined as
\begin{equation}
    \hat{M} =
    \begin{pmatrix}
        0 &   M  \\ 
        M^T &   \mu
    \end{pmatrix}.
\end{equation}
Since we are in the seesaw limit, the above block matrix can be diagonalized by the full unitary mixing matrix $\mathbb{U}$, which, keeping terms up to $\mathcal{O}\big({\hat{M}}_D^2/\hat{M}^2\big)$, is given by~\cite{Grimus:2000vj}
\begin{equation}
    \mathbb{U} = WT =
    \begin{pmatrix}
        \left(1-\frac{1}{2} \epsilon\right)U_\nu &   {\hat{M}}_D^* (\hat{M}^{-1})^* U_R  \\ 
        -\hat{M}^{-1}{\hat{M}}_D^T U_\nu &   \left(1-\frac{1}{2} \epsilon'\right)U_R
    \end{pmatrix} ,
    \label{seewsawdiagonal}
\end{equation}
where $T$ is defined by $T=\textrm{diag}(U_\nu,U_R)$, $U_\nu$ and $U_R$ being the unitary matrices diagonalizing $M_{\mathrm{light}}$ and $\hat{M}$, respectively. In the above equation, the matrices $\epsilon$ and $\epsilon'$ are
\begin{equation}
    \epsilon = {\hat{M}}_D^* (\hat{M}^{-1})^* \hat{M}^{-1} {\hat{M}}_D^T = {M_D}^* ({M}^{-1})^* {M^T}^{-1} {M_D}^T ,
    \label{nonunitarity}
\end{equation}
and
\begin{equation}
    \epsilon' = \hat{M}^{-1} {\hat{M}}_D^T  {\hat{M}}_D^* (\hat{M}^{-1})^* .
\end{equation}
In Eq.~\eqref{seewsawdiagonal}, the unitary matrix $W$ brings Eq.~\eqref{mseesaw} to the block-diagonal form
\begin{equation}
    \begin{pmatrix}
        M_{\mathrm{light}} &   0  \\ 
        0 &   \hat{M} 
    \end{pmatrix},
\end{equation}
with the light-neutrino mass matrix given by
\begin{equation}
    M_{\mathrm{light}} \approx -{\hat{M}}_D \hat{M}^{-1} {\hat{M}}_D^T = M_D (M^T)^{-1} \mu M^{-1} M_D^T \,.
\end{equation}

Since we consider the minimal ISS with two generations each of $\nu_R$ and $N_R$, the lightest active neutrino is massless. The heavy-neutrinos spectrum instead exhibits two pairs of nearly degenerate pseudo-Dirac neutrinos $N^\pm_i$ ($i=1,2$), with masses of the order $M\pm \mu$, respectively. In Eq.~\eqref{seewsawdiagonal}, the analog of the Pontecorvo-Maki-Nakagawa-Sakata (PMNS) mixing matrix for active light neutrinos is given by $\left(1-\frac{1}{2} \epsilon\right)U_\nu$, where the presence of $\epsilon$ encodes the deviation from unitarity induced by mixing between light and heavy neutrinos. 

\section{Constraints}\label{constraints}
In this section, we summarize the experimental and theoretical constraints imposed in our numerical analysis. We require the model to reproduce the observed pattern of light neutrino masses and leptonic mixing, while satisfying the bounds from neutrinoless double beta decay, perturbativity, tree-level stability of the scalar potential, charged lepton flavor-violating observables, and deviation from unitarity of the effective PMNS matrix.

\subsection{Neutrino oscillation data and neutrinoless double beta decay} \label{neutrinos-parametrization}
We first require the model to reproduce the neutrino mass-squared differences and leptonic mixing angles obtained from global fits to neutrino oscillation data~\cite{Esteban:2024eli} at the $3\sigma$ level. This is implemented through a modified Casas-Ibarra parametrization~\cite{Casas:2001sr}, adapted to the ISS mechanism, with the Dirac neutrino mass matrix written as
\be
    m_D =  U_\nu \Big(M_{\rm light}^{\rm diag}\Big)^{1/2} R~ \mu^{-1/2} M^T.
\ee
In the above equation, $M_{\rm light}^{\rm diag}$ corresponds to the diagonal light neutrino mass matrix, which in the case of the normal hierarchy (NH) and the inverted hierarchy (IH) of the active light neutrino mass spectrum
is given as
\begin{align}
    \left(M_{\text{light}}^{\text{diag}}\right)_{\text{NH}} &= \operatorname{diag}\left( \, \sqrt{\Delta m_{\text{sol}}^2+\Delta m_{\text{atm}}^2},\, \sqrt{\Delta m_{\text{sol}}^2},0 \right),\\
    \left(M_{\text{light}}^{\text{diag}}\right)_{\text{IH}} &= \operatorname{diag}\left( 0,\, \sqrt{\Delta m_{\text{sol}}^2+\Delta m_{\text{atm}}^2},\, \sqrt{\Delta m_{\text{atm}}^2}\right).
\end{align}
Here, $\Delta m_{\rm sol}^2$ and $\Delta m_{\rm atm}^2$ are the solar and atmospheric mass-squared differences, respectively. The matrix $U_\nu$ is the unitary light-neutrino mixing matrix, parameterized in terms of the three mixing angles $\theta_{12}$, $\theta_{13}$ and $\theta_{23}$, the Dirac CP-violating phase $\delta_{\rm CP}$, and one physical Majorana phase $\alpha_M$. The matrix $R$ is an orthogonal complex $3 \times 2$ matrix parameterized by a complex parameter $z$ and, for NH and IH, is given by
\be
    R_{NH} =
    \begin{pmatrix}
        \textrm{cos}(z)  &  -\textrm{sin}(z)\\
        \textrm{sin}(z) & \textrm{cos}(z)\\
        0 & 0
    \end{pmatrix}
    \,\,\, ; \,\,\,  R_{IH} = 
    \begin{pmatrix}
        0 &   0  \\ 
        \textrm{cos}(z)  &  -\textrm{sin}(z)\\
        \textrm{sin}(z) & \textrm{cos}(z)
    \end{pmatrix}.
\ee

We also verify that the model accounts for the bounds on neutrinoless double beta decay effective mass. In the present ISS realization, neutrinoless double beta decay is mediated by the standard neutrino mass mechanism, with an amplitude proportional to the effective mass parameter~\cite{Kovalenko:2009td, Faessler:2014kka} 
\begin{equation} \label{mee3p2}
    m_{ee} = \bigg|\sum_{i=1}^N \mathbb{U}_{ei}^2\, p^2 \dfrac{m_i}{p^2+m_i^2}\bigg|\,,
\end{equation}
where the sum runs over the Majorana neutrino mass eigenstates of the model. Here, $p^2$ is usually interpreted as the mean square of the Fermi momentum of the nucleon in the decaying nucleus. Its value depends on the isotope and on the nuclear structure model used to calculate the corresponding nuclear matrix elements. In our analysis, we take the average value for the experimentally relevant isotopes, $p^2 \simeq -(150~\mathrm{MeV})^2$~\cite{Faessler:2014kka, Babic:2018ikc}.

The physical neutrino spectrum consists of three light active neutrinos, the lightest being massless, and two pairs of heavy pseudo-Dirac neutrinos with opposite CP parities. The two states in each pseudo-Dirac pair give contributions with opposite relative signs, which cancel in the lepton-number-conserving limit. The residual heavy-neutrino contribution is therefore controlled by the small LNV parameter $\mu$ and is negligible for the parameter space considered here. Thus, in practice, neutrinoless double beta decay does not impose an additional relevant constraint on the heavy pseudo-Dirac sector, and the dominant contribution comes from the light neutrinos. The effective mass then reduces to
\begin{equation}\label{meefp2}
    m_{ee} =\left|\sum_{i=1}^3 \mathbb{U}_{ei}^2\, p^2 \dfrac{m_i}{p^2+m_i^2}\right| \equiv \left|\sum_{i=1}^3 \mathbb{U}_{ei}^2\, m_i\right|\ .
\end{equation}
We require the effective Majorana mass to satisfy the most stringent KamLAND-Zen bound, $m_{ee}<36~\mathrm{meV}$ at 90\% C.L.~\cite{KamLAND-Zen:2022tow}.

\subsection{Perturbativity and tree-level stability of the scalar potential}
In our numerical analysis, we require all couplings to remain in the perturbative regime and, therefore, we impose
\begin{equation}
    |\lambda_i| < 4\pi\,, \hspace{1cm} |y_{ij}| < \sqrt{4\pi}\,,
\end{equation}
for all quartic scalar couplings and Yukawa couplings, respectively. In addition, the scalar potential has to be bounded from below in order to guarantee a stable electroweak vacuum at tree level. Since the large-field behavior of the potential is controlled by quartic interactions, tree-level stability conditions can be obtained by analyzing the quartic part of the scalar potential. For this purpose, we introduce the following Hermitian bilinear combinations of scalar fields:
\begin{eqnarray}
a &=&\phi ^{\dagger }\phi ,\hspace{1cm}b=\sigma ^{\ast }\sigma ,\hspace{1cm}%
c=\varphi _1^{\ast }\varphi _1,\hspace{1cm}d=\varphi _2^{\ast }\varphi
_2,\hspace{1cm}e=\eta ^{\ast }\eta ,  \notag \\
f &=&\varphi _1\varphi _2+\varphi _1^{\ast }\varphi _2^{\ast },%
\hspace{1cm}g=i\left( \varphi _1\varphi _2-\varphi _1^{\ast }\varphi
_2^{\ast }\right) , \notag\\
p &=&\eta ^2+\left( \eta ^{\ast }\right) ^2,\hspace{1cm}q=i\left[ \eta
^2-\left( \eta ^{\ast }\right) ^2\right] \notag\\
r &=&\varphi _2\sigma ^{\ast }+\varphi _2^{\ast }\sigma ,\hspace{1cm}s=i%
\left[ \varphi _2\sigma ^{\ast }-\varphi _2^{\ast }\sigma \right] ,
\end{eqnarray}
allowing to write the quartic scalar interactions in Eq.~\eqref{potential} as:
\begin{align}
V_4=& \lambda _1a^2+\lambda _2b^2+\lambda _3c^2+\lambda
_4d^2+\lambda _5e^2+\lambda _6ab+\lambda _7ac+\lambda
_8ad+\lambda _9ae  \notag \\
& +\lambda _{10}bc+\lambda _{11}bd+\lambda _{12}be+\lambda _{13}cd+\lambda
_{14}ce+\lambda _{15}de  \notag \\
& +\frac{1}{4}\lambda _{16}\left[ \left( r-is\right) ^2+\left( r+is\right)
^2\right] +\frac{1}{4}\lambda _{17}\left[ \left( f-ig\right) \left(
p-iq\right) +\left( f+ig\right) \left( p+iq\right) \right]. 
\end{align}
For the purpose of the analysis of the stability of the potential, we can write the quartic potential as follows:
\begin{align}
    V_4=& \left( \sqrt{\lambda _1}a-\sqrt{\lambda _2}b\right) ^2+\left( \sqrt{\lambda _1}a-\sqrt{\lambda _3}c\right) ^2+\left( \sqrt{\lambda_1}a-\sqrt{\lambda _4}d\right) ^2 +\left( \sqrt{\lambda _1}a-\sqrt{\lambda _5}e\right) ^2 \notag \\
    &+\left( \sqrt{\lambda _2}b-\sqrt{\lambda _3}c\right) ^2+\left( \sqrt{\lambda _2}b-\sqrt{\lambda _4}d\right) ^2 +\left( \sqrt{\lambda _2}b-\sqrt{\lambda _5}e\right) ^2+\left( \sqrt{\lambda _3}c-\sqrt{\lambda _4}d\right) ^2 \notag \\
    & +\left( \sqrt{\lambda _3}c-\sqrt{\lambda _5}e\right) ^2  +\left( \sqrt{\lambda _4}d-\sqrt{\lambda _5}e\right) ^2+\left(\lambda _6+2\sqrt{\lambda _1\lambda _2}\right) ab+\left( \lambda _7+2\sqrt{\lambda _1\lambda _3}\right) ac  \notag \\
    & +\left( \lambda _8+2\sqrt{\lambda _1\lambda _4}\right) ad+\left(\lambda _9+2\sqrt{\lambda _1\lambda _5}\right) ae+\left( \lambda_{10}+2\sqrt{\lambda _2\lambda _3}\right) bc  \notag \\
    & +\left( \lambda _{11}+2\sqrt{\lambda _2\lambda _4}\right) bd+\left(\lambda _{12}+2\sqrt{\lambda _2\lambda _5}\right) be+\left( \lambda_{13}+2\sqrt{\lambda _3\lambda _4}\right) cd  \notag \\
    & +\left( \lambda _{14}+2\sqrt{\lambda _3\lambda _5}\right) ce+\left(\lambda _{15}+2\sqrt{\lambda _4\lambda _5}\right) de  \notag \\
    & -3\left( \lambda _1a^2+\lambda _2b^2+\lambda _3c^2+\lambda_4d^2+\lambda _5e^2\right)  +\frac{1}{2}\lambda _{16}\left( r^2-s^2\right) +\frac{1}{2}\lambda_{17}\left( fp-gq\right).
\end{align}
Finally, and following the procedure used for analyzing the stability of multi-scalar potentials described in Refs.~\cite{Maniatis:2006fs, Bhattacharyya:2015nca, Abada:2021yot, Hernandez:2021kju}, we require the quartic potential to be positive along all the relevant large-field directions. This leads to the following tree-level bounded-from-below conditions:
\begin{equation}
\begin{aligned}
&\lambda_1 \geq 0,\quad
\lambda_2 \geq 0,\quad
\lambda_3 \geq 0,\quad
\lambda_4 \geq 0,\quad
\lambda_5 \geq 0,\quad
\lambda_{16} \geq 0,\quad
\lambda_{17} \geq 0, \\[2mm]
&\lambda_6+2\sqrt{\lambda_1\lambda_2} \geq 0,\quad
\lambda_7+2\sqrt{\lambda_1\lambda_3} \geq 0,\quad
\lambda_8+2\sqrt{\lambda_1\lambda_4} \geq 0, \\[2mm]
&\lambda_9+2\sqrt{\lambda_1\lambda_5} \geq 0,\quad
\lambda_{10}+2\sqrt{\lambda_2\lambda_3} \geq 0,\quad
\lambda_{11}+2\sqrt{\lambda_2\lambda_4} \geq 0, \\[2mm]
&\lambda_{12}+2\sqrt{\lambda_2\lambda_5} \geq 0,\quad
\lambda_{13}+2\sqrt{\lambda_3\lambda_4} \geq 0,\quad
\lambda_{14}+2\sqrt{\lambda_3\lambda_5} \geq 0, \\[2mm]
&\lambda_{15}+2\sqrt{\lambda_4\lambda_5} \geq 0.
\end{aligned}
\end{equation}

\subsection{Charged lepton flavor violation} \label{sec:clfv}
Despite years of dedicated experimental searches, so far no clear indication of rare lepton flavor-violating decays has been observed~\cite{Calibbi:2017uvl}. However, it should be noted that these experiments have reached impressive levels of precision~\cite{SINDRUM:1987nra, MEG:2013oxv, MEGII:2025gzr} and are expected to achieve even greater sensitivity in the coming years, improving their projected yield by several orders of magnitude in some cases~\cite{Baldini:2013ke, Blondel:2013ia}. Such transitions can exhibit significant rates in low-scale seesaw models that explain the smallness of neutrino masses, including the model proposed here. This section explores the impact of the presence of the new fields (scalars and fermion SM singlets)  on the rates of CLFV observables. Conversely, the present CLFV bounds have an important consequence as they place severe constraints on the model.

The most constraining CLFV observables are from the muon-sector: radiative decay $\mu\to e\gamma$, coherent $\mu-e$ conversion in the nuclei, and three-body decay $\mu\to eee$. These observables probe different combinations of the same flavor-violating parameters. The decay $\mu\to e\gamma$ is controlled by the dipole amplitude, whereas $\mu-e$ conversion and $\mu\to eee$ also receive important non-dipole photon-penguin, $Z$-penguin and box contributions. Their combined study is therefore useful not only to constrain the model but also to identify the dominant CLFV operator structure.

The current limits and representative future sensitivities used in our discussion and analysis are presented in Table~\ref{tab:clfv_bounds}. Note that in our analysis, we impose the current limits in the numerical scan, while projected sensitivities are shown as future benchmarks. For the coherent conversion $\mu-e$, the strongest present bound is obtained using a gold target, whereas the most relevant future projections correspond to aluminum targets for COMET and Mu2e.
\begin{table}[t!]
\begin{center}
\small
\renewcommand{\arraystretch}{1.28}
\begin{tabular}{c | cc}
\hline \hline
Observable & Current bound & Future sensitivity \\
\hline\hline
$\operatorname{Br}(\mu\to e\gamma)$
&
MEG II: $<1.5\times10^{-13}$~\cite{MEGII:2025gzr}
&
MEG II: $\sim 6\times10^{-14}$~\cite{MEGII:2025gzr}
\\ \hline
$\operatorname{CR}(\mu-e,\mathrm{Au})$
&
SINDRUM II: $<7.0\times10^{-13}$~\cite{SINDRUMII:2006dvw}
&
-- 
\\
$\operatorname{CR}(\mu-e,\mathrm{Al})$
&
--
&
\begin{tabular}{@{}c@{}}
COMET: $\sim 1.0\times10^{-16}$~\cite{COMET:2009qeh} \\
Mu2e: $\sim (0.8\text{--}6.2)\times10^{-16}$~\cite{Mu2e:2022ggl,Bernstein:2019fyh}
\end{tabular}
\\ \hline
$\operatorname{Br}(\mu\to eee)$
&
SINDRUM: $<1.0\times10^{-12}$~\cite{SINDRUM:1987nra}
&
\begin{tabular}{@{}c@{}}
Mu3e: $\sim 10^{-16}$~\cite{Mu3e:2020gyw,Hesketh:2022wgw}
\end{tabular}
\\
\hline \hline
\end{tabular}
\end{center}
\caption{Present upper bounds and projected future sensitivities for the muon-sector CLFV observables considered in this work.}
\label{tab:clfv_bounds}
\end{table}

In what follows, we present the predictions of the model for these CLFV
observables.

\subsubsection[$\ell_i\to \ell_j\gamma$]{\boldmath$\ell_i\to \ell_j\gamma$}
The admixture of the heavy sterile neutrinos in the left-handed charged-current weak interaction induces the radiative decay $\ell_i\to \ell_j\gamma$ at one-loop level. The corresponding branching ratio is given by~\cite{Langacker:1988up, Lavoura:2003xp, Hue:2017lak, CarcamoHernandez:2020pnh, Bonilla:2023egs, Bonilla:2023wok, CarcamoHernandez:2023atk, Batra:2023mds}
\begin{align}
\operatorname{Br}\!\left(\ell_i \to \ell_j \gamma\right)
&=
\frac{\alpha_W^3 s_W^2 m_{\ell_i}^5}
     {256\pi^2 m_W^4 \Gamma_i}
\left|G_{ij}\right|^2,
\label{eq:Br_ligamma}
\\[2mm]
G_{ij}
&=
\sum_{a=1}^N 
\mathbb{U}_{ja}\mathbb{U}_{ia}^{*}\,
G_\gamma\!\left(x_a\right),
\qquad
x_a \equiv \frac{m_{n_a}^2}{m_W^2},
\label{eq:Gij_ligamma}
\\[2mm]
G_\gamma(x)
&\equiv
\frac{
10-43x+78x^2-49x^3
+18x^3\ln x+4x^4
}{
12(1-x)^4
}.
\label{eq:Ggamma_ligamma}
\end{align}
Here $\ell_i$ and $\ell_j$ ($i\ne j$) denote charged leptons with flavors $i,j=e,\mu,\tau$, $\alpha_W=g_W^2/(4\pi)$ is the weak fine-structure constant, $s_W$ is the sine of the weak mixing angle, $m_W$ is the mass of the boson $W$, $m_{\ell_i}$ is the mass of the decaying charged lepton and $\Gamma_i$ is its total decay width. In particular, for the muon we use $\Gamma_\mu \simeq 3\times10^{-19}~\mathrm{GeV}$~\cite{ParticleDataGroup:2024cfk}.

\subsubsection[$\text{CR}(\mu -e, \text{N})$]{\boldmath $\text{CR}(\mu -e, \text{N})$}
The coherent $\mu-e$ conversion 
occurs when a negative muon is captured by a nucleus denoted by $N$, forms a muonic atom, and converts into an electron without changing the nuclear state. The conversion rate is defined as
\begin{equation}
    \label{eq:CR:def}
    \operatorname{CR}(\mu -e, \mathrm{N})
    \equiv
    \frac{\Gamma (\mu^- + N \to e^- +N)}
    {\Gamma (\mu^- + N \to \text{all})}\, .
\end{equation}
In contrast to $\mu\to e\gamma$, this observable receives contributions from dipole, photon-penguin, $Z$-penguin and box diagrams, leading to the following rate:
\begin{align} \label{eq:conv}
    \operatorname{CR}(\mu-e, \mathrm{N}) &= \frac{32\, G_F^2\, \alpha_W^2\, m_\mu^5}{(4\pi)^2\, \Gamma_\text{capt}(Z)} \nonumber\\
    &\qquad \times \left|V^{(p)} \left(2\, \tilde{F}_{u}^{\mu e} + \tilde{F}_{d}^{\mu e}\right) + V^{(n)} \left(\tilde{F}_u^{\mu e} + 2\, \tilde{F}_{d}^{\mu e}\right) + D\, G^{\mu e}_{\gamma} \frac{s_W^2}{8 \sqrt{4 \pi\, \alpha}}\right|^2,
\end{align}
where $\Gamma_\text{capt}(Z)$ denotes the muon capture rate for a nucleus with atomic number $Z$~\cite{Kitano:2002mt}, $G_F$ is the Fermi constant, $m_\mu$ is the muon mass and $\alpha \equiv e^2/(4\pi)$. The form factors $\tilde{F}_{q}^{\mu e}$, with $q=u,d$, are given by
\begin{equation} \label{eq:tildeFqmue}
    \tilde F_q^{\mu e} = Q_q \, s_W^2 F^{\mu e}_\gamma + F^{\mu e}_Z \left(\frac{{I}^3_q}{2}-Q_q\, s_W^2\right) + \frac14 F^{\mu eqq}_\text{box}\,,
\end{equation}
where $Q_q$ denotes the electric charge of the quark $q$, $Q_u=2/3$ and $Q_d=-1/3$, while ${I}^3_q$ is the weak isospin, ${I}^3_u=1/2$ and ${I}^3_d=-1/2$. The quantities $F^{\mu e}_\gamma$, $F^{\mu e}_Z$, $F^{\mu eqq}_\text{box}$ and $G^{\mu e}_\gamma$ denote the photon-penguin, $Z$-penguin, box and dipole form factors, respectively. Their explicit expressions are collected in Appendix~\ref{app:formfactors}. The nuclear dependence is encoded in the overlap integrals $D$, $V^{(p)}$ and $V^{(n)}$, for which we use the numerical values given in Ref.~\cite{Kitano:2002mt}.

\subsubsection[$\ell_\alpha\to \ell_\beta\ell_\beta\ell_\rho$]{\boldmath $\ell_\alpha\to \ell_\beta\ell_\beta\ell_\rho$}
Three-body charged-lepton decays provide an additional complementary test of the same flavor-violating interactions entering $\mu\to e\gamma$ and coherent $\mu-e$ conversion. In the muon sector, the most relevant channel is $\mu\to eee$, which receives photon-penguin, $Z$-penguin and box contributions. The branching ratio used in our analysis is~\cite{Ilakovac:1994kj, Alonso:2012ji}
\begin{align} \label{eq:mueee}
    \operatorname{Br}(\mu \to eee) =&\, \frac{\alpha_W^4}{24576\pi^3} \frac{m^4_\mu}{M^4_W} \frac{m_\mu}{\Gamma_\mu} \Bigg\{ 2 \left|\frac12F^{\mu eee}_{\rm Box} +F^{\mu e}_Z -2s^2_W(F^{\mu e}_Z-F^{\mu e}_\gamma) \right|^2  \nonumber \\
    &+4 s^4_W \left|F^{\mu e}_Z-F^{\mu e}_\gamma\right|^2 +16 s^2_W \operatorname{Re}\left[ \left(F^{\mu e}_Z+\frac12F^{\mu eee}_{\rm Box}\right) G^{\mu e*}_\gamma \right] \nonumber \\
    &-48 s^4_W \operatorname{Re}\left[ (F^{\mu e}_Z-F^{\mu e}_\gamma) G^{\mu e*}_\gamma \right]  +32 s^4_W |G^{\mu e}_\gamma|^2 \left[ \ln \frac{m^2_\mu}{m^2_{e}}-\frac{11}{4} \right] \Bigg\}\ .
\end{align}
This expression contains the same form factors entering $\operatorname{CR}(\mu-e,\mathrm{N})$ in Eq.~\eqref{eq:conv}, although in a different combination. Therefore, $\mu\to eee$ is sensitive to both dipole-dominated and contact-interaction-dominated regimes, as discussed, for instance, in Refs.~\cite{Abada:2023zbb, Lindner:2016bgg}.

\subsection{Bounds on the non-unitarity of the PMNS matrix}
One of the notable features of the ISS mechanism is the presence of sizable mixing between the active light neutrinos and the heavy neutrinos, which leads to potentially considerable violation of the unitarity of the PMNS matrix. This can in turn give rise to large rates for lepton-flavor-violating (LFV) decays, such as $\mu \rightarrow e \gamma$, as well as interesting collider signatures. The non-observation of LFV processes, together with constraints from other flavor and electroweak precision observables, sets bounds on the non-unitarity parameter $\epsilon$ defined in Eq.~\eqref{nonunitarity}. The global analysis performed in Ref.~\cite{Blennow:2023mqx} gives the following constraints on $\epsilon$:
\be
    |\epsilon| \leq
    \begin{pmatrix} 
        9.4 \times 10^{-6} &\quad 1.2 \times 10^{-5} &\quad 2.2 \times 10^{-5} \\
        1.2 \times 10^{-5} &\quad 1.3 \times 10^{-4} &\quad 1.3 \times 10^{-4} \\
        2.2 \times 10^{-5} &\quad 1.3 \times 10^{-4} &\quad 2.1 \times 10^{-4}
    \end{pmatrix}.
    \label{eq:nonunitarity_bound}
\ee

The applicability of these limits depends on the heavy-neutrino mass regime. The bounds of Ref.~\cite{Blennow:2023mqx} are obtained in the regime in which the heavy states are not kinematically accessible in the low-energy processes entering the fit, so that their effects can be described by non-unitarity parameters after integrating them out. In particular, the charged-lepton flavor violating constraints entering the fit are treated in the heavy-neutrino limit, with $M_N \gtrsim m_W$ and, more conservatively, are essentially saturated once the heavy-neutrino masses are a few times larger than $m_W$. Therefore, for the mass range considered in our analysis, $M_N \gtrsim \mathcal{O}(200)~{\rm GeV}$, Eq.~\eqref{eq:nonunitarity_bound} provides a good approximation.

\subsection{Collider searches} \label{sec:colliders}
Here, we briefly discuss the main collider signatures of the proposed model. At the LHC, the heavy quasi-Dirac sterile neutrinos can be produced together with a SM charged lepton through quark--antiquark annihilation via the Drell--Yan mechanism. As in the $U(1)^\prime$ scenario studied in~\cite{Deppisch:2013cya}, sterile neutrinos can undergo two-body decays
\begin{equation}
    N \rightarrow \ell_i^\pm W^\mp,
    \qquad
    N \rightarrow \nu_i Z,
    \qquad
    N \rightarrow \nu_i h,
    \qquad i=1,2,3.
\end{equation}
These decay modes are suppressed by the small active--sterile mixing angle, typically of the order $\theta\sim\mathcal{O}(10^{-3})$. Such a small mixing is required to keep CLFV processes below their current experimental bounds and to satisfy constraints arising from deviations from leptonic mixing-matrix unitarity~\cite{Abada:2018nio, Fernandez-Martinez:2016lgt}. The two-body decay modes discussed above subsequently give rise to several three-body final states, including
\begin{equation}
    N\rightarrow \ell_i^{+}\ell_j^{-}\nu_k,
    \qquad
    N\rightarrow \ell_i^{-}u_j\bar{d}_k,
    \qquad
    N\rightarrow b\bar{b}\nu_k,
    \qquad i,j,k=1,2,3,
\end{equation}
where $i$, $j$, and $k$ denote the corresponding flavor indices. Similar three-body decay channels also arise in the $U(1)^\prime$ model considered in Ref.~\cite{Deppisch:2013cya}. A detailed study of the collider signatures of our model goes beyond the scope of the present work and will be done elsewhere.

\section{Cosmological implications} \label{cosmo}
In this section, the productions of multi-component DM through the WIMP mechanism and the BAU via leptogenesis are studied in detail.

\subsection{Multi-component dark matter} \label{Sec:DM}
One of the features of our model is that the same residual symmetry $\mathbb{Z}_2 \otimes \mathbb{Z}_3$ responsible for the radiative origin of the ISS LNV Majorana mass parameter also stabilizes the dark sector. After spontaneous breaking of the $U(1)'$ symmetry by $\langle \sigma\rangle$ and of the electroweak symmetry by $\langle \phi\rangle$, the $\mathbb{Z}_2\otimes \mathbb{Z}_3$ symmetry remains unbroken. Thus, the lightest particles carrying non-trivial residual charges cannot decay into SM states and then turn out to be potential candidates for DM relic abundance. As shown in Table~\ref{tab:dark_sectors}, the unbroken $\mathbb{Z}_2\otimes\mathbb{Z}_3$ symmetry organizes the new neutral fields into three independent dark sectors, the lightest state of each sector being stable.
\begin{table}[t!]
    \renewcommand{\arraystretch}{1.3}
    \centering
    \begin{tabular}{|c||c|c|c|}
        \hline
        Dark sector & $\mathbb{Z}_2$ & $\mathbb{Z}_3$ & Fields \\
        \hline\hline
        I & $1$ & $0$ & $\varphi_2$ \\
        \hline
        II & $0$ & $2$ & $\eta$ \\
        \hline
        III & $1$ & $2$ &
        $\Psi_1,\,\Psi_2,\,\varphi_1$ \\
        \hline
    \end{tabular}
    \caption{Classification of the dark sectors according to their $\mathbb{Z}_2 \otimes \mathbb{Z}_3$ charge assignments. Fields belonging to different sectors cannot mix. The lightest state in each sector may contribute to the DM relic abundance.}
    \label{tab:dark_sectors}
\end{table}

In order to explore all the possibilities allowed within our model, we perform a scan over the fundamental parameters of the theory and identify the regions of the parameter space that can successfully reproduce the measured DM relic abundance, $\Omega h^2 \simeq 0.12$~\cite{Planck:2018vyg}, with the use of micrOMEGAs~\cite{Belanger:2001fz, Alguero:2023zol, Belanger:2026asz}. To efficiently sample this multidimensional parameter space, we employ the MultiNest algorithm~\cite{Feroz:2008xx, Feroz:2013hea}. A fully fine-grained scan over all our model parameters  would be utterly ineffective and would be extremely computationally intensive. Although such a scan could be optimized to reproduce the relic abundance, the resulting points would still have to satisfy all phenomenological requirements, among them accommodating neutrino oscillation data and flavor constraints discussed in Section~\ref{constraints}. To keep the analysis as general as possible while maintaining computational feasibility, we adopt a two-step strategy, described in detail in Appendix~\ref{app:multinest_and_importance}. First, we perform an initial scan that includes all relevant parameters of the model and carry out an importance analysis to determine which parameters have a significant impact on the relic abundance. Based on this analysis, we then select the most relevant parameters and perform a second, more focused scan, keeping the parameters with a lower impact fixed.

\begin{figure}[t!]
\centering
\includegraphics[width=0.95\textwidth]{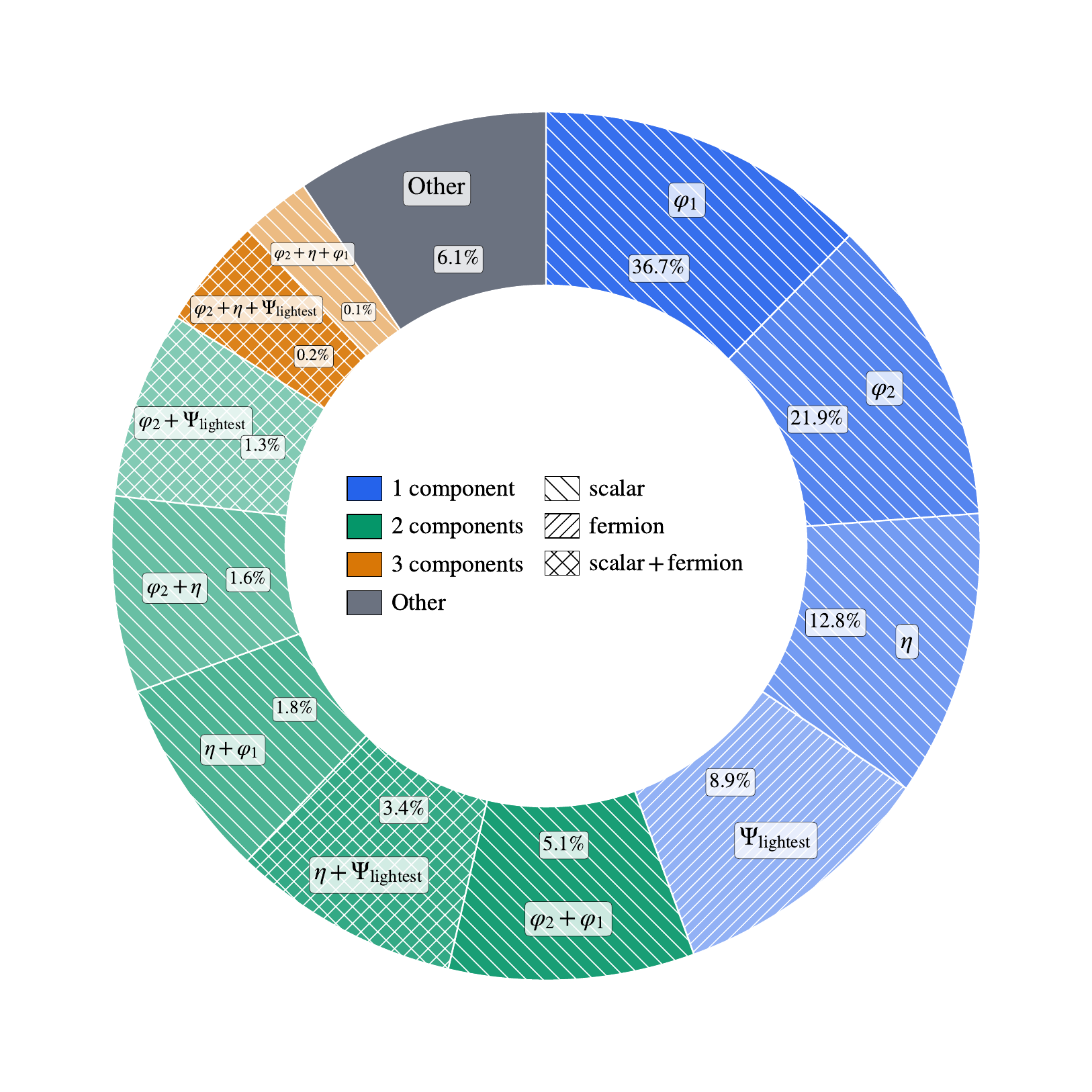}
\caption{Relative composition of the points reproducing the observed DM relic abundance, classified according to their DM candidate content. Each slice represents the fraction of scan points corresponding to a given DM configuration. A single-component-dominance scenario is defined as the one in which a single DM candidate contributes for at least $95\%$ of the total relic abundance. A co-dominance scenario is defined as the one in which all DM components present contribute for at least $10\%$ of the total relic abundance. Points not satisfying these criteria are grouped into the category ``Other''.}
\label{Fig:pie_composition}
\end{figure}

Our model accommodates a wide variety of DM scenarios, including single-component DM, co-dominance between two DM candidates, and co-dominance among three candidates. These scenarios can involve purely scalar, purely fermion 
mixed scalar--fermion, scalar--scalar, scalar--scalar--scalar, and scalar--scalar--fermion DM compositions. The relative occurrence of all viable configurations is summarized in Fig.~\ref{Fig:pie_composition}, which shows the composition of the parameter-space points that reproduce a relic abundance close to the observed value. We define a single-dominance scenario as one in which a single DM component contributes for at least $95\%$ of the total relic DM density, while a co-dominance scenario requires each contributing component to account for at least $10\%$ of the relic abundance.

\subsubsection{Single component domination}
The simplest viable possibility is that only  one stable state dominates the DM relic abundance. In the scan, we identify these points as a single-component DM when one component accounts for at least $95\%$ of the total relic density. From the sector classification in Table~\ref{tab:dark_sectors}, the dominant state can be one of the neutral scalars $\varphi_1$, $\varphi_2$, or $\eta$, or the lightest among the fermions $\Psi_i$. In what follows, we discuss in greater detail the $\varphi_1$- and $\Psi_{\text{lightest}}$-dominated cases, which are representative of the scalar and fermion single-component regimes found in our scan.

\subsubsection*{Scalar dark matter}
\begin{figure}[t]
    \centering
    \includegraphics[width=.28\textwidth]{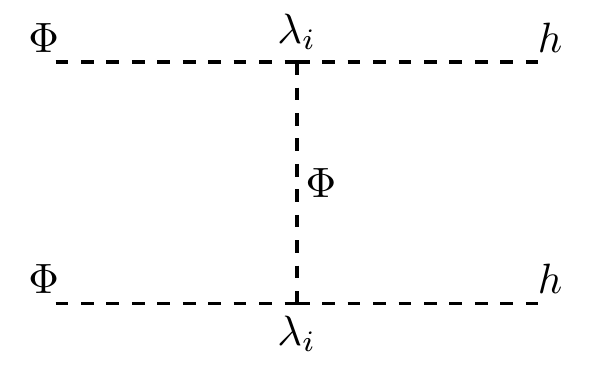}
    \hfill
    \includegraphics[width=.33\textwidth]{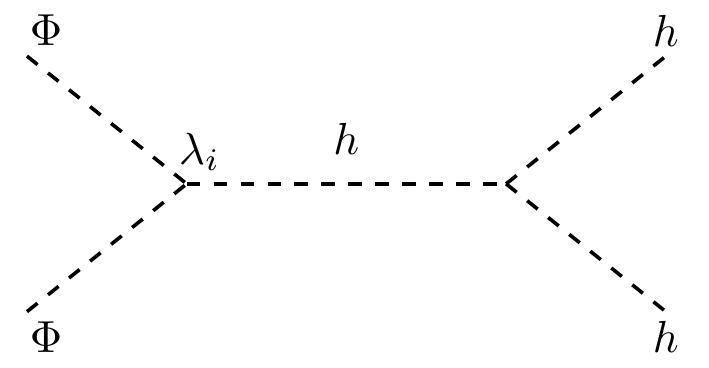}
    \hfill
    \includegraphics[width=.23\textwidth]{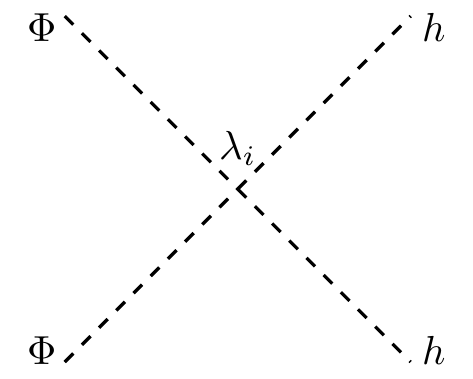}
    \vspace{0.43cm}
    \includegraphics[width=.23\textwidth]{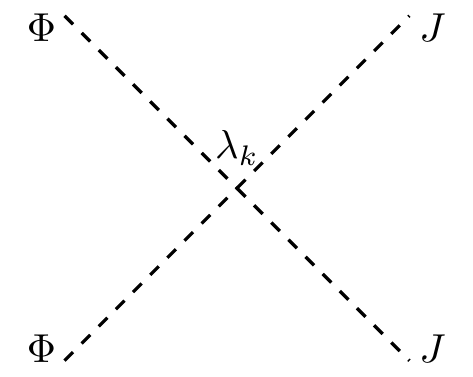}
    \hspace{2cm}
    \includegraphics[width=.36\textwidth]{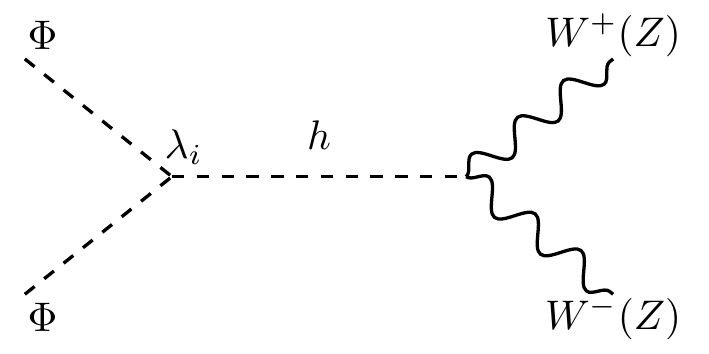}
    \caption{Dominant annihilation channels of scalar DM candidates $\Phi = \varphi_1, \varphi_2, \eta$. Here, $\lambda_i$ and $\lambda_k$ represent the Higgs-portal couplings $\lambda_i = \lambda_7, \lambda_8, \lambda_9$ and Goldstone quartic couplings $\lambda_k = \lambda_{10}, \lambda_{11}, \lambda_{12}$, corresponding to $\varphi_1$, $\varphi_2$, and $\eta$, respectively.}
    \label{fig:annihilation_scalar}
\end{figure}
\begin{figure}[t!]
    \centering
    \includegraphics[width=0.495\linewidth]{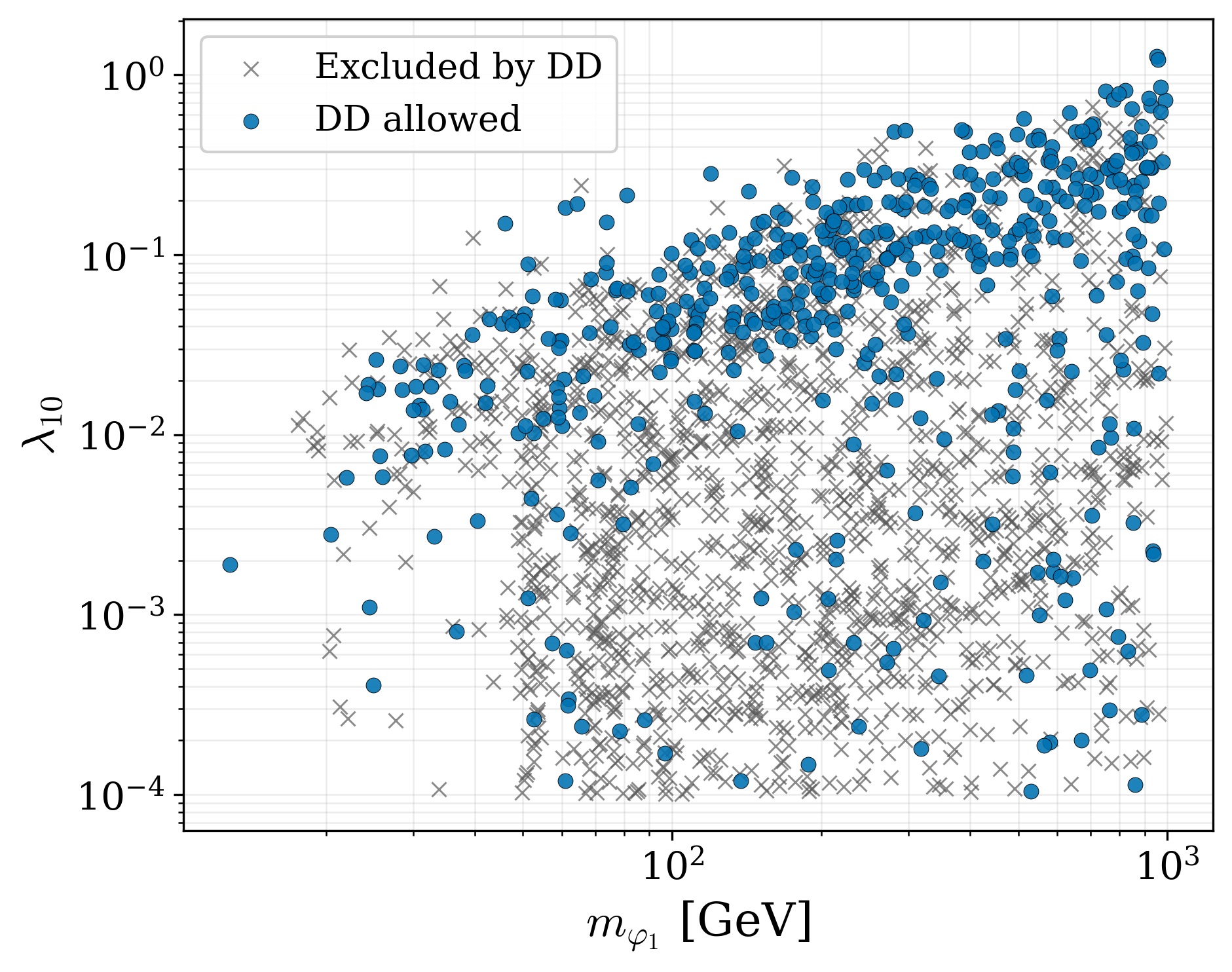}
    \includegraphics[width=0.495\linewidth]{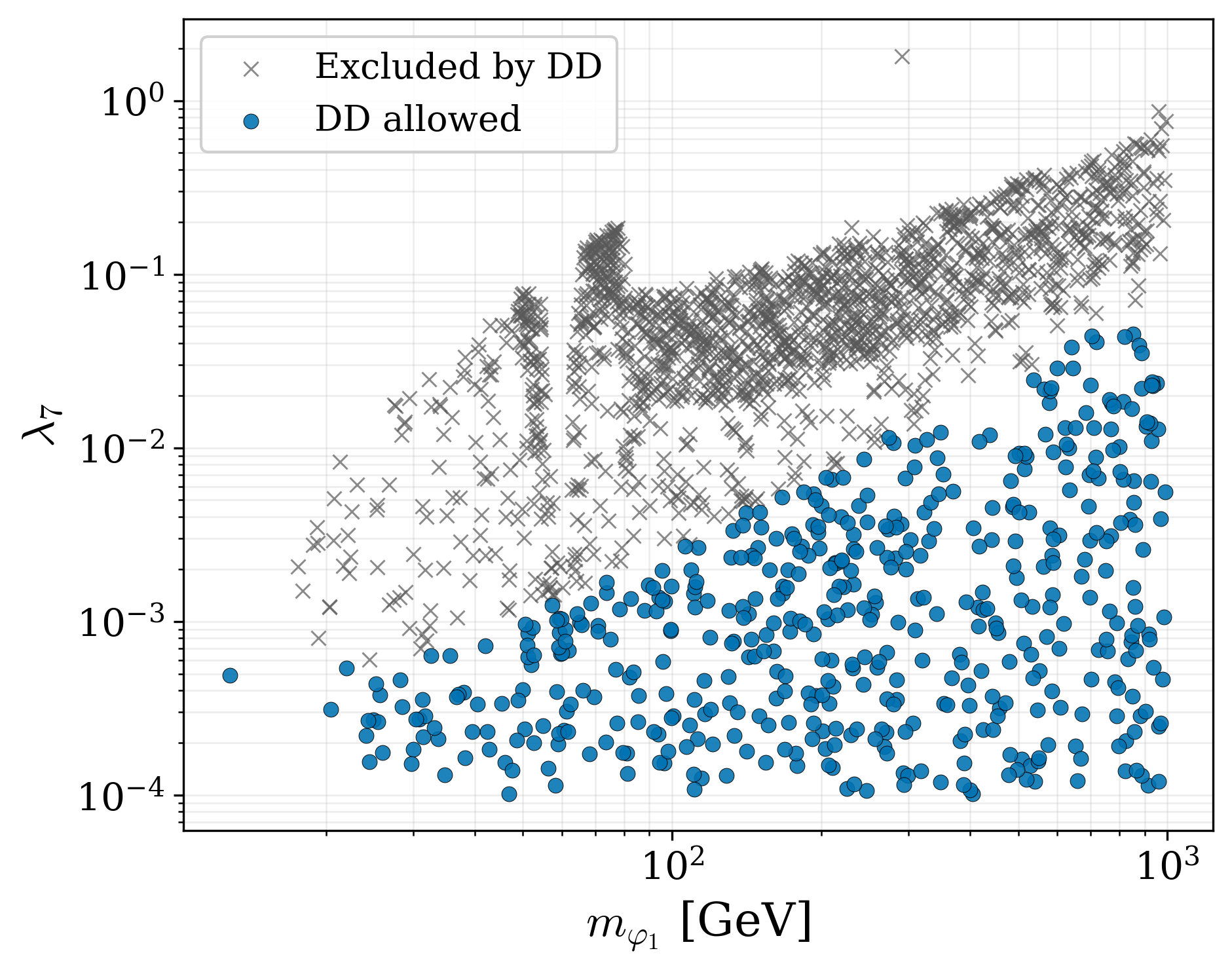}
    \includegraphics[width=0.495\linewidth]{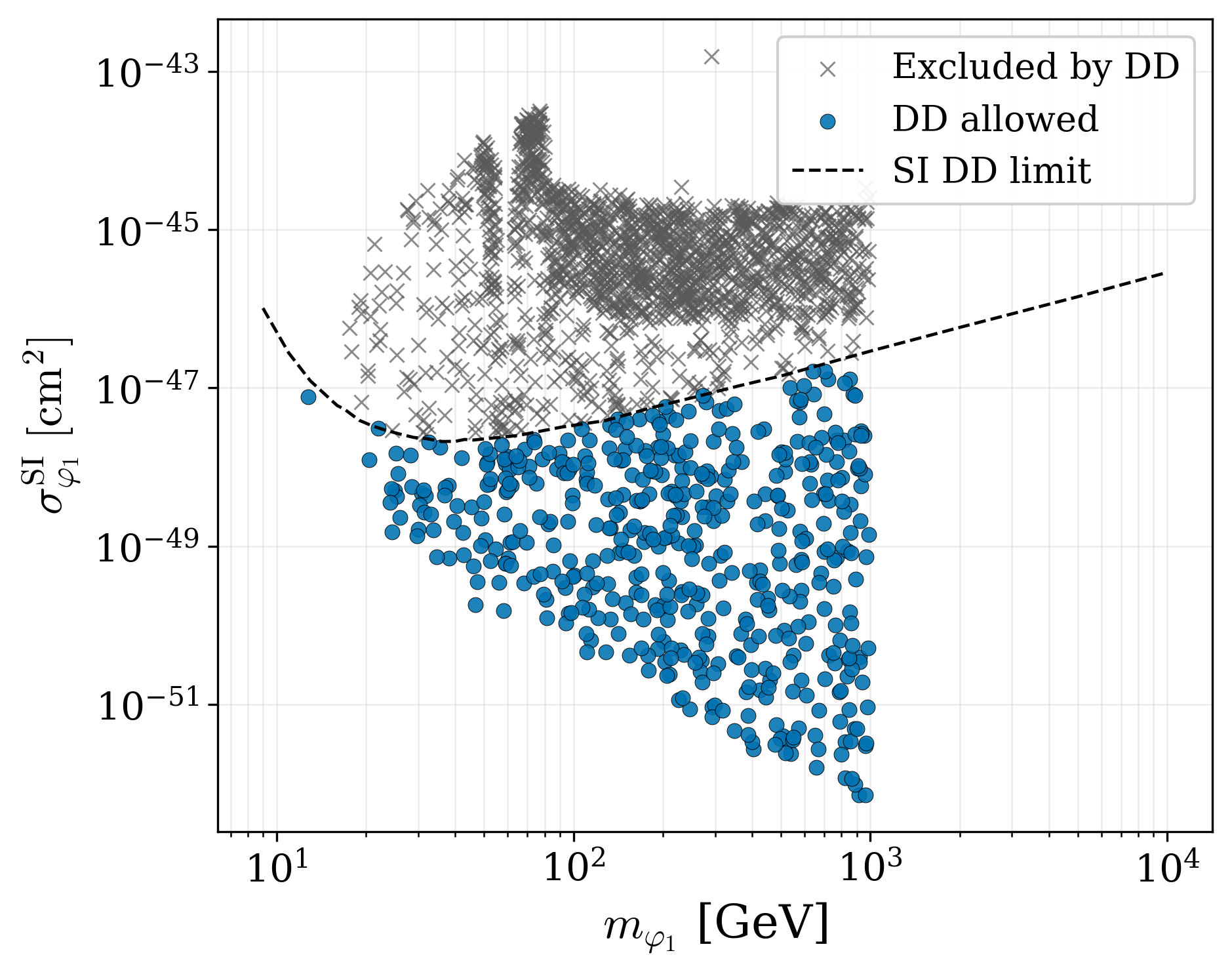}
     \caption{Scalar $\varphi_1$-dominated scenario. The three panels correspond to the projection in the planes $[m_{\varphi_1}, \lambda_{10}]$, $[m_{\varphi_1}, \lambda_7]$, and $[m_{\varphi_1}, \sigma_{\varphi_1}^{\rm SI}]$. All blue dots and gray crosses fit the whole DM abundance, while only the blue dots satisfy the bounds from direct detection searches.}
    \label{fig:varphi_domination}
\end{figure}
\begin{figure}[t]
    \centering
    \includegraphics[width=.38\textwidth]{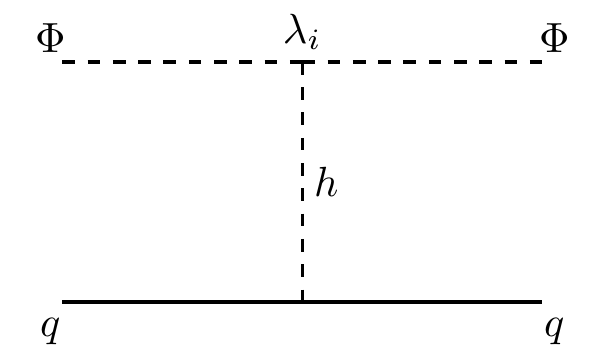}
    \caption{Higgs-mediated 
    direct detection scattering process for scalar DM candidates $\Phi = \varphi_1, \varphi_2, \eta$. As in Fig.~\ref{fig:annihilation_scalar}, the Higgs-portal coupling $\lambda_i$ represents $\lambda_i = \lambda_7, \lambda_8, \lambda_9$ for $\Phi = \varphi_1, \varphi_2, \eta$, respectively.}
    \label{fig:dd}
\end{figure}

For the scalar case, DM can be produced in the early Universe through 2-to-2 scatterings of SM particles, as shown in Fig.~\ref{fig:annihilation_scalar}. Figure~\ref{fig:varphi_domination} summarizes with dots and crosses the parameter region that fits the whole observed DM abundance, in the scenario where $\varphi_1$ constitutes  the dominant contribution. The two upper panels show the quartic couplings $\lambda_{10}$ and $\lambda_7$, respectively, as a function of $m_{\varphi_1}$. These two couplings are particularly relevant because they correspond to scalar portals for DM annihilation: once $\sigma$ and $\phi$ acquire vacuum expectation values, the interaction $\lambda_{10} (\sigma^\ast \sigma) (\varphi_1^\ast \varphi_1)$ and $\lambda_7 (\phi^\dagger \phi) (\varphi_1^\ast \varphi_1)$ generate trilinear couplings that drive the DM annihilation through the scalar sector (see the corresponding diagrams in Fig.~\ref{fig:annihilation_scalar}). Large values of $\lambda_{10}$ and $\lambda_7$ enhance in a similar way the annihilation rate, reducing the relic abundance. However, in relation to direct detection of DM (cf. Fig.~\ref{fig:dd}), they play a very different role. In the lower panel of Fig.~\ref{fig:varphi_domination} the limit on the spin-independent elastic scattering cross-section $\sigma_{\varphi_1}^{\rm SI}$ from Ref.~\cite{LZ:2024zvo} is shown. In the three panels, the blue dots correspond to the points satisfying the direct detection bound, while the gray crosses violate it. The direct detection signal, as shown in Fig.~\ref{fig:dd}, is mediated by the interaction with the Higgs boson $\phi$ (and not with $\sigma$), small values for $\lambda_7$ are required, which, in turn, implies larger values for $\lambda_{10}$.

\subsubsection*{Fermionic dark matter}
\begin{figure}[t!]
    \centering
    \includegraphics[width=.34\textwidth]{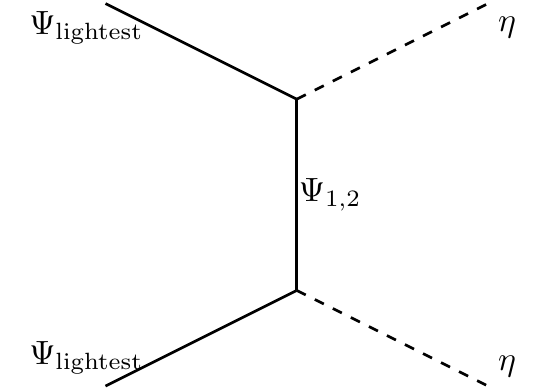}
    \hspace{1.7cm}
    \includegraphics[width=.34\textwidth]{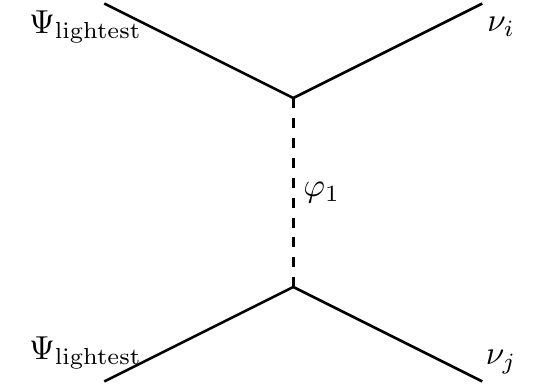}
    \caption{Annihilation channels of the fermion DM candidate $\Psi_{\rm lightest}$. The indices $i, j$ for the physical neutrinos in the final state run in the range $i,j = 1, ..., 7$.}
    \label{fig:annihilation_fermion}   
\end{figure}
\medskip
\begin{figure}[t!]
    \centering
    \includegraphics[width=0.59\textwidth]{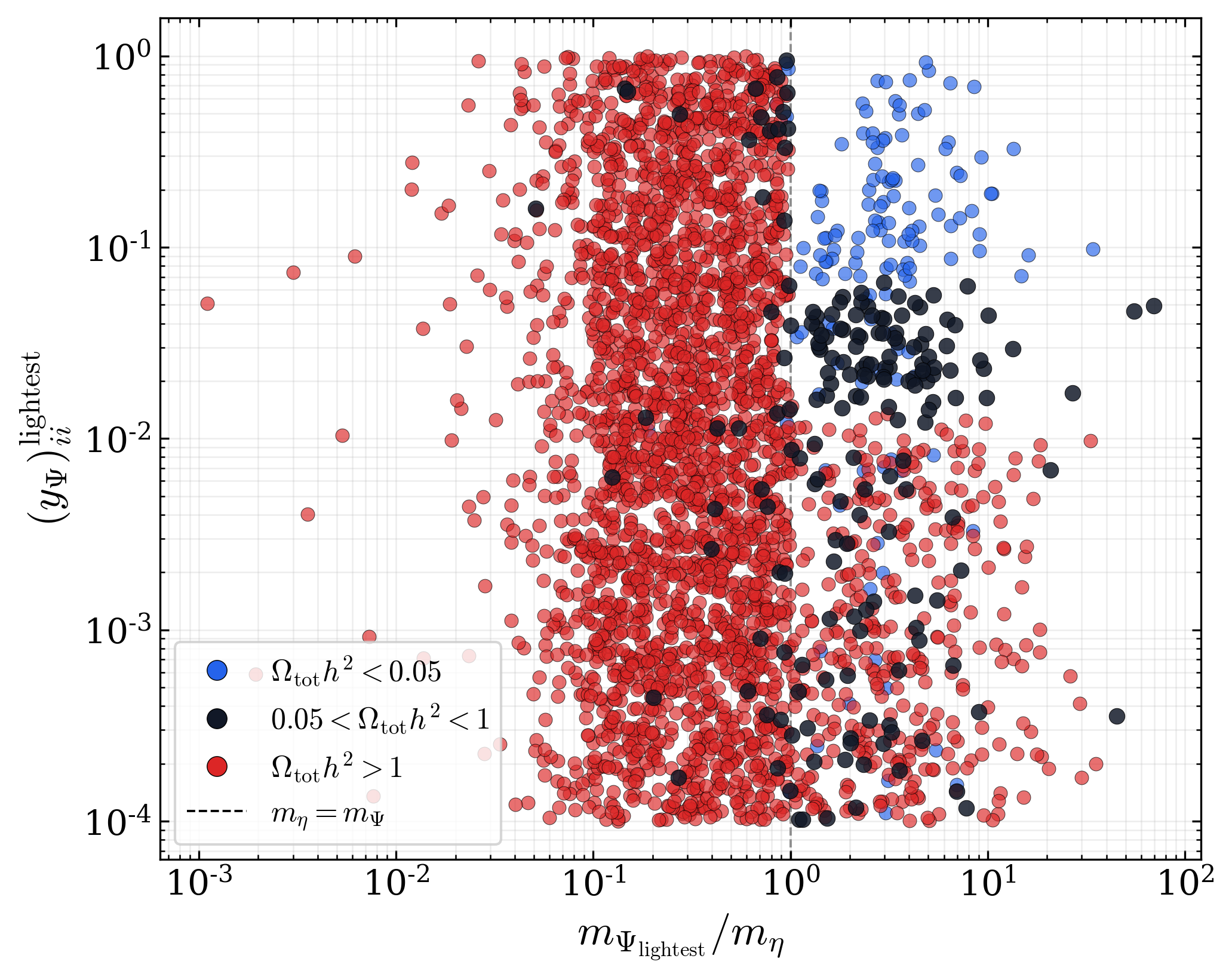}
    \caption{Fermion $\Psi$-dominated scenario in the plane $[m_{\Psi_{\rm lightest}}/m_\eta,\, (y_\Psi)_{ii}^{\rm lightest}]$. Blue, black, and red points correspond to under-abundant ($\Omega_{\rm tot}h^2<0.05$), close-to-observed ($0.05<\Omega_{\rm tot}h^2<1$), and over-abundant ($\Omega_{\rm tot}h^2>1$) scenarios, respectively. The vertical dashed line indicates the change in mass hierarchy between $\eta$ and $\Psi_{\rm lightest}$}.
    \label{Fig:psi_dominant}
\end{figure}
Alternatively, fermionic DM could be produced in the early Universe through annihilations into $\eta$ or neutrinos, as shown in Fig.~\ref{fig:annihilation_fermion}. In the single-component fermionic scenario shown in Fig.~\ref{Fig:psi_dominant}, the relic abundance is predominantly controlled by the interplay between lightest vector-like Dirac fermion, $\Psi_{\rm lightest}$, and the scalar $\eta$. From Fig.~\ref{Fig:psi_dominant} one can identify  two distinct regimes: for $m_{\Psi_{\rm lightest}}/m_\eta<1$, the annihilation of $\Psi_{\rm lightest}$ in $\eta$ is kinematically forbidden and therefore the relic abundance tends to be too large. By contrast, for $m_{\Psi_{\rm lightest}}/m_\eta>1$, this annihilation channel becomes kinematically accessible. The resulting relic abundance then depends strongly on the couplings entering the interaction $\left(y_\Psi\right)_{nk} \overline{\Psi_{kR}^{C}} \eta\Psi_{nR}$. For $(y_\Psi)_{ii}^{\rm lightest}<10^{-2}$, the annihilation rate is typically too small, leading to an over-abundant scenario. For $10^{-2}<(y_\Psi)_{ii}^{\rm lightest}<10^{-1}$, the annihilation rate can be enough to fit the observed relic abundance. Finally, for $(y_\Psi)_{ii}^{\rm lightest}>10^{-1}$, the annihilation becomes too efficient, resulting in an under-abundant scenario.

It is important to note that for large values of the coupling, there is an efficient conversion to $\eta$. However, in the $\Psi$-dominated scenario considered here, the abundance of $\eta$ must itself be depleted through annihilations into SM particles. The main connection between $\eta$ and the visible sector is provided by the Higgs-portal interaction $\lambda_9(\phi^\dagger\phi)(\eta^\ast\eta)$, see Eq.~\eqref{potential}. After electroweak symmetry breaking, this interaction induces the vertices $h\eta^\ast\eta$ and $hh\eta^\ast\eta$, allowing $\eta\eta^\ast$ annihilation into SM fermions, $W^+W^-$, $ZZ$, and, when kinematically accessible, into Higgs-boson pairs. In addition, the residual $\mathbb{Z}_3$ structure and the trilinear interaction $A_\eta(\eta^3 + \eta^{\ast 3})$ may allow semi-annihilation processes~\cite{Hambye:2008bq, DEramo:2010keq} such as $\eta\eta\rightarrow\eta^\ast h$, depending on masses and couplings.

We also note that $\Psi$ can annihilate into neutrinos through the $\left( y_N  \right)_{nk}\, \overline{\Psi}_{kL}\, \varphi_1\, N_{nR}$ interaction. This process is mediated by $\varphi_1$ in the $t$-channel and depends on the mixing between the right-handed neutrinos $N_{nR}$ and the physical neutrinos $\nu _i$. The corresponding diagram is shown in the right panel of Fig.~\ref{fig:annihilation_fermion}. However, this annihilation channel provides a subleading contribution to the relic abundance, compared to the annihilation in $\eta$, shown in the left panel of Fig.~\ref{fig:annihilation_fermion}. Therefore, the successful $\Psi_{\rm lightest}$-dominated points rely on a two-step depletion mechanism: first, $\Psi_{\rm lightest}$ is converted into $\eta$ through the interaction controlled by $y_\Psi$; second, the resulting $\eta$ population is depleted through Higgs-portal annihilation or semi-annihilation. If the first process is inefficient, $\Psi_{\rm lightest}$ freezes out with an excessively large abundance. Conversely, if the conversion is efficient but the subsequent depletion of $\eta$ is too weak, the abundance is merely transferred from $\Psi_{\rm lightest}$ to $\eta$, instead leading to a scenario dominated by $\eta$.

\subsubsection{Two-component domination}
\begin{figure}[t!]
    \centering
    \includegraphics[width=0.77\textwidth]{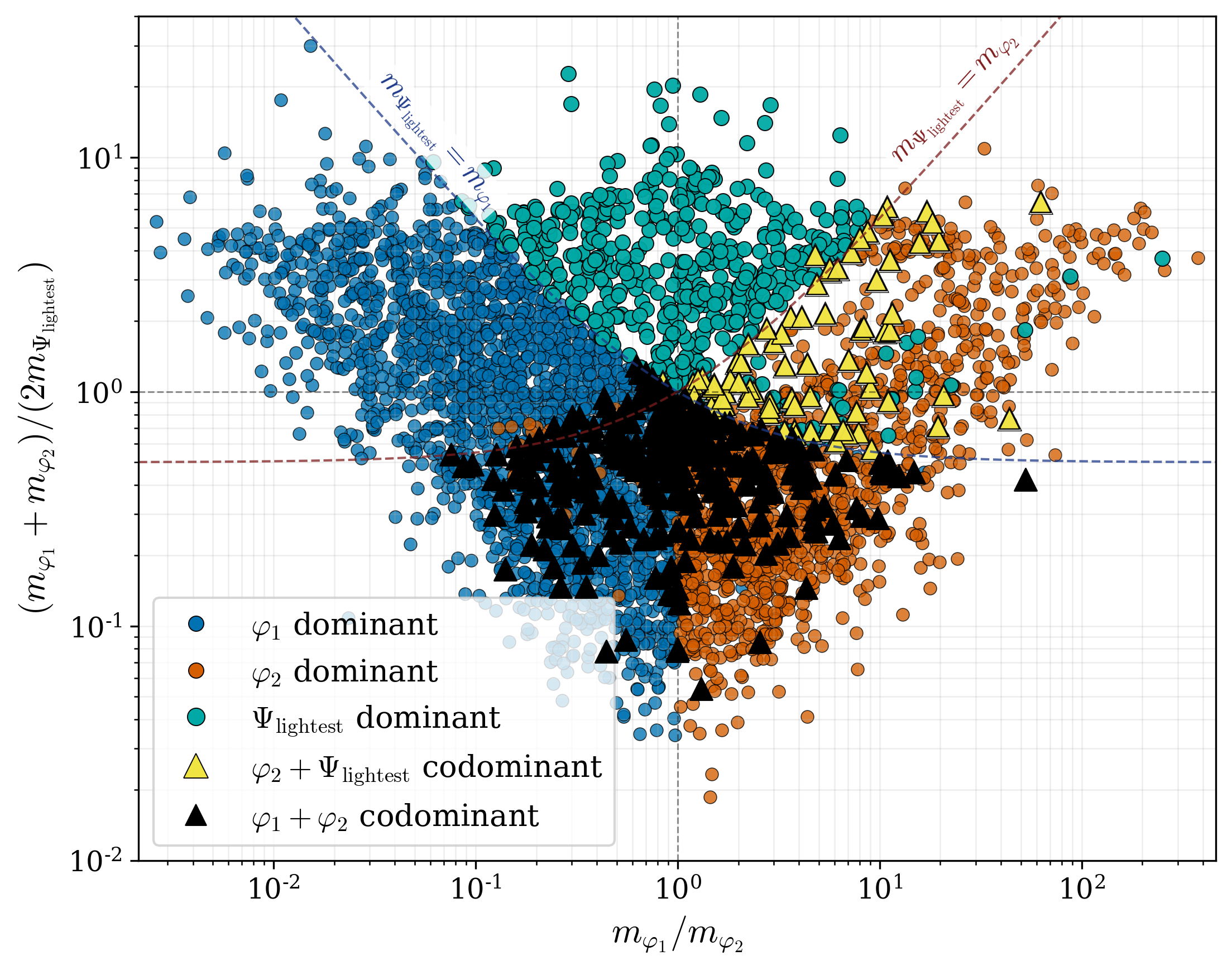}
    \caption{Co-dominant versus single-component-dominated scenarios. We show the distribution of viable scan points in the plane defined by the scalar mass ratio $m_{\varphi_1}/m_{\varphi_2}$ and the ratio $(m_{\varphi_1}+m_{\varphi_2})/(2m_{\Psi_{\rm lightest}})$ between the average scalar mass and the lightest fermion mass. The colors indicate the dominant or co-dominant DM composition. All points satisfy the direct-detection constraints and lie within the broad relic-abundance interval $0.05 < \Omega_{\rm tot} h^2 < 1$. The diagonal boundaries correspond to $m_{\Psi_{\rm lightest}}=m_{\varphi_1}$ and $m_{\Psi_{\rm lightest}}=m_{\varphi_2}$.}
    \label{fig:codomination_cases}
\end{figure}

In addition to the single-component regimes discussed in the previous section, the scan contains viable configurations in which two stable particles contribute a significant amount to the total DM relic abundance. We consider two representative cases: the scalar--scalar pair $\varphi_1+\varphi_2$ and the scalar--fermion pair $\varphi_2+\Psi_1$. Figure~\ref{fig:codomination_cases} shows the mass-hierarchy regions in which each co-dominant scenario can occur. All points shown satisfy direct-detection constraints and yield a relic abundance close to the observed value, for which, for numerical purposes, we adopt the broad interval $0.05<\Omega_{\rm tot}h^2<1$. The horizontal axis represents the hierarchy between the two scalars, $\varphi_1$ and $\varphi_2$, while the vertical one compares the characteristic scalar mass scale with that of the mass scale of the fermionic sector through $(m_{\varphi_1}+m_{\varphi_2})/(2m_{\Psi_{\rm lightest}})$.

When the fermionic sector is heavier, $(m_{\varphi_1} + m_{\varphi_2}) / (2 m_{\Psi_{\rm lightest}}) < 1$, the DM tends to be dominated by scalars. In this region, the lighter of the two scalars typically provides the dominant contribution, while the scalar--scalar co-dominance becomes more common near $m_{\varphi_1}\simeq m_{\varphi_2}$, as can be seen in Fig.~\ref{fig:codomination_cases}. In contrast, for $(m_{\varphi_1}+m_{\varphi_2})/(2m_{\Psi_{\rm lightest}})>1$, the boundaries defined by $m_{\Psi_{\rm lightest}}=m_{\varphi_1}$ and $m_{\Psi_{\rm lightest}}=m_{\varphi_2}$ correspond to the two visible diagonals in the upper part of the plot. Since $\Psi$ and $\varphi_1$ belong to the same dark sector, and the DM candidate within each sector is necessarily its lightest state, there is no overlap between the blue $\varphi_1$-dominated region and the green $\Psi$-dominated region. Between the two diagonals, $\Psi_{\rm lightest}$ is lighter than both scalars, giving rise to a $\Psi$-dominated scenario. Close to the boundary $m_{\Psi_{\rm lightest}}\simeq m_{\varphi_2}$, a sizable number of mixed scalar--fermion co-dominant points appears, corresponding to the $\varphi_2+\Psi$ scenario. When $m_{\varphi_2}$ becomes substantially smaller than $m_{\Psi_{\rm lightest}}$, the system returns to a $\varphi_2$-dominated regime.

Lastly, the characteristic inverted-triangular shape can be understood from the variables used in the plot. The lower vertex corresponds to the minimum value of $(m_{\varphi_1}+m_{\varphi_2})/(2m_{\Psi_{\rm lightest}})$. This minimum is reached near $m_{\varphi_1} / m_{\varphi_2} = 1$, where both scalar masses can simultaneously take small values. By contrast, a pronounced hierarchy between the scalar masses requires one of them to be heavier, thereby increasing the average scalar mass and, consequently, the ratio between the scalar and fermionic mass scales.

\subsubsection{Three-component domination}
Finally, the residual $\mathbb{Z}_2 \otimes \mathbb{Z}_3$ symmetry allows the relic abundance to be shared among three stable states, one of each independent dark sector. Their total abundance is determined by the combined scalar and fermion annihilation processes. As shown in Fig.~\ref{Fig:pie_composition} (cf.~orange regions), these configurations are less frequent than single- and two-component scenarios but demonstrate that the model naturally accommodates three-component DM without requiring additional stabilizing symmetries.
 
\subsection{Implications for baryon asymmetry of the Universe}
In this section, we consider the possibility of having successful baryogenesis via leptogenesis, through a dynamical mechanism that gives rise to the structure of the neutrino mass matrix in Eq.~\eqref{nuLang} and is capable of explaining the neutrino data (light masses and oscillation frequencies with the measured lepton mixings). In the presence of fermion singlets $\nu_{kR}$ and $N_{kR}$ ($k=1,2$), it is natural to consider the possibility of generating the BAU through leptogenesis. In this section, we consider its viability in the presence of out-of-equilibrium CP- and LNV processes. We then discuss the regimes explaining mixing and light neutrino masses that also satisfy the first necessary conditions for successful leptogenesis. 

To simplify our analysis, we consider the case where $\left\vert M_{11}\right\vert \ll \left\vert M_{22}\right\vert$ in Eq.~\eqref{mseesaw}. We further assume that the gauge-singlet neutral leptons $\Psi_{k}$ as well as the dark scalar singlet $\varphi_1$ are heavier than the lightest pseudo-Dirac fermions $N_1^\pm\equiv N^\pm$, while for simplicity we work on the basis of a diagonal SM charged lepton mass matrix. In the scenario mentioned above, lepton asymmetry is induced only via the decay of the first generation of the two pairs $N_k^{\pm}$ ($k=1,2$). Then, the CP asymmetry obtained from the latter decay into an active lepton and a Higgs is~\cite{Gu:2010xc, Pilaftsis:1997jf}
\begin{equation}
    \varepsilon_{CP}^\pm \equiv \sum_{i=1}^{3}\frac{\left[ \Gamma \left(N^{\pm}\to \nu_i h\right) -\Gamma \left(N^\pm\to \overline{\nu }_i h\right) \right] }{\left[ \Gamma \left(N^\pm\to \nu_i h\right) +\Gamma \left(N^\pm\to \overline{\nu }_i h\right) \right] }\simeq \frac{\text{Im}\left\{ \left(\left[ \left(y_{N_{+}}\right)^\dagger \left(y_{N_{-}}\right) \right]^2\right)_{11}\right\} }{8\pi A_\pm}\frac{r}{r^2+\frac{\Gamma_\pm^2}{m_{N^\pm}^2}} \,,
\end{equation}
with
\begin{align}
    r = \frac{m_{N^{+}}^2-m_{N^{-}}^2}{m_{N^{+}} m_{N^{-}}},\hspace{4.7cm} &A_\pm=\left[ \left(y_{N^\pm}\right)^\dagger y_{N^\pm}\right]_{11}, \\
    y_{N^\pm} = y_\nu \left(1\mp S\right) =y_\nu \left(1\pm \frac14 M^{-1}\mu \right) \hspace{1.4cm} &\Gamma_\pm=\frac{A_\pm m_{N^\pm}}{8\pi}\,.
\end{align}

Neglecting the interference terms involving the two different sterile neutrinos of the lightest  pseudo-Dirac pair $N^\pm$, the washout parameter $K_{N^{+}}+K_{N^{-}}$ is huge, as mentioned in Ref.~\cite{Dolan:2018qpy}. However, small mass splitting between pseudo-Dirac neutrinos leads to destructive interference in the scattering process~\cite{Blanchet:2009kk}. The washout parameter including the interference term has the form
\begin{equation}
    K^\text{eff} \simeq K_{N^{+}}\, \delta_{+}^2 + K_{N^{-}}\, \delta_{-}^2\,,
\end{equation}
where
\begin{equation}
    \delta_\pm=\frac{m_{N^{+}}-m_{N^{-}}}{\Gamma_{\pm}}\,,\hspace{0.7cm} \hspace{0.7cm}K_{N^\pm}=\frac{\Gamma_\pm}{H(m_{N^\pm})}\,,
\end{equation}
and $H$ corresponds to the Hubble expansion rate of the universe. In the case of a standard cosmological scenario where the total energy density is dominated by SM radiation,
\begin{equation} \label{eq: Hubble}
    H(T) = \frac{\pi}{3}\, \sqrt{\frac{g_\star}{10}}\, \frac{T^2}{M_P}\,,
\end{equation}
where $g_\star$ corresponds to the number of relativistic degrees of freedom in the SM bath, and $M_P \simeq 2.4 \times 10^{18}$~GeV is the reduced Planck mass.

In weak ($K^\text{eff}\ll 1$) and strong ($K^\text{eff}\gg 1$) washout regimes, the BAU is related to lepton asymmetry as~\cite{Pilaftsis:1997jf} 
\begin{align}
    Y_{\Delta B} &\equiv \frac{n_{B}-\overline{n}_{B}}{s}=-\frac{28}{79}\frac{\varepsilon_{CP}^++\varepsilon_{CP}^-}{g^*},\hspace*{2.1cm}\text{for}\hspace*{0.5cm}K^\text{eff}\ll 1, \\
    Y_{\Delta B} &\equiv \frac{n_{B}-\overline{n}_{B}}{s}=-\frac{28}{79}\frac{0.3\left(\varepsilon_{CP}^++\varepsilon_{CP}^-\right) }{g^*K^\text{eff}\left(\ln K^\text{eff}\right)^{0.6}}\,,\hspace*{0.9cm}\text{for}\hspace*{0.5cm}K^\text{eff}\gg 1.
\end{align}
In the analysis, we constrain the model to successfully reproduce the measured baryon density and the value of the baryon asymmetry parameter determined by the Planck Collaboration~\cite{Planck:2018vyg}.
\begin{equation}
    Y_{\Delta B} = \left(0.87\pm 0.01\right) \times 10^{-10}.
\end{equation}
In order to provide a quantitative proof, one needs to solve the Boltzmann equations and take into account all  washout processes, flavor effects, and other possible resonant effects. This deserves a dedicated study, which is beyond the scope of this work.

\begin{figure}[t!]
    \centering    \includegraphics[width=0.69\textwidth]{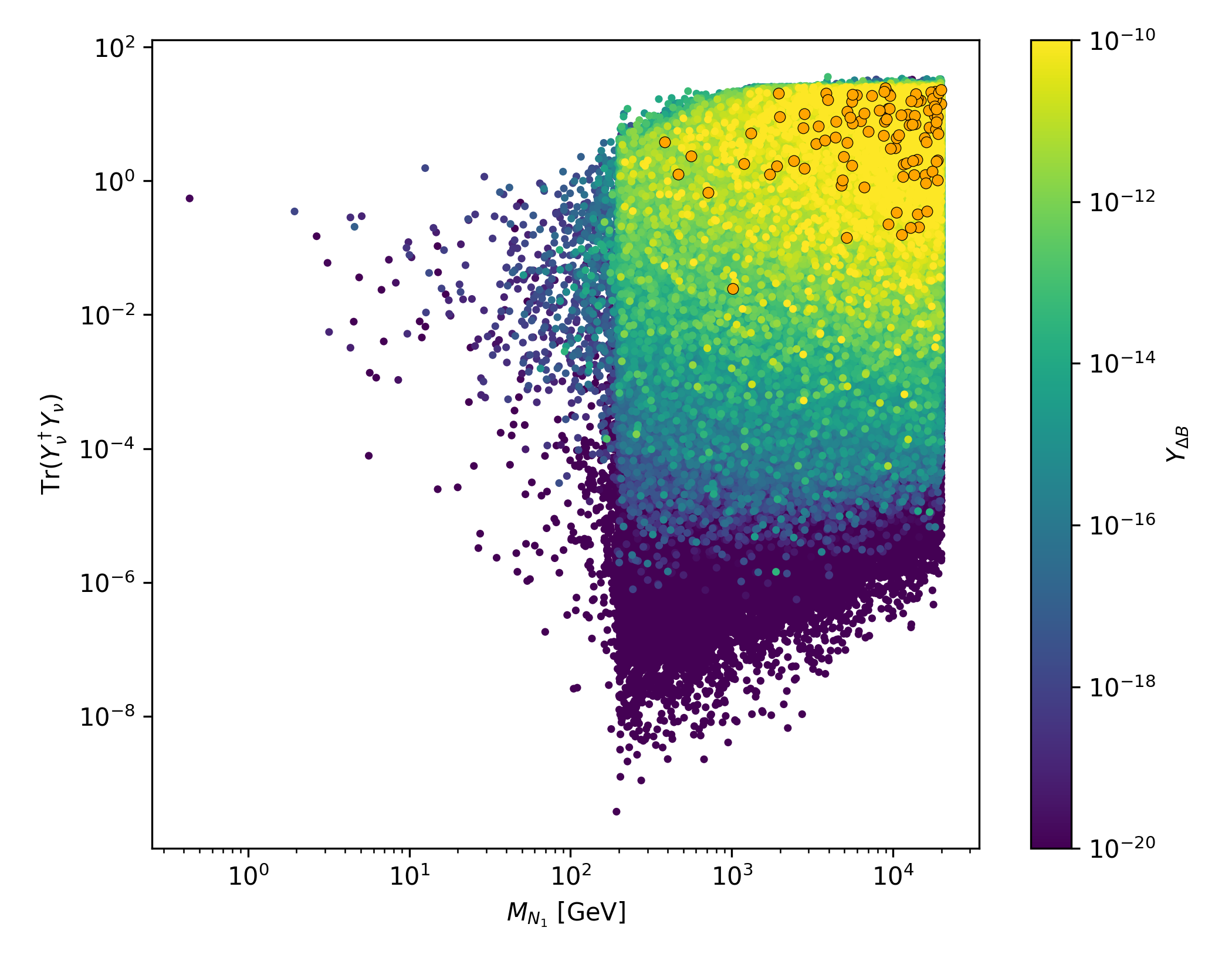}
    \caption{Correlation of Tr[$y_\nu^\dag y_\nu$] against $M_{N_1}$ with values of $Y_{\Delta B}$ given by the color scale. The orange circles correspond to the points that satisfy observed baryon asymmetry.}
    \label{leptogenesis1}
\end{figure}
In Fig.~\ref{leptogenesis1}, we have plotted  Tr[$y_\nu^\dag y_\nu$] against the lightest heavy Majorana neutrino mass $M_{N_1}$ with the values of $Y_{\Delta B}$ given by the color scale.  The orange circles correspond to the points that exactly satisfy the observed baryon asymmetry. For this analysis, we used the set of parameter points from the DM scan pointing to the couplings and parameters whose importance was found to be significant in the DM analysis (see Appendix~\ref{app:multinest_and_importance} and  Fig.~\ref{Fig:shap_importance}). The remaining parameters were scanned over the following ranges: $M_{11}\in[0.2,20]$ TeV, $M_{22}\in[10,500]\times M_{11}$, the neutrino mass-squared differences, mixing angles, and the Dirac CP phase were varied within their experimental $3\sigma$ ranges, the physical Majorana phase $\alpha_M$ was varied in the range $[0,2\pi]$, the complex parameter $z$, which parametrizes the orthogonal matrix in the Casas--Ibarra parametrization, was varied such that $|z|\in[10^{-4},10]$, the absolute values of the unfixed entries of the Yukawa coupling matrices $y_\Psi$ and $y_N$ were varied in the range $[0,\sqrt{4\pi}]$, and the scalar cubic terms $A_\eta $ and $A_1$ were varied in the range $[1,20]$ TeV. We have only considered those points which satisfy all the constraints discussed in Section~\ref{constraints}. From Fig.~\ref{leptogenesis1}, we can see that the observed baryon asymmetry is satisfied for heavy neutrino masses as low as $\sim 500$ GeV with Tr$(y_\nu^\dag y_\nu)$ lying in the range $[0.1-10]$ for most of the points.

\section{Interplay between charged-lepton flavor violation\\and cosmological observables} \label{sec:interplay}
\begin{figure}[t!]
    \centering    \includegraphics[width=0.69\textwidth]{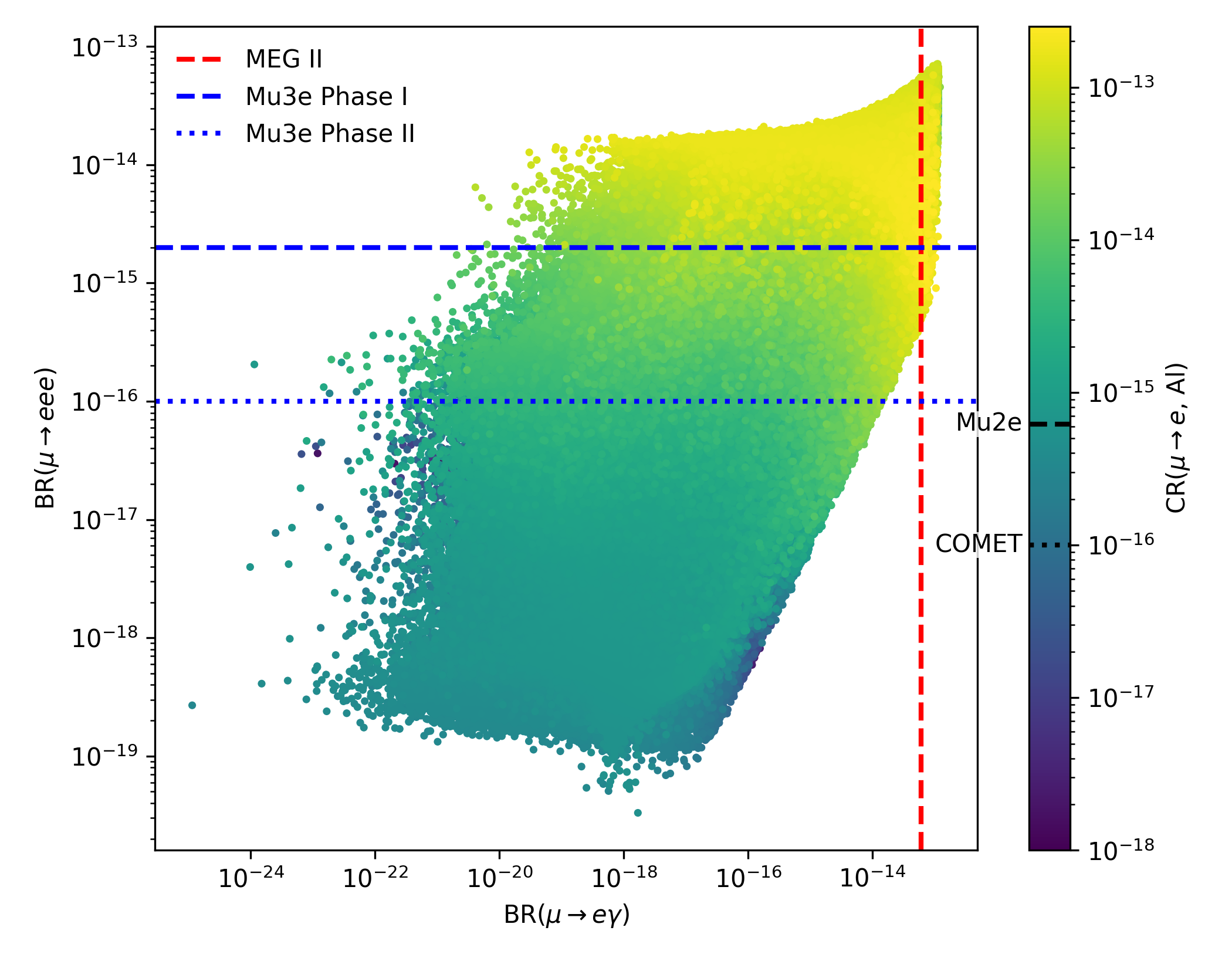}    
    \caption{Correlation among rates of cLFV processes: BR$(\mu \rightarrow e e e)$, BR$(\mu \rightarrow e \gamma)$ and CR($(\mu \rightarrow e, N)$. All the points satisfy the constraints discussed in Section~\ref{constraints}, whereas the projected sensitivities are indicated by the dotted and dashed lines.}
    \label{fig_lfv}
\end{figure}
As we have already seen, the parameters, in particular, the heavy neutrino masses and their corresponding Yukawa couplings that affect leptogenesis, also affect the CLFV processes. So, it is worth exploring the correlations among the predictions for the BAU as well as various CLFV observables. In Fig.~\ref{fig_lfv}, we have first shown the correlations between the predictions for the three CLFV processes that we have considered in this work. The BR$(\mu \rightarrow e e e)$ is plotted against the BR$(\mu \rightarrow e \gamma)$. The maximum value of CR$(\mu \rightarrow e, Al)$ corresponding to each point obtained in our scan is indicated by the color scale. All the points satisfy the existing bounds (and are solution for DM),  whereas the projected sensitivities are indicated by the dotted and dashed lines. The various parameters were scanned in the same ranges as discussed in the previous section. From this figure, we can see that all three processes offer complementary probes of the model through CLFV.

\begin{figure}[t!]
    \centering    \includegraphics[width=0.69\textwidth]{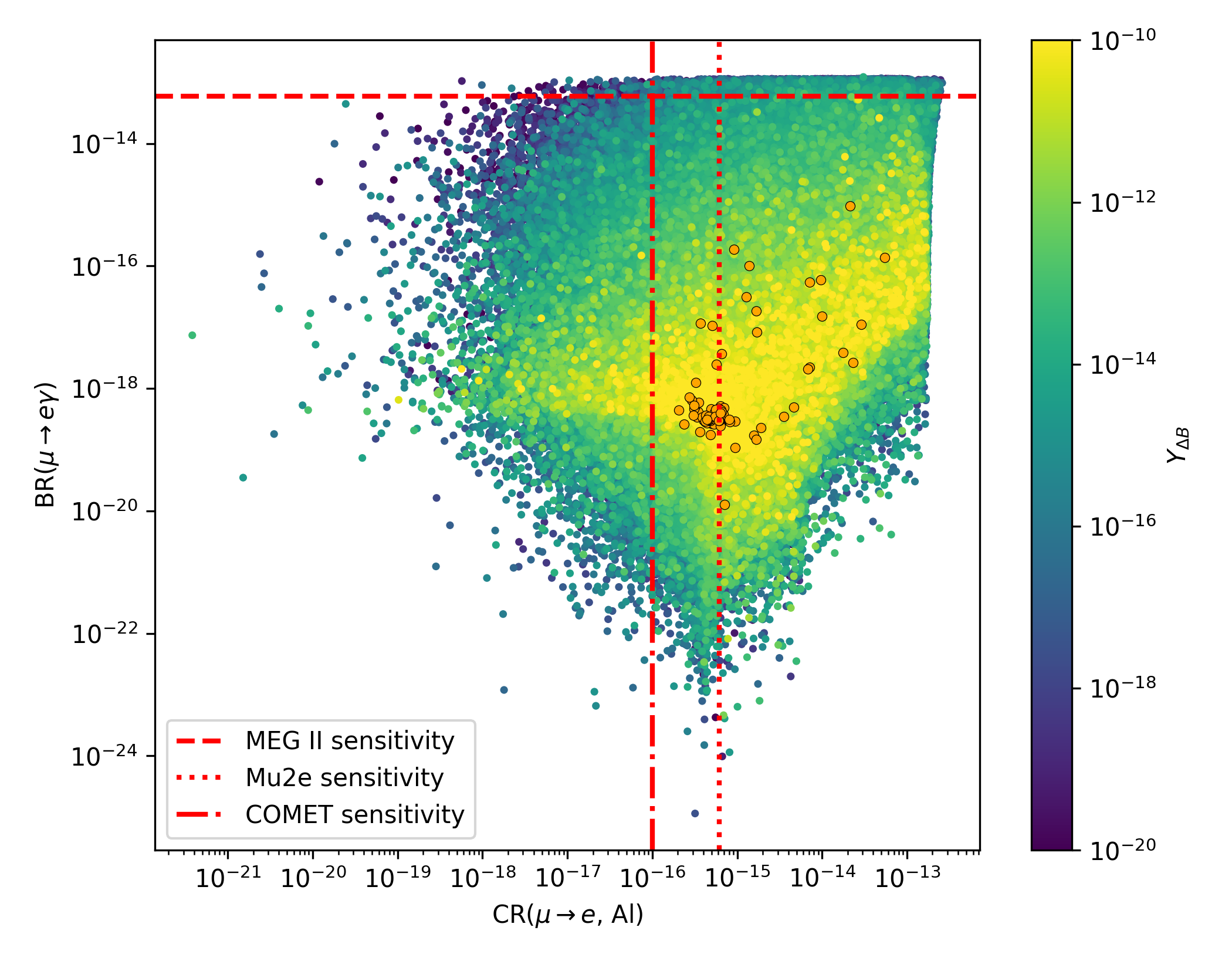}    
    \caption{The correlation of BR$(\mu \rightarrow e \gamma)$ against CR($(\mu \rightarrow e, Al)$. The value of the baryon asymmetry corresponding to each point is indicated by the color graph with the orange points satisfying the observed baryon asymmetry. All the points satisfy the existing bounds whereas the projected sensitivities are indicated by the red lines.}
    \label{leptogenesis2}
\end{figure}

The correlation between the baryon asymmetry parameter $Y_{\Delta B}$ and the CLFV observables that we considered is then explored. An example  is shown in Fig.~\ref{leptogenesis2}, where we have plotted BR$(\mu \rightarrow e \gamma)$ against CR($\mu \rightarrow e$, Al), with the value of the baryon asymmetry corresponding to each point indicated by the color scale. The orange points satisfy the observed baryon asymmetry. As before, all points satisfy the existing bounds (including DM), while the red lines indicate the projected sensitivities.

From this figure, we can see that many points that satisfy the correct baryon asymmetry in our scan lie within the future sensitivity of the Mu2e experiment, while all the points with the correct baryon asymmetry lie within the sensitivity range of COMET. It is interesting that all of these points escape the sensitivity of MEG II. Thus, $\mu \rightarrow e$ conversion experiments offer a better reach for the parameter space that gives the correct BAU.

It should be noted that this scan does not conclusively show that a null result at the COMET experiment can rule out this model as a solution to the baryon asymmetry of the universe. To establish such a conclusion, one has to consider the full Boltzmann equations, including all scattering processes and flavor effects, which is beyond the scope of this work. Moreover, we can also see that there are many points with $Y_{\Delta B} \sim 10^{-10}$ (yellow) that escape the sensitivity of COMET.

\section{Conclusions} \label{Sec:conclusions}
We have constructed a novel model in which the tiny masses of the active neutrinos arise through an ISS mechanism whose LNV Majorana mass submatrix $\mu$ is dynamically generated at the three-loop level through a topology not previously employed to induce the inverse seesaw. In our proposed model, the SM particle content is extended by the gauge-singlet scalar fields $\varphi_1$, $\varphi_2$, $\eta$, and $\sigma$, as well as by two generations of right-handed neutrinos $\nu_{1R},\nu_{2R}$, sterile fermions $N_{1R},N_{2R}$, and two pairs of vector-like neutral leptons $\Psi_{kR},\Psi_{kL}$, with $k=1,2$. The SM gauge symmetry is supplemented by a spontaneously broken global $U(1)'$ symmetry and a preserved $\mathbb{Z}_2\otimes\mathbb{Z}_3$ discrete symmetry.

The preserved $\mathbb{Z}_2\otimes\mathbb{Z}_3$ symmetry enforces the radiative origin of the LNV Majorana mass submatrix $\mu$, stabilizes the scalar and fermionic DM candidates, and allows for multi-component DM composed of the lightest states of each independent dark sector. We have performed a thorough analysis of the implications of the model for neutrino data, charged lepton flavor violation, DM, and the viability of leptogenesis. To efficiently explore its high-dimensional parameter space, we carried out a two-stage numerical scan based on the MultiNest algorithm, using an importance analysis to identify the parameters with the largest impact on the DM relic abundance. We found viable single-, two-, and three-component scenarios involving scalar, fermionic, and mixed scalar--fermion DM scenarios. Their relic abundances are determined by a rich interplay of annihilation, coannihilation, conversion, and semi-annihilation processes. These solutions are compatible with neutrino oscillation data and with the constraints from CLFV processes, DM relic abundance, and direct-detection searches.

Our results also reveal a strong complementarity between cosmological and charged lepton flavor-violating observables. The pseudo-Dirac heavy-neutrino spectrum provides the necessary ingredients for low-scale resonant leptogenesis, while the same masses and couplings relevant for the generation of the baryon asymmetry also determine the predicted CLFV rates. In the parameter space explored in our scan, all points that reproduce the observed baryon asymmetry lie within the projected sensitivity of COMET, and many of them can also be probed by Mu2e, while remaining beyond the expected sensitivity of MEG~II. However, a conclusive assessment of leptogenesis will require solving the complete set of flavored Boltzmann equations, which lies beyond the scope of this work.

\section*{Acknowledgments}
AA and SU acknowledge support from the IN2P3 (CNRS) Master Project, ``Hunting for Heavy Neutral Leptons'' (12-PH-0100). AA acknowledges support from the Institut Universitaire de France (IUF). NB received funding from the grants PID2023-151418NB-I00 funded by MCIU/AEI/10.13039 /501100011033/ FEDER and PID2022-139841NB-I00 of MICIU/AEI/10.13039/501100011033 and FEDER, UE. AECH is supported by ANID-Chile FONDECYT 1261103, 1241855, ANID CCTVal CIA250027, ANID – Millennium Science Initiative Program code ICN2019$\_$044, ICTP through the Associates Programme (2026-2031). SK is supported by ANID-Chile FONDECYT 1230160, and Milenio-ANID-ICN2019$\_$044. This project has received funding and support from the European Union's Horizon 2020  research and innovation programme under the Marie Skłodowska-Curie grant agreement N$^\circ$~101086085-ASYMMETRY. 

\appendix
\section{Neutrino Mass Loop Function} \label{Appendix:mu-loop}
The loop function in Eq. (\ref{eq:mu-loop}) for the neutrino $\mu$-mass parameter 
is given by
\begin{align}
    F(x_1,x_2,x_3,x_4,x_5)  \equiv & \int d^3x \, \frac{\delta(x+y+z-1)}{y(y-1)+z(z-1)+2yz}  \nonumber \\
    & \times \int d^4 x' \, \frac{\delta(\alpha+\beta+\gamma+\delta - 1)} {\left[(\alpha Y + \delta)^2 - \delta - \alpha Y^2 - \alpha X\right]^2} \nonumber \\
    & \times \int d^3 x'' \, \frac{\rho~ \delta(\rho + \sigma + \omega - 1)} {\left[\rho A(x_1,x_2,x_3,x_5) - \sigma x_4 - \omega x_1 \right]^2}\,,
\end{align}
where
\begin{align}
    A(x_1,x_2,x_3,x_5) = & - \frac{\alpha\left[(x+y)x_2 + z x_5\right]} {\left[(\alpha Y + \delta)^2 - \delta - \alpha Y^2 - \alpha X\right] \left[y(y-1)+z(z-1)+2yz\right]} \nonumber \\
    & + \frac{\beta x_1 + \gamma x_3 + \delta x_2} {\left[(\alpha Y + \delta)^2 - \delta - \alpha Y^2 - \alpha X\right]}\,,
\end{align}
and
\begin{equation}
    X = -\left(\frac{y}{y+z} \right)^2 + \frac{y(y-1)}{y(y-1)+z(z-1)+2 y z}\,, \,\,\, \, \,\,\, Y =\frac{y}{y+z}\,.
\end{equation}
The integrals are defined as 
\begin{align}
    \int d^3 x \delta(x+y+z-1) &= \int_0^1 dx \int_0^{1-x}dy, \\
    \int d^4x' \delta(\alpha+\beta+\gamma+\delta - 1) &=
    \int_0^1 d\alpha \int_0^{1-\alpha} d\beta \int_0^{1-\alpha-\beta} d\gamma, \\
    \int d^3 x'' \delta(\rho + \sigma + \omega -1) &= \int_0^1 d\rho \int_0^{1-\rho}d\sigma.
\end{align}

\section{CLFV Loop Functions and Form Factors}
\label{app:formfactors}
In this appendix, we collect the explicit expressions for the form factors
and loop functions entering the charged-lepton-flavor-violating observables
discussed in Section~\ref{sec:clfv}. The form factors
$F^{\mu e}_{\gamma}$, $G_\gamma^{\mu e}$, $F_Z^{\mu e}$, and
$F_{\rm Box}^{\mu e\alpha\alpha}$ are given by
Refs.~\cite{Alonso:2012ji,Ilakovac:1994kj,Abada:2021zcm}.

\begingroup
\allowdisplaybreaks[2]
\begin{subequations}
\label{eq:FF}
\begin{align}
F^{\mu e}_\gamma
&=
\sum_{j=1}^N
\mathbb{U}_{ej}\mathbb{U}^*_{\mu j}F_\gamma(x_j)\,,
\label{Fmue}
\\
G^{\mu e}_\gamma
&=
\sum_{j=1}^N
\mathbb{U}_{ej}\mathbb{U}^*_{\mu j}G_\gamma(x_j)\,,
\label{Ggammamue}
\\
F^{\mu e}_Z
&=
\sum_{j,k=1}^N
\mathbb{U}_{ej}\mathbb{U}^*_{\mu k}
\biggl[
\delta_{jk}F_Z(x_j)
+C_{jk}G_Z(x_j,x_k)
+C^*_{jk}H_Z(x_j,x_k)
\biggr]\,,
\\
F^{\mu e uu}_{\rm Box}
&=
\sum_{j=1}^N\sum_{d_\alpha=d,s,b}
\mathbb{U}_{ej}\mathbb{U}^*_{\mu j}
V_{u d_\alpha}V^*_{u d_\alpha}
F_{\rm Box}(x_j,x_{d_\alpha})\,,
\\
F^{\mu e dd}_{\rm Box}
&=
\sum_{j=1}^N\sum_{u_\alpha=u,c,t}
\mathbb{U}_{ej}\mathbb{U}^*_{\mu j}
V_{d u_\alpha}V^*_{d u_\alpha}
F_{\rm XBox}(x_j,x_{u_\alpha})\,,
\\
F^{\mu eee}_{\rm Box}
&=
\sum_{j,k=1}^N
\mathbb{U}_{ej}\mathbb{U}^*_{\mu k}
\biggl[
\mathbb{U}_{ej}\mathbb{U}^*_{ek}G_{\rm Box}(x_j,x_k)
-2\,\mathbb{U}^*_{ej}\mathbb{U}_{ek}
F_{\rm XBox}(x_j,x_k)
\biggr]\,.
\label{Fmueee}
\end{align}
\end{subequations}
\endgroup

Here, $\mathbb{U}$ is the full neutral-fermion mixing matrix defined in
Eq.~\eqref{seewsawdiagonal}, $V_{qq^\prime}$ denotes the CKM matrix, and
\begin{equation}
C_{jk}
\equiv
\sum_{\alpha=e,\mu,\tau}
\mathbb{U}_{\alpha j}\mathbb{U}^*_{\alpha k}\,.
\end{equation}
The dimensionless mass ratios are defined as $x_i=m_i^2/M_W^2$ for neutral
fermions and $x_q=m_q^2/M_W^2$ for quarks. Neglecting terms containing four
active--sterile mixing insertions, their leading behavior for $x_j\ll1$ is
\begin{equation}
\label{eq:FF-light}
\begin{aligned}
F^{\mu e}_\gamma
&\xrightarrow[x_j\ll1]{}
-\sum_{j=1}^N
\mathbb{U}_{ej}\mathbb{U}^*_{\mu j}x_j\,,
\\
G^{\mu e}_\gamma
&\xrightarrow[x_j\ll1]{}
\frac{1}{4}\sum_{j=1}^N
\mathbb{U}_{ej}\mathbb{U}^*_{\mu j}x_j\,,
\\
F^{\mu e}_Z
&\xrightarrow[x_j\ll1]{}
\sum_{j=1}^N
\mathbb{U}_{ej}\mathbb{U}^*_{\mu j}
x_j\left(-\frac{5}{2}-\ln x_j\right)\,,
\\
F^{\mu eee}_{\rm Box}
&\xrightarrow[x_j\ll1]{}
2\sum_{j=1}^N
\mathbb{U}_{ej}\mathbb{U}^*_{\mu j}
x_j\left(1+\ln x_j\right)\,.
\end{aligned}
\end{equation}

The loop functions entering Eq.~\eqref{eq:FF} are
\begin{equation}
\label{eqn:lfun:gbox}
\begin{aligned}
F_Z(x)
&=
-\frac{5x}{2(1-x)}
-\frac{5x^2}{2(1-x)^2}\ln x\,,
\\
G_Z(x,y)
&=
-\frac{1}{2(x-y)}
\left[
\frac{x^2(1-y)}{1-x}\ln x
-\frac{y^2(1-x)}{1-y}\ln y
\right]\,,
\\
H_Z(x,y)
&=
\frac{\sqrt{xy}}{4(x-y)}
\left[
\frac{x^2-4x}{1-x}\ln x
-\frac{y^2-4y}{1-y}\ln y
\right]\,,
\\
F_\gamma(x)
&=
\frac{x(7x^2-x-12)}{12(1-x)^3}
-\frac{x^2(x^2-10x+12)}{6(1-x)^4}\ln x\,,
\\
G_\gamma(x)
&=
-\frac{x(2x^2+5x-1)}{4(1-x)^3}
-\frac{3x^3}{2(1-x)^4}\ln x\,,
\\
F_{\rm Box}(x,y)
&=
\frac{1}{x-y}
\biggl\{
\left(4+\frac{xy}{4}\right)
\left[
\frac{1}{1-x}
+\frac{x^2}{(1-x)^2}\ln x
\right]
\\
&\hspace{2.2cm}
-2xy
\left[
\frac{1}{1-x}
+\frac{x}{(1-x)^2}\ln x
\right]
-(x\leftrightarrow y)
\biggr\}\,,
\\
F_{\rm XBox}(x,y)
&=
-\frac{1}{x-y}
\biggl\{
\left(1+\frac{xy}{4}\right)
\left[
\frac{1}{1-x}
+\frac{x^2}{(1-x)^2}\ln x
\right]
\\
&\hspace{2.2cm}
-2xy
\left[
\frac{1}{1-x}
+\frac{x}{(1-x)^2}\ln x
\right]
-(x\leftrightarrow y)
\biggr\}\,,
\\
G_{\rm Box}(x,y)
&=
-\frac{\sqrt{xy}}{x-y}
\biggl\{
(4+xy)
\left[
\frac{1}{1-x}
+\frac{x}{(1-x)^2}\ln x
\right]
\\
&\hspace{2.2cm}
-2
\left[
\frac{1}{1-x}
+\frac{x^2}{(1-x)^2}\ln x
\right]
-(x\leftrightarrow y)
\biggr\}\,.
\end{aligned}
\end{equation}

For degenerate arguments, the relevant exact expressions and light-mass
limits are
\begin{equation}
\label{limitval2}
\begin{aligned}
F_Z(x)
&\xrightarrow[x\ll1]{}
-\frac{5x}{2}\,,
\\
G_Z(x,x)
&=
-\frac{x}{2}
-\frac{x\ln x}{1-x}
\xrightarrow[x\ll1]{}
-x\left(\frac{1}{2}+\ln x\right)\,,
\\
H_Z(x,x)
&=
\frac{3}{4}
-\frac{x}{4}
-\frac{3}{4(1-x)}
-\frac{x^3-2x^2+4x}{4(1-x)^2}\ln x
\\
&\xrightarrow[x\ll1]{}
-x\left(1+\ln x\right)\,,
\\
F_\gamma(x)
&\xrightarrow[x\ll1]{}
-x\,,
\\
G_\gamma(x)
&\xrightarrow[x\ll1]{}
\frac{x}{4}\,,
\\
F_{\rm Box}(x,x)
&=
-\frac{
x^4-16x^3+31x^2-16
+2x(3x^2+4x-16)\ln x
}{
4(1-x)^3
}\,,
\\
F_{\rm XBox}(x,x)
&=
\frac{x^4-16x^3+19x^2-4}{4(1-x)^3}
+\frac{3x^3+4x^2-4x}{2(1-x)^3}\ln x\,.
\end{aligned}
\end{equation}

\section{Multinest and importance} \label{app:multinest_and_importance}
To identify the parameters that most strongly affect the DM relic-density prediction, we performed a machine-learning-based feature-importance analysis. Starting from an initial scan of the full parameter space, we trained a Random Forest regressor~\cite{Breiman:2001RandomForests}, using the fundamental model parameters as input variables. The target variable was defined so that the regressor learns which regions of parameter space reproduce the observed relic abundance, $\Omega_{\rm tot}\, h^2 \simeq 0.12$~\cite{Planck:2018vyg}. After training, we used SHAP (SHapley Additive exPlanations) values~\cite{Lundberg:2017SHAP, Lundberg:2018TreeSHAP}, a game-theoretic method for interpreting machine-learning predictions, to quantify the contribution of each input parameter to the model output.

\begin{figure}[t!]
    \centering
    \includegraphics[width=0.86\textwidth]{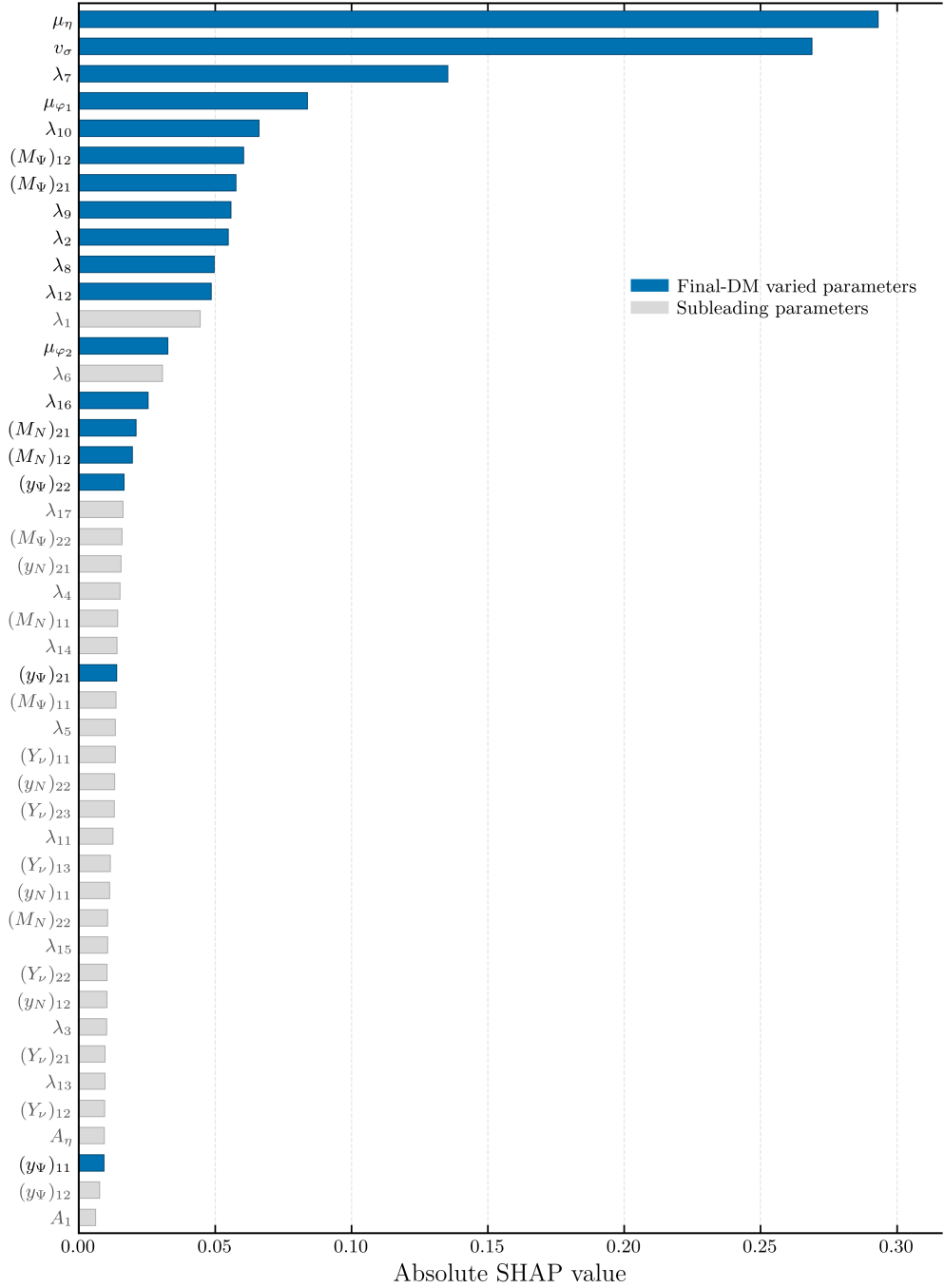}
    \caption{Ranking of parameter importance obtained from the mean absolute SHAP values. The relevant input parameters varied in the final DM scan are highlighted in blue, while the remaining parameters are shown in grey. We select the 16 most relevant parameters and the $y_{\Psi}$ couplings, which are relevant for the production and annihilation of the $\Psi$-dominated scenario, with the exception of $\lambda_1$ and $\lambda_6$, which are kept fixed in order to satisfy Higgs-sector constraints.}
    \label{Fig:shap_importance}
\end{figure}
The hierarchy found in the SHAP analysis, shown in Fig.~\ref{Fig:shap_importance}, has a clear physical interpretation. The most relevant parameters are those that directly determine the masses, portal interactions, and annihilation rates of the dark-sector states. In particular, $\mu_\eta$, $\mu_{\varphi_1}$ and $\mu_{\varphi_2}$ enter the tree-level masses of the neutral scalar candidates, while $v_\sigma$ controls the scale of $U(1)'$ breaking and contributes to the scalar masses through the singlet-portal terms. The large importance of $\lambda_7$, $\lambda_8$ and $\lambda_9$ is expected because these are the Higgs-portal couplings of $\varphi_1$, $\varphi_2$ and $\eta$, respectively; after electroweak symmetry breaking they generate both trilinear interactions with the Higgs-like scalar and quartic interactions with Higgs pairs, thereby controlling annihilation into Standard Model final states and spin-independent direct-detection rates. Similarly, $\lambda_{10}$, $\lambda_{11}$ and $\lambda_{12}$ are important because they couple the dark scalars to the singlet field $\sigma$, opening annihilation channels through the singlet-like scalar sector and modifying the corresponding scalar masses. The parameter $\lambda_2$ affects the singlet-like scalar mass and therefore the mediator structure relevant for DM annihilation. Although $\lambda_1$ and $\lambda_6$ also affect the CP-even scalar spectrum and the Higgs--singlet mixing, we keep them fixed in the second scan in order to satisfy Higgs data, with $\lambda_1$ fixed by the measured SM-like Higgs mass and $\lambda_6$ chosen to suppress Higgs--singlet mixing. The significant importance of $\lambda_{16}$ follows from its role in the splitting of the CP-even and CP-odd components of $\varphi_2$, affecting which component becomes stable and how efficiently it annihilates. Finally, the fermionic parameters $M_{\Psi_{12}}$, $M_{\Psi_{21}}$, $M_{N_{12}}$ and $M_{N_{21}}$ enter the neutral-fermion spectrum and interactions, and therefore affect the regions where a fermionic state contributes significantly to the relic abundance.

Since one of our benchmark scenarios will be $\Psi$-dominated, we will also vary the couplings $y_{\Psi}$, even though they have relatively low SHAP importance. At this point, it is important to emphasize that a low SHAP importance does not necessarily imply that a parameter is physically irrelevant. A clear counterexample is provided by couplings that connect different DM sectors, such as $y_{\Psi}$, which couples the $\Psi$ and $\eta$ sectors. Varying $y_{\Psi}$ can significantly affect the conversion between these sectors. However, if a change in the dominant DM scenario can be compensated for by other parameters while still reproducing the observed relic abundance, then $y_{\Psi}$ effectively corresponds to a flat direction in the relic-density prediction. Since the SHAP analysis only quantifies the importance of parameters for predicting the relic abundance, independently of which DM scenario realizes it, such a parameter can receive a low SHAP value despite having a non-trivial physical impact, such as changing the dominant DM candidate. The parameters retained for the final DM scan were sampled logarithmically over the ranges
\begin{align}
v_\sigma
&\in \left[10^2,\,2\times10^4\right]~\mathrm{GeV},
&
\mu_\eta,\;\mu_{\varphi_1},\;\mu_{\varphi_2}
&\in \left[10^2,\,10^8\right]~\mathrm{GeV},
\\
(M_\Psi)_{ij},\;(M_N)_{ij}
&\in \left[10^{-4},\,10^4\right]~\mathrm{GeV},
&
\lambda_i
&\in \left[10^{-4},\,3\right],
\\
(y_\Psi)_{ij}
&\in \left[10^{-4},\,1\right].
\end{align}

\bibliographystyle{utphys}
\bibliography{bibliography}

\providecommand{\href}[2]{#2}\begingroup\raggedright\begin{thebibliography}{100}

\bibitem{Minkowski:1977sc}
P.~Minkowski, ``{$\mu \to e\gamma$ at a Rate of One Out of $10^{9}$ Muon
  Decays?},'' \href{http://dx.doi.org/10.1016/0370-2693(77)90435-X}{{\em
  Phys.Lett.B} {\bfseries 67} (1977) 421--428}.

\bibitem{Yanagida:1979as}
T.~Yanagida, ``{Horizontal gauge symmetry and masses of neutrinos},'' {\em
  Conf. Proc. C} {\bfseries 7902131} (1979) 95--99.

\bibitem{Glashow:1979nm}
S.~L. Glashow, ``{The Future of Elementary Particle Physics},''
  \href{http://dx.doi.org/10.1007/978-1-4684-7197-7_15}{{\em NATO Sci. Ser. B}
  {\bfseries 61} (1980) 687}.

\bibitem{Mohapatra:1979ia}
R.~N. Mohapatra and G.~Senjanovic, ``{Neutrino Mass and Spontaneous Parity
  Nonconservation},'' \href{http://dx.doi.org/10.1103/PhysRevLett.44.912}{{\em
  Phys. Rev. Lett.} {\bfseries 44} (1980) 912}.

\bibitem{Gell-Mann:1979vob}
M.~Gell-Mann, P.~Ramond, and R.~Slansky, ``{Complex Spinors and Unified
  Theories},'' {\em Conf. Proc. C} {\bfseries 790927} (1979) 315--321,
  \href{http://arxiv.org/abs/1306.4669}{{\ttfamily arXiv:1306.4669 [hep-th]}}.

\bibitem{Schechter:1980gr}
J.~Schechter and J.~W.~F. Valle, ``{Neutrino Masses in $SU(2) \otimes U(1)$
  Theories},'' \href{http://dx.doi.org/10.1103/PhysRevD.22.2227}{{\em
  Phys.Rev.D} {\bfseries 22} (1980) 2227}.

\bibitem{Schechter:1981cv}
J.~Schechter and J.~W.~F. Valle, ``{Neutrino Decay and Spontaneous Violation of
  Lepton Number},'' \href{http://dx.doi.org/10.1103/PhysRevD.25.774}{{\em
  Phys.Rev.D} {\bfseries 25} (1982) 774}.

\bibitem{Wyler:1982dd}
D.~Wyler and L.~Wolfenstein, ``{Massless Neutrinos in Left-Right Symmetric
  Models},'' \href{http://dx.doi.org/10.1016/0550-3213(83)90482-0}{{\em Nucl.
  Phys. B} {\bfseries 218} (1983) 205--214}.

\bibitem{Mohapatra:1986bd}
R.~Mohapatra and J.~W.~F. Valle, ``{Neutrino Mass and Baryon Number
  Nonconservation in Superstring Models},''
  \href{http://dx.doi.org/10.1103/PhysRevD.34.1642}{{\em Phys.Rev.D} {\bfseries
  34} (1986) 1642}.

\bibitem{Gonzalez-Garcia:1988okv}
M.~González-García and J.~W.~F. Valle, ``{Fast Decaying Neutrinos and
  Observable Flavor Violation in a New Class of Majoron Models},''
  \href{http://dx.doi.org/10.1016/0370-2693(89)91131-3}{{\em Phys.Lett.B}
  {\bfseries 216} (1989) 360--366}.

\bibitem{GonzalezGarcia:1988rw}
M.~C. González-García and J.~W.~F. Valle, ``{Fast Decaying Neutrinos and
  Observable Flavor Violation in a New Class of Majoron Models},''
  \href{http://dx.doi.org/10.1016/0370-2693(89)91131-3}{{\em Phys. Lett. B}
  {\bfseries 216} (1989) 360--366}.

\bibitem{Akhmedov:1995ip}
E.~K. Akhmedov {\em et~al.}, ``{Left-right symmetry breaking in NJL
  approach},'' \href{http://dx.doi.org/10.1016/0370-2693(95)01504-3}{{\em
  Phys.Lett.B} {\bfseries 368} (1996) 270--280},
  \href{http://arxiv.org/abs/hep-ph/9507275}{{\ttfamily arXiv:hep-ph/9507275
  [hep-ph]}}.

\bibitem{Akhmedov:1995vm}
E.~K. Akhmedov {\em et~al.}, ``{Dynamical left-right symmetry breaking},''
  \href{http://dx.doi.org/10.1103/PhysRevD.53.2752}{{\em Phys.Rev.D} {\bfseries
  53} (1996) 2752--2780}, \href{http://arxiv.org/abs/hep-ph/9509255}{{\ttfamily
  arXiv:hep-ph/9509255 [hep-ph]}}.

\bibitem{Malinsky:2005bi}
M.~Malinsky, J.~Rom\~ao, and J.~W.~F. Valle, ``{Novel supersymmetric $SO(10)$
  seesaw mechanism},''
  \href{http://dx.doi.org/10.1103/PhysRevLett.95.161801}{{\em Phys.Rev.Lett.}
  {\bfseries 95} (2005) 161801},
  \href{http://arxiv.org/abs/hep-ph/0506296}{{\ttfamily arXiv:hep-ph/0506296
  [hep-ph]}}.

\bibitem{Malinsky:2009df}
M.~Malinsky, T.~Ohlsson, Z.-z. Xing, and H.~Zhang, ``{Non-unitary neutrino
  mixing and CP violation in the minimal inverse seesaw model},''
  \href{http://dx.doi.org/10.1016/j.physletb.2009.07.038}{{\em Phys. Lett. B}
  {\bfseries 679} (2009) 242--248},
  \href{http://arxiv.org/abs/0905.2889}{{\ttfamily arXiv:0905.2889 [hep-ph]}}.

\bibitem{Abada:2014vea}
A.~Abada and M.~Lucente, ``{Looking for the minimal inverse seesaw
  realisation},'' \href{http://dx.doi.org/10.1016/j.nuclphysb.2014.06.003}{{\em
  Nucl. Phys. B} {\bfseries 885} (2014) 651--678},
  \href{http://arxiv.org/abs/1401.1507}{{\ttfamily arXiv:1401.1507 [hep-ph]}}.

\bibitem{tHooft:1979rat}
G.~'t~Hooft, ``{Naturalness, chiral symmetry, and spontaneous chiral symmetry
  breaking},'' \href{http://dx.doi.org/10.1007/978-1-4684-7571-5_9}{{\em NATO
  Sci. Ser. B} {\bfseries 59} (1980) 135--157}.

\bibitem{Balakrishna:1988ks}
B.~S. Balakrishna, A.~L. Kagan, and R.~N. Mohapatra, ``{Quark Mixings and Mass
  Hierarchy From Radiative Corrections},''
  \href{http://dx.doi.org/10.1016/0370-2693(88)91676-0}{{\em Phys. Lett. B}
  {\bfseries 205} (1988) 345--352}.

\bibitem{Ma:1988fp}
E.~Ma, ``{Radiative Quark and Lepton Masses Through Soft Supersymmetry
  Breaking},'' \href{http://dx.doi.org/10.1103/PhysRevD.39.1922}{{\em Phys.
  Rev. D} {\bfseries 39} (1989) 1922}.

\bibitem{Ma:1989ys}
E.~Ma, D.~Ng, J.~T. Pantaleone, and G.-G. Wong, ``{One Loop Induced Fermion
  Masses and Exotic Interactions in a Standard Model Context},''
  \href{http://dx.doi.org/10.1103/PhysRevD.40.1586}{{\em Phys. Rev. D}
  {\bfseries 40} (1989) 1586}.

\bibitem{Ma:1990ce}
E.~Ma, ``{Hierarchical Radiative Quark and Lepton Mass Matrices},''
  \href{http://dx.doi.org/10.1103/PhysRevLett.64.2866}{{\em Phys. Rev. Lett.}
  {\bfseries 64} (1990) 2866--2869}.

\bibitem{Ma:1998dn}
E.~Ma, ``{Pathways to naturally small neutrino masses},''
  \href{http://dx.doi.org/10.1103/PhysRevLett.81.1171}{{\em Phys. Rev. Lett.}
  {\bfseries 81} (1998) 1171--1174},
  \href{http://arxiv.org/abs/hep-ph/9805219}{{\ttfamily arXiv:hep-ph/9805219}}.

\bibitem{Tao:1996vb}
Z.-j. Tao, ``{Radiative seesaw mechanism at weak scale},''
  \href{http://dx.doi.org/10.1103/PhysRevD.54.5693}{{\em Phys. Rev. D}
  {\bfseries 54} (1996) 5693--5697},
  \href{http://arxiv.org/abs/hep-ph/9603309}{{\ttfamily arXiv:hep-ph/9603309}}.

\bibitem{Ma:2006km}
E.~Ma, ``{Verifiable radiative seesaw mechanism of neutrino mass and dark
  matter},'' \href{http://dx.doi.org/10.1103/PhysRevD.73.077301}{{\em
  Phys.Rev.D} {\bfseries 73} (2006) 077301},
  \href{http://arxiv.org/abs/hep-ph/0601225}{{\ttfamily arXiv:hep-ph/0601225
  [hep-ph]}}.

\bibitem{Gu:2007ug}
P.-H. Gu and U.~Sarkar, ``{Radiative Neutrino Mass, Dark Matter and
  Leptogenesis},'' \href{http://dx.doi.org/10.1103/PhysRevD.77.105031}{{\em
  Phys. Rev. D} {\bfseries 77} (2008) 105031},
  \href{http://arxiv.org/abs/0712.2933}{{\ttfamily arXiv:0712.2933 [hep-ph]}}.

\bibitem{Ma:2008cu}
E.~Ma and D.~Suematsu, ``{Fermion Triplet Dark Matter and Radiative Neutrino
  Mass},'' \href{http://dx.doi.org/10.1142/S021773230903059X}{{\em Mod. Phys.
  Lett. A} {\bfseries 24} (2009) 583--589},
  \href{http://arxiv.org/abs/0809.0942}{{\ttfamily arXiv:0809.0942 [hep-ph]}}.

\bibitem{Hirsch:2013ola}
M.~Hirsch {\em et~al.}, ``{WIMP dark matter as radiative neutrino mass
  messenger},'' \href{http://dx.doi.org/10.1007/JHEP10(2013)149}{{\em JHEP}
  {\bfseries 10} (2013) 149}, \href{http://arxiv.org/abs/1307.8134}{{\ttfamily
  arXiv:1307.8134 [hep-ph]}}.

\bibitem{Aranda:2015xoa}
A.~Aranda and E.~Peinado, ``{A new radiative neutrino mass generation mechanism
  with higher dimensional scalar representations and custodial symmetry},''
  \href{http://dx.doi.org/10.1016/j.physletb.2016.01.007}{{\em Phys. Lett. B}
  {\bfseries 754} (2016) 11--13},
  \href{http://arxiv.org/abs/1508.01200}{{\ttfamily arXiv:1508.01200
  [hep-ph]}}.

\bibitem{Restrepo:2015ura}
D.~Restrepo, A.~Rivera, M.~S\'anchez-Pel\'aez, {\'O}.~Zapata, and W.~Tangarife,
  ``{Radiative Neutrino Masses in the Singlet-Doublet Fermion Dark Matter Model
  with Scalar Singlets},''
  \href{http://dx.doi.org/10.1103/PhysRevD.92.013005}{{\em Phys. Rev. D}
  {\bfseries 92} no.~1, (2015) 013005},
  \href{http://arxiv.org/abs/1504.07892}{{\ttfamily arXiv:1504.07892
  [hep-ph]}}.

\bibitem{Longas:2015sxk}
R.~Longas, D.~Portillo, D.~Restrepo, and {\'O}.~Zapata, ``{The Inert Zee
  Model},'' \href{http://dx.doi.org/10.1007/JHEP03(2016)162}{{\em JHEP}
  {\bfseries 03} (2016) 162}, \href{http://arxiv.org/abs/1511.01873}{{\ttfamily
  arXiv:1511.01873 [hep-ph]}}.

\bibitem{Fraser:2015zed}
S.~Fraser, E.~Ma, and M.~Zakeri, ``{Verifiable Associated Processes from
  Radiative Lepton Masses with Dark Matter},''
  \href{http://dx.doi.org/10.1103/PhysRevD.93.115019}{{\em Phys. Rev. D}
  {\bfseries 93} no.~11, (2016) 115019},
  \href{http://arxiv.org/abs/1511.07458}{{\ttfamily arXiv:1511.07458
  [hep-ph]}}.

\bibitem{Fraser:2015mhb}
S.~Fraser, C.~Kownacki, E.~Ma, and O.~Popov, ``{Type II Radiative Seesaw Model
  of Neutrino Mass with Dark Matter},''
  \href{http://dx.doi.org/10.1103/PhysRevD.93.013021}{{\em Phys. Rev. D}
  {\bfseries 93} no.~1, (2016) 013021},
  \href{http://arxiv.org/abs/1511.06375}{{\ttfamily arXiv:1511.06375
  [hep-ph]}}.

\bibitem{Wang:2015saa}
W.~Wang and Z.-L. Han, ``{Radiative linear seesaw model, dark matter, and
  $U(1)_{B-L}$},'' \href{http://dx.doi.org/10.1103/PhysRevD.92.095001}{{\em
  Phys. Rev. D} {\bfseries 92} (2015) 095001},
  \href{http://arxiv.org/abs/1508.00706}{{\ttfamily arXiv:1508.00706
  [hep-ph]}}.

\bibitem{Arbelaez:2016mhg}
C.~Arbel\'aez, A.~E. C\'arcamo~Hern\'andez, S.~Kovalenko, and I.~Schmidt,
  ``{Radiative Seesaw-type Mechanism of Fermion Masses and Non-trivial Quark
  Mixing},'' \href{http://dx.doi.org/10.1140/epjc/s10052-017-4948-9}{{\em Eur.
  Phys. J. C} {\bfseries 77} no.~6, (2017) 422},
  \href{http://arxiv.org/abs/1602.03607}{{\ttfamily arXiv:1602.03607
  [hep-ph]}}.

\bibitem{vonderPahlen:2016cbw}
F.~von~der Pahlen, G.~Palacio, D.~Restrepo, and {\'O}.~Zapata, ``{Radiative
  Type III Seesaw Model and its collider phenomenology},''
  \href{http://dx.doi.org/10.1103/PhysRevD.94.033005}{{\em Phys. Rev. D}
  {\bfseries 94} no.~3, (2016) 033005},
  \href{http://arxiv.org/abs/1605.01129}{{\ttfamily arXiv:1605.01129
  [hep-ph]}}.

\bibitem{Nomura:2016emz}
T.~Nomura and H.~Okada, ``{Radiatively induced Quark and Lepton Mass Model},''
  \href{http://dx.doi.org/10.1016/j.physletb.2016.08.023}{{\em Phys. Lett. B}
  {\bfseries 761} (2016) 190--196},
  \href{http://arxiv.org/abs/1606.09055}{{\ttfamily arXiv:1606.09055
  [hep-ph]}}.

\bibitem{Kownacki:2016hpm}
C.~Kownacki and E.~Ma, ``{Gauge $U(1)$ dark symmetry and radiative light
  fermion masses},''
  \href{http://dx.doi.org/10.1016/j.physletb.2016.06.024}{{\em Phys. Lett. B}
  {\bfseries 760} (2016) 59--62},
  \href{http://arxiv.org/abs/1604.01148}{{\ttfamily arXiv:1604.01148
  [hep-ph]}}.

\bibitem{Abada:2025ozu}
A.~Abada, A.~E. C{\'a}rcamo~Hern{\'a}ndez, and S.~Urrea, ``{Dynamical
  scotogenic generation of the linear and inverse seesaws},''
  \href{http://dx.doi.org/10.1007/JHEP05(2026)086}{{\em JHEP} {\bfseries 05}
  (2026) 086}, \href{http://arxiv.org/abs/2512.12029}{{\ttfamily
  arXiv:2512.12029 [hep-ph]}}.

\bibitem{Nomura:2017emk}
T.~Nomura and H.~Okada, ``{Loop induced type-II seesaw model and GeV dark
  matter with $U(1)_{B-L}$ gauge symmetry},''
  \href{http://dx.doi.org/10.1016/j.physletb.2017.10.033}{{\em Phys. Lett. B}
  {\bfseries 774} (2017) 575--581},
  \href{http://arxiv.org/abs/1704.08581}{{\ttfamily arXiv:1704.08581
  [hep-ph]}}.

\bibitem{Nomura:2017vzp}
T.~Nomura and H.~Okada, ``{Radiative neutrino mass in an alternative
  $U(1)_{B-L}$ gauge symmetry},''
  \href{http://dx.doi.org/10.1016/j.nuclphysb.2019.02.025}{{\em Nucl. Phys. B}
  {\bfseries 941} (2019) 586--599},
  \href{http://arxiv.org/abs/1705.08309}{{\ttfamily arXiv:1705.08309
  [hep-ph]}}.

\bibitem{Bernal:2017xat}
N.~Bernal, A.~E. C\'arcamo~Hern\'andez, I.~de~Medeiros~Varzielas, and
  S.~Kovalenko, ``{Fermion masses and mixings and dark matter constraints in a
  model with radiative seesaw mechanism},''
  \href{http://dx.doi.org/10.1007/JHEP05(2018)053}{{\em JHEP} {\bfseries 05}
  (2018) 053}, \href{http://arxiv.org/abs/1712.02792}{{\ttfamily
  arXiv:1712.02792 [hep-ph]}}.

\bibitem{Wang:2017mcy}
W.~Wang, R.~Wang, Z.-L. Han, and J.-Z. Han, ``{The $B-L$ Scotogenic Models for
  Dirac Neutrino Masses},''
  \href{http://dx.doi.org/10.1140/epjc/s10052-017-5446-9}{{\em Eur. Phys. J. C}
  {\bfseries 77} no.~12, (2017) 889},
  \href{http://arxiv.org/abs/1705.00414}{{\ttfamily arXiv:1705.00414
  [hep-ph]}}.

\bibitem{Bonilla:2018ynb}
C.~Bonilla, S.~Centelles-Chuli\'a, R.~Cepedello, E.~Peinado, and R.~Srivastava,
  ``{Dark matter stability and Dirac neutrinos using only Standard Model
  symmetries},'' \href{http://dx.doi.org/10.1103/PhysRevD.101.033011}{{\em
  Phys. Rev. D} {\bfseries 101} no.~3, (2020) 033011},
  \href{http://arxiv.org/abs/1812.01599}{{\ttfamily arXiv:1812.01599
  [hep-ph]}}.

\bibitem{Calle:2018ovc}
J.~Calle, D.~Restrepo, C.~E. Yaguna, and {\'O}.~Zapata, ``{Minimal radiative
  Dirac neutrino mass models},''
  \href{http://dx.doi.org/10.1103/PhysRevD.99.075008}{{\em Phys. Rev. D}
  {\bfseries 99} no.~7, (2019) 075008},
  \href{http://arxiv.org/abs/1812.05523}{{\ttfamily arXiv:1812.05523
  [hep-ph]}}.

\bibitem{Avila:2019hhv}
I.~M. {\'A}vila, V.~De~Romeri, L.~Duarte, and J.~W.~F. Valle, ``{Phenomenology
  of scotogenic scalar dark matter},''
  \href{http://dx.doi.org/10.1140/epjc/s10052-020-08480-z}{{\em Eur.Phys.J.C}
  {\bfseries 80} (2020) 908}, \href{http://arxiv.org/abs/1910.08422}{{\ttfamily
  arXiv:1910.08422 [hep-ph]}}.

\bibitem{CarcamoHernandez:2018aon}
A.~E. C\'arcamo~Hern\'andez and S.~F. King, ``{Muon anomalies and the $SU(5)$
  Yukawa relations},'' \href{http://dx.doi.org/10.1103/PhysRevD.99.095003}{{\em
  Phys. Rev. D} {\bfseries 99} no.~9, (2019) 095003},
  \href{http://arxiv.org/abs/1803.07367}{{\ttfamily arXiv:1803.07367
  [hep-ph]}}.

\bibitem{Alvarado:2021fbw}
C.~Alvarado, C.~Bonilla, J.~Leite, and J.~W.~F. Valle, ``{Phenomenology of
  fermion dark matter as neutrino mass mediator with gauged B-L},''
  \href{http://dx.doi.org/10.1016/j.physletb.2021.136292}{{\em Phys. Lett. B}
  {\bfseries 817} (2021) 136292},
  \href{http://arxiv.org/abs/2102.07216}{{\ttfamily arXiv:2102.07216
  [hep-ph]}}.

\bibitem{Mandal:2021yph}
S.~Mandal, R.~Srivastava, and J.~W.~F. Valle, ``{The simplest scoto-seesaw
  model: WIMP dark matter phenomenology and Higgs vacuum stability},''
  \href{http://dx.doi.org/10.1016/j.physletb.2021.136458}{{\em Phys.Lett.B}
  {\bfseries 819} (2021) 136458},
  \href{http://arxiv.org/abs/2104.13401}{{\ttfamily arXiv:2104.13401
  [hep-ph]}}.

\bibitem{Arbelaez:2022ejo}
C.~Arbel\'aez, R.~Cepedello, J.~C. Helo, M.~Hirsch, and S.~Kovalenko, ``{How
  many 1-loop neutrino mass models are there?},''
  \href{http://dx.doi.org/10.1007/JHEP08(2022)023}{{\em JHEP} {\bfseries 08}
  (2022) 023}, \href{http://arxiv.org/abs/2205.13063}{{\ttfamily
  arXiv:2205.13063 [hep-ph]}}.

\bibitem{Cepedello:2022xgb}
R.~Cepedello, P.~Escribano, and A.~Vicente, ``{Neutrino masses, flavor
  anomalies, and muon $g-2$ from dark loops},''
  \href{http://dx.doi.org/10.1103/PhysRevD.107.035034}{{\em Phys. Rev. D}
  {\bfseries 107} no.~3, (2023) 035034},
  \href{http://arxiv.org/abs/2209.02730}{{\ttfamily arXiv:2209.02730
  [hep-ph]}}.

\bibitem{CarcamoHernandez:2022vjk}
A.~E. C{\'a}rcamo~Hern{\'a}ndez, C.~Espinoza, J.~C. G{\'o}mez-Izquierdo,
  J.~Marchant~Gonz{\'a}lez, and M.~Mondrag{\'o}n, ``{Phenomenology of extended
  multiHiggs doublet models with $S_4$ family symmetry},''
  \href{http://dx.doi.org/10.1140/epjc/s10052-024-13633-5}{{\em Eur. Phys. J.
  C} {\bfseries 84} no.~11, (2024) 1239},
  \href{http://arxiv.org/abs/2212.12000}{{\ttfamily arXiv:2212.12000
  [hep-ph]}}.

\bibitem{Leite:2023gzl}
J.~Leite, S.~Sadhukhan, and J.~W.~F. Valle, ``{Dynamical scoto-seesaw mechanism
  with gauged B-L symmetry},''
  \href{http://dx.doi.org/10.1103/PhysRevD.109.035023}{{\em Phys. Rev. D}
  {\bfseries 109} no.~3, (2024) 035023},
  \href{http://arxiv.org/abs/2307.04840}{{\ttfamily arXiv:2307.04840
  [hep-ph]}}.

\bibitem{Kumar:2025cte}
R.~Kumar, N.~Nath, R.~Srivastava, and S.~Yadav, ``{Dirac Scoto inverse-seesaw
  from A$_{4}$ flavor symmetry},''
  \href{http://dx.doi.org/10.1007/JHEP10(2025)088}{{\em JHEP} {\bfseries 10}
  (2025) 088}, \href{http://arxiv.org/abs/2505.01407}{{\ttfamily
  arXiv:2505.01407 [hep-ph]}}.

\bibitem{Kumar:2025zvv}
R.~Kumar, H.~K. Prajapati, R.~Srivastava, and S.~Yadav, ``{Flavor imprints on
  novel low mass dark matter},''
  \href{http://dx.doi.org/10.1007/JHEP11(2025)094}{{\em JHEP} {\bfseries 11}
  (2025) 094}, \href{http://arxiv.org/abs/2510.02972}{{\ttfamily
  arXiv:2510.02972 [hep-ph]}}.

\bibitem{Krauss:2002px}
L.~M. Krauss, S.~Nasri, and M.~Trodden, ``{A Model for neutrino masses and dark
  matter},'' \href{http://dx.doi.org/10.1103/PhysRevD.67.085002}{{\em Phys.
  Rev. D} {\bfseries 67} (2003) 085002},
  \href{http://arxiv.org/abs/hep-ph/0210389}{{\ttfamily arXiv:hep-ph/0210389}}.

\bibitem{Aoki:2008av}
M.~Aoki, S.~Kanemura, and O.~Seto, ``{Neutrino mass, Dark Matter and Baryon
  Asymmetry via TeV-Scale Physics without Fine-Tuning},''
  \href{http://dx.doi.org/10.1103/PhysRevLett.102.051805}{{\em Phys. Rev.
  Lett.} {\bfseries 102} (2009) 051805},
  \href{http://arxiv.org/abs/0807.0361}{{\ttfamily arXiv:0807.0361 [hep-ph]}}.

\bibitem{Kajiyama:2013lja}
Y.~Kajiyama, H.~Okada, and K.~Yagyu, ``{$T_7$ Flavor Model in Three Loop Seesaw
  and Higgs Phenomenology},''
  \href{http://dx.doi.org/10.1007/JHEP10(2013)196}{{\em JHEP} {\bfseries 10}
  (2013) 196}, \href{http://arxiv.org/abs/1307.0480}{{\ttfamily arXiv:1307.0480
  [hep-ph]}}.

\bibitem{Ahriche:2014cda}
A.~Ahriche, C.-S. Chen, K.~L. McDonald, and S.~Nasri, ``{Three-loop model of
  neutrino mass with dark matter},''
  \href{http://dx.doi.org/10.1103/PhysRevD.90.015024}{{\em Phys. Rev. D}
  {\bfseries 90} (2014) 015024},
  \href{http://arxiv.org/abs/1404.2696}{{\ttfamily arXiv:1404.2696 [hep-ph]}}.

\bibitem{Ahriche:2014oda}
A.~Ahriche, K.~L. McDonald, and S.~Nasri, ``{A Model of Radiative Neutrino
  Mass: with or without Dark Matter},''
  \href{http://dx.doi.org/10.1007/JHEP10(2014)167}{{\em JHEP} {\bfseries 10}
  (2014) 167}, \href{http://arxiv.org/abs/1404.5917}{{\ttfamily arXiv:1404.5917
  [hep-ph]}}.

\bibitem{Hatanaka:2014tba}
H.~Hatanaka, K.~Nishiwaki, H.~Okada, and Y.~Orikasa, ``{A Three-Loop Neutrino
  Model with Global $U(1)$ Symmetry},''
  \href{http://dx.doi.org/10.1016/j.nuclphysb.2015.03.006}{{\em Nucl. Phys. B}
  {\bfseries 894} (2015) 268--283},
  \href{http://arxiv.org/abs/1412.8664}{{\ttfamily arXiv:1412.8664 [hep-ph]}}.

\bibitem{Chen:2014ska}
C.-S. Chen, K.~L. McDonald, and S.~Nasri, ``{A Class of Three-Loop Models with
  Neutrino Mass and Dark Matter},''
  \href{http://dx.doi.org/10.1016/j.physletb.2014.05.082}{{\em Phys. Lett. B}
  {\bfseries 734} (2014) 388--393},
  \href{http://arxiv.org/abs/1404.6033}{{\ttfamily arXiv:1404.6033 [hep-ph]}}.

\bibitem{Jin:2014glp}
L.-G. Jin, R.~Tang, and F.~Zhang, ``{A three-loop radiative neutrino mass model
  with dark matter},''
  \href{http://dx.doi.org/10.1016/j.physletb.2014.12.034}{{\em Phys. Lett. B}
  {\bfseries 741} (2015) 163--167},
  \href{http://arxiv.org/abs/1501.02020}{{\ttfamily arXiv:1501.02020
  [hep-ph]}}.

\bibitem{Okada:2015hia}
H.~Okada and K.~Yagyu, ``{Three-loop neutrino mass model with doubly charged
  particles from isodoublets},''
  \href{http://dx.doi.org/10.1103/PhysRevD.93.013004}{{\em Phys. Rev. D}
  {\bfseries 93} no.~1, (2016) 013004},
  \href{http://arxiv.org/abs/1508.01046}{{\ttfamily arXiv:1508.01046
  [hep-ph]}}.

\bibitem{Nishiwaki:2015iqa}
K.~Nishiwaki, H.~Okada, and Y.~Orikasa, ``{Three loop neutrino model with
  isolated $k^{\pm\pm}$},''
  \href{http://dx.doi.org/10.1103/PhysRevD.92.093013}{{\em Phys. Rev. D}
  {\bfseries 92} no.~9, (2015) 093013},
  \href{http://arxiv.org/abs/1507.02412}{{\ttfamily arXiv:1507.02412
  [hep-ph]}}.

\bibitem{Ahriche:2015wha}
A.~Ahriche, K.~L. McDonald, S.~Nasri, and T.~Toma, ``{A Model of Neutrino Mass
  and Dark Matter with an Accidental Symmetry},''
  \href{http://dx.doi.org/10.1016/j.physletb.2015.05.031}{{\em Phys. Lett. B}
  {\bfseries 746} (2015) 430--435},
  \href{http://arxiv.org/abs/1504.05755}{{\ttfamily arXiv:1504.05755
  [hep-ph]}}.

\bibitem{CarcamoHernandez:2015hjp}
A.~E. C{\'a}rcamo~Hern{\'a}ndez, ``{A novel and economical explanation for SM
  fermion masses and mixings},''
  \href{http://dx.doi.org/10.1140/epjc/s10052-016-4351-y}{{\em Eur. Phys. J. C}
  {\bfseries 76} no.~9, (2016) 503},
  \href{http://arxiv.org/abs/1512.09092}{{\ttfamily arXiv:1512.09092
  [hep-ph]}}.

\bibitem{Gu:2016xno}
P.-H. Gu, ``{High-scale leptogenesis with three-loop neutrino mass generation
  and dark matter},'' \href{http://dx.doi.org/10.1007/JHEP04(2017)159}{{\em
  JHEP} {\bfseries 04} (2017) 159},
  \href{http://arxiv.org/abs/1611.03256}{{\ttfamily arXiv:1611.03256
  [hep-ph]}}.

\bibitem{Cheung:2017efc}
K.~Cheung, T.~Nomura, and H.~Okada, ``{A Three-loop Neutrino Model with
  Leptoquark Triplet Scalars},''
  \href{http://dx.doi.org/10.1016/j.physletb.2017.03.021}{{\em Phys. Lett. B}
  {\bfseries 768} (2017) 359--364},
  \href{http://arxiv.org/abs/1701.01080}{{\ttfamily arXiv:1701.01080
  [hep-ph]}}.

\bibitem{Dutta:2018qei}
B.~Dutta, S.~Ghosh, I.~Gogoladze, and T.~Li, ``{Three-loop neutrino masses via
  new massive gauge bosons from $E_6$ GUT},''
  \href{http://dx.doi.org/10.1103/PhysRevD.98.055028}{{\em Phys. Rev. D}
  {\bfseries 98} no.~5, (2018) 055028},
  \href{http://arxiv.org/abs/1805.01866}{{\ttfamily arXiv:1805.01866
  [hep-ph]}}.

\bibitem{CarcamoHernandez:2019cbd}
A.~E. C{\'a}rcamo~Hern{\'a}ndez, S.~Kovalenko, R.~Pasechnik, and I.~Schmidt,
  ``{Sequentially loop-generated quark and lepton mass hierarchies in an
  extended Inert Higgs Doublet model},''
  \href{http://dx.doi.org/10.1007/JHEP06(2019)056}{{\em JHEP} {\bfseries 06}
  (2019) 056}, \href{http://arxiv.org/abs/1901.02764}{{\ttfamily
  arXiv:1901.02764 [hep-ph]}}.

\bibitem{Cepedello:2020lul}
R.~Cepedello, M.~Hirsch, P.~Rocha-Mor{\'a}n, and A.~Vicente, ``{Minimal 3-loop
  neutrino mass models and charged lepton flavor violation},''
  \href{http://dx.doi.org/10.1007/JHEP08(2020)067}{{\em JHEP} {\bfseries 08}
  (2020) 067}, \href{http://arxiv.org/abs/2005.00015}{{\ttfamily
  arXiv:2005.00015 [hep-ph]}}.

\bibitem{Hernandez:2021iss}
A.~E.~C. Hern\'andez, S.~Kovalenko, M.~Maniatis, and I.~Schmidt, ``{Fermion
  mass hierarchy and g \ensuremath{-} 2 anomalies in an extended 3HDM Model},''
  \href{http://dx.doi.org/10.1007/JHEP10(2021)036}{{\em JHEP} {\bfseries 10}
  (2021) 036}, \href{http://arxiv.org/abs/2104.07047}{{\ttfamily
  arXiv:2104.07047 [hep-ph]}}.

\bibitem{Abada:2022dvm}
A.~Abada, N.~Bernal, A.~E. C{\'a}rcamo~Hern{\'a}ndez, S.~Kovalenko, T.~B.
  de~Melo, and T.~Toma, ``{Phenomenological and cosmological implications of a
  scotogenic three-loop neutrino mass model},''
  \href{http://dx.doi.org/10.1007/JHEP03(2023)035}{{\em JHEP} {\bfseries 03}
  (2023) 035}, \href{http://arxiv.org/abs/2212.06852}{{\ttfamily
  arXiv:2212.06852 [hep-ph]}}.

\bibitem{Bonilla:2023wok}
C.~Bonilla, A.~E. Cárcamo~Hernández, S.~Kovalenko, H.~Lee, R.~Pasechnik, and
  I.~Schmidt, ``{Fermion mass hierarchy in an extended left-right symmetric
  model},'' \href{http://dx.doi.org/10.1007/JHEP12(2023)075}{{\em JHEP}
  {\bfseries 12} (2023) 075}, \href{http://arxiv.org/abs/2305.11967}{{\ttfamily
  arXiv:2305.11967 [hep-ph]}}.

\bibitem{Abada:2023znk}
T.~B. de~Melo, A.~Abada, N.~Bernal, A.~E. C{\'a}rcamo~Hern{\'a}ndez,
  S.~Kovalenko, and T.~Toma, ``{Phenomenology of a scotogenic neutrino mass
  model at 3-loops},'' \href{http://dx.doi.org/10.22323/1.441.0255}{{\em PoS}
  {\bfseries TAUP2023} (2024) 255},
  \href{http://arxiv.org/abs/2311.14716}{{\ttfamily arXiv:2311.14716
  [hep-ph]}}.

\bibitem{Abada:2023zbb}
A.~Abada, N.~Bernal, A.~E. C{\'a}rcamo~Hern{\'a}ndez, S.~Kovalenko, and T.~B.
  de~Melo, ``{Three-loop inverse scotogenic seesaw models},''
  \href{http://dx.doi.org/10.1007/JHEP05(2024)035}{{\em JHEP} {\bfseries 05}
  (2024) 035}, \href{http://arxiv.org/abs/2312.14105}{{\ttfamily
  arXiv:2312.14105 [hep-ph]}}.

\bibitem{Loualidi:2025jqn}
M.~A. Loualidi, S.~Nasri, and M.~A. Rivera, ``{Radiative neutrino mass
  generation and dark matter through vectorlike leptons},''
  \href{http://dx.doi.org/10.1103/w9js-sdgh}{{\em Phys. Rev. D} {\bfseries 113}
  no.~7, (2026) 075008}, \href{http://arxiv.org/abs/2511.16762}{{\ttfamily
  arXiv:2511.16762 [hep-ph]}}.

\bibitem{Ma:2009gu}
E.~Ma, ``{Radiative inverse seesaw mechanism for nonzero neutrino mass},''
  \href{http://dx.doi.org/10.1103/PhysRevD.80.013013}{{\em Phys. Rev. D}
  {\bfseries 80} (2009) 013013},
  \href{http://arxiv.org/abs/0904.4450}{{\ttfamily arXiv:0904.4450 [hep-ph]}}.

\bibitem{Bazzocchi:2010dt}
F.~Bazzocchi, ``{Minimal Dynamical Inverse See Saw},''
  \href{http://dx.doi.org/10.1103/PhysRevD.83.093009}{{\em Phys. Rev. D}
  {\bfseries 83} (2011) 093009},
  \href{http://arxiv.org/abs/1011.6299}{{\ttfamily arXiv:1011.6299 [hep-ph]}}.

\bibitem{Law:2012mj}
S.~S.~C. Law and K.~L. McDonald, ``{Inverse seesaw and dark matter in models
  with exotic lepton triplets},''
  \href{http://dx.doi.org/10.1016/j.physletb.2012.06.044}{{\em Phys. Lett. B}
  {\bfseries 713} (2012) 490--494},
  \href{http://arxiv.org/abs/1204.2529}{{\ttfamily arXiv:1204.2529 [hep-ph]}}.

\bibitem{Das:2012ze}
A.~Das and N.~Okada, ``{Inverse seesaw neutrino signatures at the LHC and
  ILC},'' \href{http://dx.doi.org/10.1103/PhysRevD.88.113001}{{\em Phys. Rev.
  D} {\bfseries 88} (2013) 113001},
  \href{http://arxiv.org/abs/1207.3734}{{\ttfamily arXiv:1207.3734 [hep-ph]}}.

\bibitem{Okada:2012np}
H.~Okada and T.~Toma, ``{Fermionic Dark Matter in Radiative Inverse Seesaw
  Model with $U(1)_{B-L}$},''
  \href{http://dx.doi.org/10.1103/PhysRevD.86.033011}{{\em Phys. Rev. D}
  {\bfseries 86} (2012) 033011},
  \href{http://arxiv.org/abs/1207.0864}{{\ttfamily arXiv:1207.0864 [hep-ph]}}.

\bibitem{Fraser:2014yha}
S.~Fraser, E.~Ma, and O.~Popov, ``{Scotogenic Inverse Seesaw Model of Neutrino
  Mass},'' \href{http://dx.doi.org/10.1016/j.physletb.2014.08.069}{{\em Phys.
  Lett. B} {\bfseries 737} (2014) 280--282},
  \href{http://arxiv.org/abs/1408.4785}{{\ttfamily arXiv:1408.4785 [hep-ph]}}.

\bibitem{Ahriche:2016acx}
A.~Ahriche, S.~M. Boucenna, and S.~Nasri, ``{Dark Radiative Inverse Seesaw
  Mechanism},'' \href{http://dx.doi.org/10.1103/PhysRevD.93.075036}{{\em Phys.
  Rev. D} {\bfseries 93} no.~7, (2016) 075036},
  \href{http://arxiv.org/abs/1601.04336}{{\ttfamily arXiv:1601.04336
  [hep-ph]}}.

\bibitem{CarcamoHernandez:2013krw}
A.~E. C{\'a}rcamo~Hern{\'a}ndez, R.~Martínez, and F.~Ochoa, ``{Fermion masses
  and mixings in the 3-3-1 model with right-handed neutrinos based on the $S_3$
  flavor symmetry},''
  \href{http://dx.doi.org/10.1140/epjc/s10052-016-4480-3}{{\em Eur. Phys. J. C}
  {\bfseries 76} no.~11, (2016) 634},
  \href{http://arxiv.org/abs/1309.6567}{{\ttfamily arXiv:1309.6567 [hep-ph]}}.

\bibitem{Das:2017ski}
A.~Das, T.~Nomura, H.~Okada, and S.~Roy, ``{Generation of a radiative neutrino
  mass in the linear seesaw framework, charged lepton flavor violation, and
  dark matter},'' \href{http://dx.doi.org/10.1103/PhysRevD.96.075001}{{\em
  Phys. Rev. D} {\bfseries 96} no.~7, (2017) 075001},
  \href{http://arxiv.org/abs/1704.02078}{{\ttfamily arXiv:1704.02078
  [hep-ph]}}.

\bibitem{CarcamoHernandez:2017owh}
A.~E. C{\'a}rcamo~Hern{\'a}ndez, S.~Kovalenko, J.~W.~F. Valle, and
  C.~Vaquera-Araujo, ``{Predictive Pati-Salam theory of fermion masses and
  mixing},'' \href{http://dx.doi.org/10.1007/JHEP07(2017)118}{{\em JHEP}
  {\bfseries 07} (2017) 118}, \href{http://arxiv.org/abs/1705.06320}{{\ttfamily
  arXiv:1705.06320 [hep-ph]}}.

\bibitem{CarcamoHernandez:2018hst}
A.~E. C\'arcamo~Hern\'andez, S.~Kovalenko, J.~W.~F. Valle, and C.~A.
  Vaquera-Araujo, ``{Neutrino predictions from a left-right symmetric flavored
  extension of the standard model},''
  \href{http://dx.doi.org/10.1007/JHEP02(2019)065}{{\em JHEP} {\bfseries 02}
  (2019) 065}, \href{http://arxiv.org/abs/1811.03018}{{\ttfamily
  arXiv:1811.03018 [hep-ph]}}.

\bibitem{CarcamoHernandez:2018iel}
A.~E. C{\'a}rcamo~Hern{\'a}ndez, H.~N. Long, and V.~V. Vien, ``{The first
  $\Delta(27)$ flavor 3-3-1 model with low scale seesaw mechanism},''
  \href{http://dx.doi.org/10.1140/epjc/s10052-018-6284-0}{{\em Eur. Phys. J. C}
  {\bfseries 78} no.~10, (2018) 804},
  \href{http://arxiv.org/abs/1803.01636}{{\ttfamily arXiv:1803.01636
  [hep-ph]}}.

\bibitem{Bertuzzo:2018ftf}
E.~Bertuzzo, S.~Jana, P.~A.~N. Machado, and R.~Zukanovich~Funchal, ``{Neutrino
  Masses and Mixings Dynamically Generated by a Light Dark Sector},''
  \href{http://dx.doi.org/10.1016/j.physletb.2019.02.023}{{\em Phys. Lett. B}
  {\bfseries 791} (2019) 210--214},
  \href{http://arxiv.org/abs/1808.02500}{{\ttfamily arXiv:1808.02500
  [hep-ph]}}.

\bibitem{Mandal:2019oth}
S.~Mandal, N.~Rojas, R.~Srivastava, and J.~W.~F. Valle, ``{Dark matter as the
  origin of neutrino mass in the inverse seesaw mechanism},''
  \href{http://dx.doi.org/10.1016/j.physletb.2021.136609}{{\em Phys.Lett.B}
  {\bfseries 821} (2021) 136609},
  \href{http://arxiv.org/abs/1907.07728}{{\ttfamily arXiv:1907.07728
  [hep-ph]}}.

\bibitem{Das:2019pua}
A.~Das, S.~Goswami, K.~N. Vishnudath, and T.~Nomura, ``{Constraining a general
  U(1)$^\prime$ inverse seesaw model from vacuum stability, dark matter and
  collider},'' \href{http://dx.doi.org/10.1103/PhysRevD.101.055026}{{\em Phys.
  Rev. D} {\bfseries 101} no.~5, (2020) 055026},
  \href{http://arxiv.org/abs/1905.00201}{{\ttfamily arXiv:1905.00201
  [hep-ph]}}.

\bibitem{CarcamoHernandez:2019eme}
A.~E. C\'arcamo~Hern\'andez and S.~F. King, ``{Littlest Inverse Seesaw
  Model},'' \href{http://dx.doi.org/10.1016/j.nuclphysb.2020.114950}{{\em Nucl.
  Phys. B} {\bfseries 953} (2020) 114950},
  \href{http://arxiv.org/abs/1903.02565}{{\ttfamily arXiv:1903.02565
  [hep-ph]}}.

\bibitem{CarcamoHernandez:2019pmy}
A.~E. C\'arcamo~Hern\'andez, J.~Marchant~Gonz\'alez, and U.~J. Salda\~na
  Salazar, ``{Viable low-scale model with universal and inverse seesaw
  mechanisms},'' \href{http://dx.doi.org/10.1103/PhysRevD.100.035024}{{\em
  Phys. Rev. D} {\bfseries 100} no.~3, (2019) 035024},
  \href{http://arxiv.org/abs/1904.09993}{{\ttfamily arXiv:1904.09993
  [hep-ph]}}.

\bibitem{CarcamoHernandez:2019vih}
A.~E. C\'arcamo~Hern\'andez, Y.~Hidalgo~Vel\'asquez, and N.~A. P\'erez-Julve,
  ``{A 3-3-1 model with low scale seesaw mechanisms},''
  \href{http://dx.doi.org/10.1140/epjc/s10052-019-7325-z}{{\em Eur. Phys. J. C}
  {\bfseries 79} no.~10, (2019) 828},
  \href{http://arxiv.org/abs/1905.02323}{{\ttfamily arXiv:1905.02323
  [hep-ph]}}.

\bibitem{CarcamoHernandez:2019lhv}
A.~E. C\'arcamo~Hern\'andez, D.~T. Huong, and H.~N. Long, ``{Minimal model for
  the fermion flavor structure, mass hierarchy, dark matter, leptogenesis, and
  the electron and muon anomalous magnetic moments},''
  \href{http://dx.doi.org/10.1103/PhysRevD.102.055002}{{\em Phys. Rev. D}
  {\bfseries 102} no.~5, (2020) 055002},
  \href{http://arxiv.org/abs/1910.12877}{{\ttfamily arXiv:1910.12877
  [hep-ph]}}.

\bibitem{Hernandez:2021uxx}
A.~E.~C. Hern\'andez and I.~Schmidt, ``{A renormalizable left-right symmetric
  model with low scale seesaw mechanisms},''
  \href{http://dx.doi.org/10.1016/j.nuclphysb.2022.115696}{{\em Nucl. Phys. B}
  {\bfseries 976} (2022) 115696},
  \href{http://arxiv.org/abs/2101.02718}{{\ttfamily arXiv:2101.02718
  [hep-ph]}}.

\bibitem{Hernandez:2021xet}
A.~E.~C. Hern\'andez, D.~T. Huong, and I.~Schmidt, ``{Universal inverse seesaw
  mechanism as a source of the SM fermion mass hierarchy},''
  \href{http://dx.doi.org/10.1140/epjc/s10052-022-10011-x}{{\em Eur. Phys. J.
  C} {\bfseries 82} no.~1, (2022) 63},
  \href{http://arxiv.org/abs/2109.12118}{{\ttfamily arXiv:2109.12118
  [hep-ph]}}.

\bibitem{Hernandez:2021kju}
A.~E.~C. Hern\'andez, C.~Espinoza, J.~C. G\'omez-Izquierdo, and M.~Mondrag\'on,
  ``{Fermion masses and mixings, dark matter, leptogenesis and $g-2$ muon
  anomaly in an extended 2HDM with inverse seesaw},''
  \href{http://dx.doi.org/10.1140/epjp/s13360-022-03432-w}{{\em Eur. Phys. J.
  Plus} {\bfseries 137} no.~11, (2022) 1224},
  \href{http://arxiv.org/abs/2104.02730}{{\ttfamily arXiv:2104.02730
  [hep-ph]}}.

\bibitem{Nomura:2021adf}
T.~Nomura, H.~Okada, and P.~Sanyal, ``{A radiatively induced inverse seesaw
  model with hidden U(1) gauge symmetry},''
  \href{http://dx.doi.org/10.1140/epjc/s10052-022-10662-w}{{\em Eur. Phys. J.
  C} {\bfseries 82} no.~8, (2022) 697},
  \href{http://arxiv.org/abs/2103.09494}{{\ttfamily arXiv:2103.09494
  [hep-ph]}}.

\bibitem{Hernandez:2021mxo}
A.~E.~C. Hern\'andez, H.~N. Long, M.~L. Mora-Urrutia, N.~H. Thao, and V.~V.
  Vien, ``{Fermion masses and mixings and $g-2$ muon anomaly in a 3-3-1 model
  with $D_4$ family symmetry},''
  \href{http://dx.doi.org/10.1140/epjc/s10052-022-10639-9}{{\em Eur. Phys. J.
  C} {\bfseries 82} no.~8, (2022) 769},
  \href{http://arxiv.org/abs/2104.04559}{{\ttfamily arXiv:2104.04559
  [hep-ph]}}.

\bibitem{Abada:2021yot}
A.~Abada, N.~Bernal, A.~E.~C. Hern\'andez, X.~Marcano, and G.~Piazza, ``{Gauged
  inverse seesaw from dark matter},''
  \href{http://dx.doi.org/10.1140/epjc/s10052-021-09535-5}{{\em Eur. Phys. J.
  C} {\bfseries 81} no.~8, (2021) 758},
  \href{http://arxiv.org/abs/2107.02803}{{\ttfamily arXiv:2107.02803
  [hep-ph]}}.

\bibitem{Bonilla:2023egs}
C.~Bonilla, A.~E. Cárcamo~Hernández, B.~Saez~Díaz, S.~Kovalenko, and
  J.~Marchant~González, ``{Dark matter from a radiative inverse seesaw majoron
  model},'' \href{http://dx.doi.org/10.1016/j.physletb.2023.138282}{{\em Phys.
  Lett. B} {\bfseries 847} (2023) 138282},
  \href{http://arxiv.org/abs/2306.08453}{{\ttfamily arXiv:2306.08453
  [hep-ph]}}.

\bibitem{Binh:2024lez}
V.~H. Binh, C.~Bonilla, A.~E. C{\'a}rcamo~Hern{\'a}ndez, D.~T. Huong, V.~K.~N.,
  H.~N. Long, P.~N. Thu, and I.~Schmidt, ``{Phenomenology of 3-3-1 models with
  a radiative inverse seesaw mechanism},''
  \href{http://dx.doi.org/10.1103/PhysRevD.110.075022}{{\em Phys. Rev. D}
  {\bfseries 110} no.~7, (2024) 075022},
  \href{http://arxiv.org/abs/2404.13373}{{\ttfamily arXiv:2404.13373
  [hep-ph]}}.

\bibitem{Abada:2024xpf}
A.~Abada, N.~Bernal, A.~E. C{\'a}rcamo~Hern{\'a}ndez, S.~Kovalenko, T.~B.
  de~Melo, and T.~Toma, ``{Pheno {\&} Cosmo Implications of Scotogenic 3-loop
  Neutrino Mass Models},'' in {\em {$58^{th}$ Rencontres de Moriond on
  Electroweak Interactions and Unified Theories}}.
\newblock 5, 2024.
\newblock \href{http://arxiv.org/abs/2405.12242}{{\ttfamily arXiv:2405.12242
  [hep-ph]}}.

\bibitem{deMelo:2024lux}
T.~B. de~Melo, A.~Abada, N.~Bernal, A.~E. C{\'a}rcamo~Hern{\'a}ndez,
  S.~Kovalenko, and T.~Toma, ``{Phenomenology of scotogenic-like 3-loop
  neutrino mass models},'' \href{http://dx.doi.org/10.22323/1.476.0183}{{\em
  PoS} {\bfseries ICHEP2024} (2025) 183},
  \href{http://arxiv.org/abs/2411.00020}{{\ttfamily arXiv:2411.00020
  [hep-ph]}}.

\bibitem{Pathak:2024sei}
G.~Pathak, P.~Das, and M.~K. Das, ``{Neutrino mass genesis in scoto-inverse
  seesaw with modular $A_4$},''
  \href{http://dx.doi.org/10.1140/epjc/s10052-025-14263-1}{{\em Eur. Phys. J.
  C} {\bfseries 85} no.~5, (2025) 569},
  \href{http://arxiv.org/abs/2411.13895}{{\ttfamily arXiv:2411.13895
  [hep-ph]}}.

\bibitem{Wang:2024qhe}
Z.~Wang, Y.~Reyimuaji, and N.~Yalikun, ``{Z4 symmetric inverse seesaw model for
  neutrino masses and FIMP dark matter},''
  \href{http://dx.doi.org/10.1103/3tvj-qmld}{{\em Phys. Rev. D} {\bfseries 112}
  no.~5, (2025) 055041}, \href{http://arxiv.org/abs/2412.15672}{{\ttfamily
  arXiv:2412.15672 [hep-ph]}}.

\bibitem{CarcamoHernandez:2024hll}
A.~E. C{\'a}rcamo~Hern{\'a}ndez, D.~T. Huong, H.~N. Long, and
  D.~Salinas-Arizmendi, ``{Fermion masses and mixings and charged lepton flavor
  violation in a 3-3-1 model with inverse seesaw},''
  \href{http://dx.doi.org/10.1093/ptep/ptaf067}{{\em PTEP} {\bfseries 6} (2025)
  063}, \href{http://arxiv.org/abs/2412.18550}{{\ttfamily arXiv:2412.18550
  [hep-ph]}}.

\bibitem{Huong:2025uwx}
D.~T. Huong, A.~E. C{\'a}rcamo~Hern{\'a}ndez, H.~T. Hung, T.~T. Hieu, N.~A.
  P{\'e}rez-Julve, and N.~T. Duy, ``{Extended IDM theory with low scale seesaw
  mechanisms},'' \href{http://arxiv.org/abs/2502.19488}{{\ttfamily
  arXiv:2502.19488 [hep-ph]}}.

\bibitem{Planck:2018vyg}
{\bfseries Planck} Collaboration, N.~Aghanim {\em et~al.}, ``{Planck 2018
  results. VI. Cosmological parameters},''
  \href{http://dx.doi.org/10.1051/0004-6361/201833910}{{\em Astron. Astrophys.}
  {\bfseries 641} (2020) A6}, \href{http://arxiv.org/abs/1807.06209}{{\ttfamily
  arXiv:1807.06209 [astro-ph.CO]}}. [Erratum: Astron.Astrophys. 652, C4
  (2021)].

\bibitem{Belanger:2001fz}
G.~Bélanger, F.~Boudjema, A.~Pukhov, and A.~Semenov, ``{MicrOMEGAs: A Program
  for calculating the relic density in the MSSM},''
  \href{http://dx.doi.org/10.1016/S0010-4655(02)00596-9}{{\em Comput. Phys.
  Commun.} {\bfseries 149} (2002) 103--120},
  \href{http://arxiv.org/abs/hep-ph/0112278}{{\ttfamily arXiv:hep-ph/0112278}}.

\bibitem{Alguero:2023zol}
G.~Alguero, G.~Bélanger, F.~Boudjema, S.~Chakraborti, A.~Goudelis, S.~Kraml,
  A.~Mjallal, and A.~Pukhov, ``{micrOMEGAs 6.0: N-component dark matter},''
  \href{http://dx.doi.org/10.1016/j.cpc.2024.109133}{{\em Comput. Phys.
  Commun.} {\bfseries 299} (2024) 109133},
  \href{http://arxiv.org/abs/2312.14894}{{\ttfamily arXiv:2312.14894
  [hep-ph]}}.

\bibitem{Belanger:2026asz}
G.~Bélanger, A.~Belyaev, N.~Bernal, F.~Boudjema, S.~Chakraborti, A.~Goudelis,
  and A.~Pukhov, ``{micrOMEGAs 7: Beyond standard cosmology},''
  \href{http://arxiv.org/abs/2606.06645}{{\ttfamily arXiv:2606.06645
  [hep-ph]}}.

\bibitem{Feroz:2008xx}
F.~Feroz, M.~P. Hobson, and M.~Bridges, ``{MultiNest: an efficient and robust
  Bayesian inference tool for cosmology and particle physics},''
  \href{http://dx.doi.org/10.1111/j.1365-2966.2009.14548.x}{{\em Mon. Not. Roy.
  Astron. Soc.} {\bfseries 398} (2009) 1601--1614},
  \href{http://arxiv.org/abs/0809.3437}{{\ttfamily arXiv:0809.3437
  [astro-ph]}}.

\bibitem{Feroz:2013hea}
F.~Feroz, M.~P. Hobson, E.~Cameron, and A.~N. Pettitt, ``{Importance Nested
  Sampling and the MultiNest Algorithm},''
  \href{http://dx.doi.org/10.21105/astro.1306.2144}{{\em Open J. Astrophys.}
  {\bfseries 2} no.~1, (2019) 10},
  \href{http://arxiv.org/abs/1306.2144}{{\ttfamily arXiv:1306.2144
  [astro-ph.IM]}}.

\bibitem{ATLAS:2022vkf}
{\bfseries ATLAS} Collaboration, G.~Aad {\em et~al.}, ``{A detailed map of
  Higgs boson interactions by the ATLAS experiment ten years after the
  discovery},'' \href{http://dx.doi.org/10.1038/s41586-022-04893-w}{{\em
  Nature} {\bfseries 607} no.~7917, (2022) 52--59},
  \href{http://arxiv.org/abs/2207.00092}{{\ttfamily arXiv:2207.00092
  [hep-ex]}}. [Erratum: Nature 612, E24 (2022)].

\bibitem{CMS:2022dwd}
{\bfseries CMS} Collaboration, A.~Tumasyan {\em et~al.}, ``{A portrait of the
  Higgs boson by the CMS experiment ten years after the discovery.},''
  \href{http://dx.doi.org/10.1038/s41586-022-04892-x}{{\em Nature} {\bfseries
  607} no.~7917, (2022) 60--68},
  \href{http://arxiv.org/abs/2207.00043}{{\ttfamily arXiv:2207.00043
  [hep-ex]}}. [Erratum: Nature 623, (2023)].

\bibitem{ParticleDataGroup:2024cfk}
{\bfseries Particle Data Group} Collaboration, S.~Navas {\em et~al.}, ``{Review
  of particle physics},''
  \href{http://dx.doi.org/10.1103/PhysRevD.110.030001}{{\em Phys. Rev. D}
  {\bfseries 110} no.~3, (2024) 030001}.

\bibitem{Raffelt:1996wa}
G.~G. Raffelt, {\em {Stars as laboratories for fundamental physics: The
  astrophysics of neutrinos, axions, and other weakly interacting particles}}.
\newblock University of Chicago Press, Chicago, 1996.

\bibitem{Kachelriess:2000qc}
M.~Kachelriess, R.~Tomas, and J.~W.~F. Valle, ``{Supernova bounds on Majoron
  emitting decays of light neutrinos},''
  \href{http://dx.doi.org/10.1103/PhysRevD.62.023004}{{\em Phys. Rev. D}
  {\bfseries 62} (2000) 023004},
  \href{http://arxiv.org/abs/hep-ph/0001039}{{\ttfamily arXiv:hep-ph/0001039}}.

\bibitem{Farzan:2002wx}
Y.~Farzan, ``{Bounds on the coupling of the Majoron to light neutrinos from
  supernova cooling},''
  \href{http://dx.doi.org/10.1103/PhysRevD.67.073015}{{\em Phys. Rev. D}
  {\bfseries 67} (2003) 073015},
  \href{http://arxiv.org/abs/hep-ph/0211375}{{\ttfamily arXiv:hep-ph/0211375}}.

\bibitem{Keung:2013mfa}
W.-Y. Keung, K.-W. Ng, H.~Tu, and T.-C. Yuan, ``{Supernova Bounds on Weinberg's
  Goldstone Bosons},'' \href{http://dx.doi.org/10.1103/PhysRevD.90.075014}{{\em
  Phys. Rev. D} {\bfseries 90} no.~7, (2014) 075014},
  \href{http://arxiv.org/abs/1312.3488}{{\ttfamily arXiv:1312.3488 [hep-ph]}}.

\bibitem{Weinberg:2013kea}
S.~Weinberg, ``{Goldstone Bosons as Fractional Cosmic Neutrinos},''
  \href{http://dx.doi.org/10.1103/PhysRevLett.110.241301}{{\em Phys. Rev.
  Lett.} {\bfseries 110} no.~24, (2013) 241301},
  \href{http://arxiv.org/abs/1305.1971}{{\ttfamily arXiv:1305.1971
  [astro-ph.CO]}}.

\bibitem{Bonilla:2015uwa}
C.~Bonilla, J.~W.~F. Valle, and J.~C. Rom{\~a}o, ``{Neutrino mass and invisible
  Higgs decays at the LHC},''
  \href{http://dx.doi.org/10.1103/PhysRevD.91.113015}{{\em Phys.Rev.D}
  {\bfseries 91} (2015) 113015},
  \href{http://arxiv.org/abs/1502.01649}{{\ttfamily arXiv:1502.01649
  [hep-ph]}}.

\bibitem{Peccei:1977hh}
R.~D. Peccei and H.~R. Quinn, ``{CP Conservation in the Presence of
  Instantons},'' \href{http://dx.doi.org/10.1103/PhysRevLett.38.1440}{{\em
  Phys. Rev. Lett.} {\bfseries 38} (1977) 1440--1443}.

\bibitem{Peccei:1977ur}
R.~D. Peccei and H.~R. Quinn, ``{Constraints Imposed by CP Conservation in the
  Presence of Instantons},''
  \href{http://dx.doi.org/10.1103/PhysRevD.16.1791}{{\em Phys. Rev. D}
  {\bfseries 16} (1977) 1791--1797}.

\bibitem{Weinberg:1977ma}
S.~Weinberg, ``{A New Light Boson?},''
  \href{http://dx.doi.org/10.1103/PhysRevLett.40.223}{{\em Phys. Rev. Lett.}
  {\bfseries 40} (1978) 223--226}.

\bibitem{Wilczek:1977pj}
F.~Wilczek, ``{Problem of Strong P and T Invariance in the Presence of
  Instantons},'' \href{http://dx.doi.org/10.1103/PhysRevLett.40.279}{{\em Phys.
  Rev. Lett.} {\bfseries 40} (1978) 279--282}.

\bibitem{DiLuzio:2020wdo}
L.~Di~Luzio, M.~Giannotti, E.~Nardi, and L.~Visinelli, ``{The landscape of QCD
  axion models},'' \href{http://dx.doi.org/10.1016/j.physrep.2020.06.002}{{\em
  Phys. Rept.} {\bfseries 870} (2020) 1--117},
  \href{http://arxiv.org/abs/2003.01100}{{\ttfamily arXiv:2003.01100
  [hep-ph]}}.

\bibitem{Barr:1992qq}
S.~M. Barr and D.~Seckel, ``{Planck scale corrections to axion models},''
  \href{http://dx.doi.org/10.1103/PhysRevD.46.539}{{\em Phys. Rev. D}
  {\bfseries 46} (1992) 539--549}.

\bibitem{Holman:1992us}
R.~Holman, S.~D.~H. Hsu, T.~W. Kephart, E.~W. Kolb, R.~Watkins, and L.~M.
  Widrow, ``{Solutions to the strong CP problem in a world with gravity},''
  \href{http://dx.doi.org/10.1016/0370-2693(92)90491-L}{{\em Phys. Lett. B}
  {\bfseries 282} (1992) 132--136},
  \href{http://arxiv.org/abs/hep-ph/9203206}{{\ttfamily arXiv:hep-ph/9203206}}.

\bibitem{Kamionkowski:1992mf}
M.~Kamionkowski and J.~March-Russell, ``{Planck scale physics and the
  Peccei-Quinn mechanism},''
  \href{http://dx.doi.org/10.1016/0370-2693(92)90492-M}{{\em Phys. Lett. B}
  {\bfseries 282} (1992) 137--141},
  \href{http://arxiv.org/abs/hep-th/9202003}{{\ttfamily arXiv:hep-th/9202003}}.

\bibitem{Akhmedov:1992hi}
E.~K. Akhmedov, Z.~G. Berezhiani, R.~N. Mohapatra, and G.~Senjanovic, ``{Planck
  scale effects on the majoron},''
  \href{http://dx.doi.org/10.1016/0370-2693(93)90887-N}{{\em Phys. Lett. B}
  {\bfseries 299} (1993) 90--93},
  \href{http://arxiv.org/abs/hep-ph/9209285}{{\ttfamily arXiv:hep-ph/9209285}}.

\bibitem{Rothstein:1992rh}
I.~Z. Rothstein, K.~S. Babu, and D.~Seckel, ``{Planck scale symmetry breaking
  and majoron physics},''
  \href{http://dx.doi.org/10.1016/0550-3213(93)90368-Y}{{\em Nucl. Phys. B}
  {\bfseries 403} (1993) 725--748},
  \href{http://arxiv.org/abs/hep-ph/9301213}{{\ttfamily arXiv:hep-ph/9301213}}.

\bibitem{Gu:2010ys}
P.-H. Gu, E.~Ma, and U.~Sarkar, ``{Pseudo-Majoron as Dark Matter},''
  \href{http://dx.doi.org/10.1016/j.physletb.2010.05.012}{{\em Phys. Lett. B}
  {\bfseries 690} (2010) 145--148},
  \href{http://arxiv.org/abs/1004.1919}{{\ttfamily arXiv:1004.1919 [hep-ph]}}.

\bibitem{deGiorgi:2023tvn}
A.~de~Giorgi, L.~Merlo, X.~Ponce~D{\'\i}az, and S.~Rigolin, ``{The minimal
  massive Majoron Seesaw Model},''
  \href{http://dx.doi.org/10.1007/JHEP03(2024)094}{{\em JHEP} {\bfseries 03}
  (2024) 094}, \href{http://arxiv.org/abs/2312.13417}{{\ttfamily
  arXiv:2312.13417 [hep-ph]}}.

\bibitem{Georis:2025kzv}
Y.~Georis, J.~Sheng, S.~Urrea, and T.~T. Yanagida, ``{Testable inverse seesaw
  motivated from a high quality QCD axion},''
  \href{http://dx.doi.org/10.1007/JHEP04(2026)004}{{\em JHEP} {\bfseries 04}
  (2026) 004}, \href{http://arxiv.org/abs/2512.13158}{{\ttfamily
  arXiv:2512.13158 [hep-ph]}}.

\bibitem{Sheng:2025sou}
J.~Sheng and T.~T. Yanagida, ``{High quality QCD axion in the Standard
  Model},'' \href{http://dx.doi.org/10.1103/h9ws-xgst}{{\em Phys. Rev. D}
  {\bfseries 113} no.~5, (2026) 055010},
  \href{http://arxiv.org/abs/2510.17370}{{\ttfamily arXiv:2510.17370
  [hep-ph]}}.

\bibitem{Fischer:2026bka}
{\bfseries Gfitter Group} Collaboration, Y.~Fischer, J.~Haller, A.~Hoecker,
  R.~Kogler, F.~Labe, K.~M{\"o}nig, D.~Schwarz, and J.~Stelzer, ``{The Higgs
  boson through the lens of electroweak precision data},''
  \href{http://arxiv.org/abs/2607.09861}{{\ttfamily arXiv:2607.09861
  [hep-ph]}}.

\bibitem{Chikashige:1980ui}
Y.~Chikashige, R.~N. Mohapatra, and R.~Peccei, ``{Are There Real Goldstone
  Bosons Associated with Broken Lepton Number?},''
  \href{http://dx.doi.org/10.1016/0370-2693(81)90011-3}{{\em Phys.Lett.B}
  {\bfseries 98} (1981) 265--268}.

\bibitem{Ivanez-Ballesteros:2024nwsz}
P.~Iv{\'a}{\~n}ez-Ballesteros and M.~C. Volpe, ``{Constraints on
  neutrino-Majoron coupling using SN1987A data},''
  \href{http://dx.doi.org/10.1103/d4vp-m261}{{\em Phys. Rev. D} {\bfseries 112}
  no.~10, (2025) L101301}, \href{http://arxiv.org/abs/2410.11517}{{\ttfamily
  arXiv:2410.11517 [hep-ph]}}.

\bibitem{Law:2013gma}
S.~S.~C. Law and K.~L. McDonald, ``{Generalized inverse seesaw mechanisms},''
  \href{http://dx.doi.org/10.1103/PhysRevD.87.113003}{{\em Phys. Rev. D}
  {\bfseries 87} no.~11, (2013) 113003},
  \href{http://arxiv.org/abs/1303.4887}{{\ttfamily arXiv:1303.4887 [hep-ph]}}.

\bibitem{Grimus:2000vj}
W.~Grimus and L.~Lavoura, ``{The Seesaw mechanism at arbitrary order:
  Disentangling the small scale from the large scale},''
  \href{http://dx.doi.org/10.1088/1126-6708/2000/11/042}{{\em JHEP} {\bfseries
  11} (2000) 042}, \href{http://arxiv.org/abs/hep-ph/0008179}{{\ttfamily
  arXiv:hep-ph/0008179}}.

\bibitem{Esteban:2024eli}
I.~Esteban, M.~C. González-García, M.~Maltoni, I.~Martínez-Soler, J.~P.
  Pinheiro, and T.~Schwetz, ``{NuFit-6.0: updated global analysis of
  three-flavor neutrino oscillations},''
  \href{http://dx.doi.org/10.1007/JHEP12(2024)216}{{\em JHEP} {\bfseries 12}
  (2024) 216}, \href{http://arxiv.org/abs/2410.05380}{{\ttfamily
  arXiv:2410.05380 [hep-ph]}}.

\bibitem{Casas:2001sr}
J.~Casas and A.~Ibarra, ``{Oscillating neutrinos and $\mu \to e, \gamma$},''
  \href{http://dx.doi.org/10.1016/S0550-3213(01)00475-8}{{\em Nucl.Phys.B}
  {\bfseries 618} (2001) 171--204},
  \href{http://arxiv.org/abs/hep-ph/0103065}{{\ttfamily arXiv:hep-ph/0103065
  [hep-ph]}}.

\bibitem{Kovalenko:2009td}
S.~Kovalenko, Z.~Lu, and I.~Schmidt, ``{Lepton Number Violating Processes
  Mediated by Majorana Neutrinos at Hadron Colliders},''
  \href{http://dx.doi.org/10.1103/PhysRevD.80.073014}{{\em Phys. Rev. D}
  {\bfseries 80} (2009) 073014},
  \href{http://arxiv.org/abs/0907.2533}{{\ttfamily arXiv:0907.2533 [hep-ph]}}.

\bibitem{Faessler:2014kka}
A.~Faessler, M.~Gonz\'alez, S.~Kovalenko, and F.~\v{S}imkovic, ``{Arbitrary
  mass Majorana neutrinos in neutrinoless double beta decay},''
  \href{http://dx.doi.org/10.1103/PhysRevD.90.096010}{{\em Phys. Rev. D}
  {\bfseries 90} no.~9, (2014) 096010},
  \href{http://arxiv.org/abs/1408.6077}{{\ttfamily arXiv:1408.6077 [hep-ph]}}.

\bibitem{Babic:2018ikc}
A.~Babi\v{c}, S.~Kovalenko, M.~I. Krivoruchenko, and F.~\v{S}imkovic,
  ``{Interpolating formula for the $0\nu\beta\beta$-decay half-life in the case
  of light and heavy neutrino mass mechanisms},''
  \href{http://dx.doi.org/10.1103/PhysRevD.98.015003}{{\em Phys. Rev. D}
  {\bfseries 98} no.~1, (2018) 015003},
  \href{http://arxiv.org/abs/1804.04218}{{\ttfamily arXiv:1804.04218
  [hep-ph]}}.

\bibitem{KamLAND-Zen:2022tow}
{\bfseries KamLAND-Zen} Collaboration, S.~Abe {\em et~al.}, ``{Search for the
  Majorana Nature of Neutrinos in the Inverted Mass Ordering Region with
  KamLAND-Zen},'' \href{http://dx.doi.org/10.1103/PhysRevLett.130.051801}{{\em
  Phys. Rev. Lett.} {\bfseries 130} no.~5, (2023) 051801},
  \href{http://arxiv.org/abs/2203.02139}{{\ttfamily arXiv:2203.02139
  [hep-ex]}}.

\bibitem{Maniatis:2006fs}
M.~Maniatis, A.~von Manteuffel, O.~Nachtmann, and F.~Nagel, ``{Stability and
  symmetry breaking in the general two-Higgs-doublet model},''
  \href{http://dx.doi.org/10.1140/epjc/s10052-006-0016-6}{{\em Eur. Phys. J. C}
  {\bfseries 48} (2006) 805--823},
  \href{http://arxiv.org/abs/hep-ph/0605184}{{\ttfamily arXiv:hep-ph/0605184}}.

\bibitem{Bhattacharyya:2015nca}
G.~Bhattacharyya and D.~Das, ``{Scalar sector of two-Higgs-doublet models: A
  minireview},'' \href{http://dx.doi.org/10.1007/s12043-016-1252-4}{{\em
  Pramana} {\bfseries 87} no.~3, (2016) 40},
  \href{http://arxiv.org/abs/1507.06424}{{\ttfamily arXiv:1507.06424
  [hep-ph]}}.

\bibitem{Calibbi:2017uvl}
L.~Calibbi and G.~Signorelli, ``{Charged Lepton Flavour Violation: An
  Experimental and Theoretical Introduction},''
  \href{http://dx.doi.org/10.1393/ncr/i2018-10144-0}{{\em Riv. Nuovo Cim.}
  {\bfseries 41} no.~2, (2018) 71--174},
  \href{http://arxiv.org/abs/1709.00294}{{\ttfamily arXiv:1709.00294
  [hep-ph]}}.

\bibitem{SINDRUM:1987nra}
{\bfseries SINDRUM} Collaboration, U.~Bellgardt {\em et~al.}, ``{Search for the
  Decay $\mu^+ \to e^+ e^+ e^-$},''
  \href{http://dx.doi.org/10.1016/0550-3213(88)90462-2}{{\em Nucl. Phys. B}
  {\bfseries 299} (1988) 1--6}.

\bibitem{MEG:2013oxv}
{\bfseries MEG} Collaboration, J.~Adam {\em et~al.}, ``{New constraint on the
  existence of the $\mu^+ \to e^+\gamma$ decay},''
  \href{http://dx.doi.org/10.1103/PhysRevLett.110.201801}{{\em Phys.Rev.Lett.}
  {\bfseries 110} (2013) 201801},
  \href{http://arxiv.org/abs/1303.0754}{{\ttfamily arXiv:1303.0754 [hep-ex]}}.

\bibitem{MEGII:2025gzr}
{\bfseries MEG II} Collaboration, K.~Afanaciev {\em et~al.}, ``{New limit on
  the ${\mu^+ \rightarrow e^+ \gamma}$ decay with the MEG II experiment},''
  \href{http://dx.doi.org/10.1140/epjc/s10052-025-14906-3}{{\em Eur. Phys. J.
  C} {\bfseries 85} no.~10, (2025) 1177},
  \href{http://arxiv.org/abs/2504.15711}{{\ttfamily arXiv:2504.15711
  [hep-ex]}}. [Erratum: Eur.Phys.J.C 85, 1317 (2025)].

\bibitem{Baldini:2013ke}
A.~M. Baldini {\em et~al.}, ``{MEG Upgrade Proposal},''
  \href{http://arxiv.org/abs/1301.7225}{{\ttfamily arXiv:1301.7225
  [physics.ins-det]}}.

\bibitem{Blondel:2013ia}
A.~Blondel {\em et~al.}, ``{Research Proposal for an Experiment to Search for
  the Decay $\mu \to eee$},'' \href{http://arxiv.org/abs/1301.6113}{{\ttfamily
  arXiv:1301.6113 [physics.ins-det]}}.

\bibitem{SINDRUMII:2006dvw}
{\bfseries SINDRUM II} Collaboration, W.~H. Bertl {\em et~al.}, ``{A Search for
  muon to electron conversion in muonic gold},''
  \href{http://dx.doi.org/10.1140/epjc/s2006-02582-x}{{\em Eur. Phys. J. C}
  {\bfseries 47} (2006) 337--346}.

\bibitem{COMET:2009qeh}
{\bfseries COMET} Collaboration, Y.~G. Cui {\em et~al.}, ``{Conceptual design
  report for experimental search for lepton flavor violating $\mu^-$ - $e^-$
  conversion at sensitivity of $10^{-16}$ with a slow-extracted bunched proton
  beam (COMET)},''.

\bibitem{Mu2e:2022ggl}
{\bfseries Mu2e} Collaboration, F.~Abdi {\em et~al.}, ``{Mu2e Run I Sensitivity
  Projections for the Neutrinoless Conversion Search in Aluminum},''
  \href{http://dx.doi.org/10.3390/universe9010054}{{\em Universe} {\bfseries 9}
  no.~1, (2023) 54}, \href{http://arxiv.org/abs/2210.11380}{{\ttfamily
  arXiv:2210.11380 [hep-ex]}}.

\bibitem{Bernstein:2019fyh}
{\bfseries Mu2e} Collaboration, R.~H. Bernstein, ``{The Mu2e Experiment},''
  \href{http://dx.doi.org/10.3389/fphy.2019.00001}{{\em Front. in Phys.}
  {\bfseries 7} (2019) 1}, \href{http://arxiv.org/abs/1901.11099}{{\ttfamily
  arXiv:1901.11099 [physics.ins-det]}}.

\bibitem{Mu3e:2020gyw}
{\bfseries Mu3e} Collaboration, K.~Arndt {\em et~al.}, ``{Technical design of
  the phase I Mu3e experiment},''
  \href{http://dx.doi.org/10.1016/j.nima.2021.165679}{{\em Nucl. Instrum. Meth.
  A} {\bfseries 1014} (2021) 165679},
  \href{http://arxiv.org/abs/2009.11690}{{\ttfamily arXiv:2009.11690
  [physics.ins-det]}}.

\bibitem{Hesketh:2022wgw}
{\bfseries Mu3e} Collaboration, G.~Hesketh, S.~Hughes, A.-K. Perrevoort, and
  N.~Rompotis, ``{The Mu3e Experiment},'' in {\em {Snowmass 2021}}.
\newblock 4, 2022.
\newblock \href{http://arxiv.org/abs/2204.00001}{{\ttfamily arXiv:2204.00001
  [hep-ex]}}.

\bibitem{Langacker:1988up}
P.~Langacker and D.~London, ``{Lepton Number Violation and Massless
  Nonorthogonal Neutrinos},''
  \href{http://dx.doi.org/10.1103/PhysRevD.38.907}{{\em Phys.Rev.D} {\bfseries
  38} (1988) 907}.

\bibitem{Lavoura:2003xp}
L.~Lavoura, ``{General formulae for $f(1) \to f(2) + \gamma$},''
  \href{http://dx.doi.org/10.1140/epjc/s2003-01212-7}{{\em Eur. Phys. J. C}
  {\bfseries 29} (2003) 191--195},
  \href{http://arxiv.org/abs/hep-ph/0302221}{{\ttfamily arXiv:hep-ph/0302221}}.

\bibitem{Hue:2017lak}
L.~T. Hue, L.~D. Ninh, T.~T. Thuc, and N.~T.~T. Dat, ``{Exact one-loop results
  for $l_i \to l_j\gamma$ in 3-3-1 models},''
  \href{http://dx.doi.org/10.1140/epjc/s10052-018-5589-3}{{\em Eur. Phys. J. C}
  {\bfseries 78} no.~2, (2018) 128},
  \href{http://arxiv.org/abs/1708.09723}{{\ttfamily arXiv:1708.09723
  [hep-ph]}}.

\bibitem{CarcamoHernandez:2020pnh}
A.~E. C\'arcamo~Hern\'andez, L.~T. Hue, S.~Kovalenko, and H.~N. Long, ``{An
  extended 3-3-1 model with two scalar triplets and linear seesaw mechanism},''
  \href{http://dx.doi.org/10.1140/epjp/s13360-021-02146-9}{{\em Eur. Phys. J.
  Plus} {\bfseries 136} no.~11, (2021) 1158},
  \href{http://arxiv.org/abs/2001.01748}{{\ttfamily arXiv:2001.01748
  [hep-ph]}}.

\bibitem{CarcamoHernandez:2023atk}
A.~E. C{\'a}rcamo~Hern{\'a}ndez, V.~K.~N., and J.~W.~F. Valle, ``{Linear seesaw
  mechanism from dark sector},''
  \href{http://dx.doi.org/10.1007/JHEP09(2023)046}{{\em JHEP} {\bfseries 09}
  (2023) 046}, \href{http://arxiv.org/abs/2305.02273}{{\ttfamily
  arXiv:2305.02273 [hep-ph]}}.

\bibitem{Batra:2023mds}
A.~Batra, P.~Bharadwaj, S.~Mandal, R.~Srivastava, and J.~W.~F. Valle,
  ``{Phenomenology of the simplest linear seesaw mechanism},''
  \href{http://dx.doi.org/10.1007/JHEP07(2023)221}{{\em JHEP} {\bfseries 07}
  (2023) 221}, \href{http://arxiv.org/abs/2305.00994}{{\ttfamily
  arXiv:2305.00994 [hep-ph]}}.

\bibitem{Kitano:2002mt}
R.~Kitano, M.~Koike, and Y.~Okada, ``{Detailed calculation of lepton flavor
  violating muon electron conversion rate for various nuclei},''
  \href{http://dx.doi.org/10.1103/PhysRevD.76.059902}{{\em Phys. Rev. D}
  {\bfseries 66} (2002) 096002},
  \href{http://arxiv.org/abs/hep-ph/0203110}{{\ttfamily arXiv:hep-ph/0203110}}.
  [Erratum: Phys.Rev.D 76, 059902 (2007)].

\bibitem{Ilakovac:1994kj}
A.~Ilakovac and A.~Pilaftsis, ``{Flavor violating charged lepton decays in
  seesaw-type models},''
  \href{http://dx.doi.org/10.1016/0550-3213(94)00567-X}{{\em Nucl.Phys.B}
  {\bfseries 437} (1995) 491},
  \href{http://arxiv.org/abs/hep-ph/9403398}{{\ttfamily arXiv:hep-ph/9403398
  [hep-ph]}}.

\bibitem{Alonso:2012ji}
R.~Alonso, M.~Dhen, M.~Gavela, and T.~Hambye, ``{Muon conversion to electron in
  nuclei in type-I seesaw models},''
  \href{http://dx.doi.org/10.1007/JHEP01(2013)118}{{\em JHEP} {\bfseries 01}
  (2013) 118}, \href{http://arxiv.org/abs/1209.2679}{{\ttfamily arXiv:1209.2679
  [hep-ph]}}.

\bibitem{Lindner:2016bgg}
M.~Lindner, M.~Platscher, and F.~S. Queiroz, ``{A Call for New Physics : The
  Muon Anomalous Magnetic Moment and Lepton Flavor Violation},''
  \href{http://dx.doi.org/10.1016/j.physrep.2017.12.001}{{\em Phys. Rept.}
  {\bfseries 731} (2018) 1--82},
  \href{http://arxiv.org/abs/1610.06587}{{\ttfamily arXiv:1610.06587
  [hep-ph]}}.

\bibitem{Blennow:2023mqx}
M.~Blennow, E.~Fern\'andez-Mart\'\i{}nez, J.~Hern\'andez-Garc\'\i{}a,
  J.~L\'opez-Pav\'on, X.~Marcano, and D.~Naredo-Tuero, ``{Bounds on lepton
  non-unitarity and heavy neutrino mixing},''
  \href{http://dx.doi.org/10.1007/JHEP08(2023)030}{{\em JHEP} {\bfseries 08}
  (2023) 030}, \href{http://arxiv.org/abs/2306.01040}{{\ttfamily
  arXiv:2306.01040 [hep-ph]}}.

\bibitem{Deppisch:2013cya}
F.~F. Deppisch, N.~Desai, and J.~W.~F. Valle, ``{Is charged lepton flavor
  violation a high energy phenomenon?},''
  \href{http://dx.doi.org/10.1103/PhysRevD.89.051302}{{\em Phys.Rev.D}
  {\bfseries 89} (2014) 051302},
  \href{http://arxiv.org/abs/1308.6789}{{\ttfamily arXiv:1308.6789 [hep-ph]}}.

\bibitem{Abada:2018nio}
A.~Abada and A.~M. Teixeira, ``{Heavy neutral leptons and high-intensity
  observables},'' \href{http://dx.doi.org/10.3389/fphy.2018.00142}{{\em Front.
  in Phys.} {\bfseries 6} (2018) 142},
  \href{http://arxiv.org/abs/1812.08062}{{\ttfamily arXiv:1812.08062
  [hep-ph]}}.

\bibitem{Fernandez-Martinez:2016lgt}
E.~Fernández-Martínez, J.~Hernández-García, and J.~López-Pavón, ``{Global
  constraints on heavy neutrino mixing},''
  \href{http://dx.doi.org/10.1007/JHEP08(2016)033}{{\em JHEP} {\bfseries 08}
  (2016) 033}, \href{http://arxiv.org/abs/1605.08774}{{\ttfamily
  arXiv:1605.08774 [hep-ph]}}.

\bibitem{LZ:2024zvo}
{\bfseries LZ} Collaboration, J.~Aalbers {\em et~al.}, ``{Dark Matter Search
  Results from 4.2{\,}{\,}Tonne-Years of Exposure of the LUX-ZEPLIN (LZ)
  Experiment},'' \href{http://dx.doi.org/10.1103/4dyc-z8zf}{{\em Phys. Rev.
  Lett.} {\bfseries 135} no.~1, (2025) 011802},
  \href{http://arxiv.org/abs/2410.17036}{{\ttfamily arXiv:2410.17036
  [hep-ex]}}.

\bibitem{Hambye:2008bq}
T.~Hambye, ``{Hidden vector dark matter},''
  \href{http://dx.doi.org/10.1088/1126-6708/2009/01/028}{{\em JHEP} {\bfseries
  01} (2009) 028}, \href{http://arxiv.org/abs/0811.0172}{{\ttfamily
  arXiv:0811.0172 [hep-ph]}}.

\bibitem{DEramo:2010keq}
F.~D'Eramo and J.~Thaler, ``{Semi-annihilation of Dark Matter},''
  \href{http://dx.doi.org/10.1007/JHEP06(2010)109}{{\em JHEP} {\bfseries 06}
  (2010) 109}, \href{http://arxiv.org/abs/1003.5912}{{\ttfamily arXiv:1003.5912
  [hep-ph]}}.

\bibitem{Gu:2010xc}
P.-H. Gu and U.~Sarkar, ``{Leptogenesis with Linear, Inverse or Double
  Seesaw},'' \href{http://dx.doi.org/10.1016/j.physletb.2010.09.062}{{\em Phys.
  Lett. B} {\bfseries 694} (2011) 226--232},
  \href{http://arxiv.org/abs/1007.2323}{{\ttfamily arXiv:1007.2323 [hep-ph]}}.

\bibitem{Pilaftsis:1997jf}
A.~Pilaftsis, ``{CP violation and baryogenesis due to heavy Majorana
  neutrinos},'' \href{http://dx.doi.org/10.1103/PhysRevD.56.5431}{{\em Phys.
  Rev. D} {\bfseries 56} (1997) 5431--5451},
  \href{http://arxiv.org/abs/hep-ph/9707235}{{\ttfamily arXiv:hep-ph/9707235}}.

\bibitem{Dolan:2018qpy}
M.~J. Dolan, T.~P. Dutka, and R.~R. Volkas, ``{Dirac-Phase Thermal Leptogenesis
  in the extended Type-I Seesaw Model},''
  \href{http://dx.doi.org/10.1088/1475-7516/2018/06/012}{{\em JCAP} {\bfseries
  06} (2018) 012}, \href{http://arxiv.org/abs/1802.08373}{{\ttfamily
  arXiv:1802.08373 [hep-ph]}}.

\bibitem{Blanchet:2009kk}
S.~Blanchet, T.~Hambye, and F.-X. Josse-Michaux, ``{Reconciling leptogenesis
  with observable $\mu \to e \gamma$ rates},''
  \href{http://dx.doi.org/10.1007/JHEP04(2010)023}{{\em JHEP} {\bfseries 04}
  (2010) 023}, \href{http://arxiv.org/abs/0912.3153}{{\ttfamily arXiv:0912.3153
  [hep-ph]}}.

\bibitem{Abada:2021zcm}
A.~Abada, J.~Kriewald, and A.~M. Teixeira, ``{On the role of leptonic CPV
  phases in cLFV observables},''
  \href{http://dx.doi.org/10.1140/epjc/s10052-021-09754-w}{{\em Eur. Phys. J.
  C} {\bfseries 81} no.~11, (2021) 1016},
  \href{http://arxiv.org/abs/2107.06313}{{\ttfamily arXiv:2107.06313
  [hep-ph]}}.

\bibitem{Breiman:2001RandomForests}
L.~Breiman, ``Random forests,''
  \href{http://dx.doi.org/10.1023/A:1010933404324}{{\em Machine Learning}
  {\bfseries 45} no.~1, (2001) 5--32}.

\bibitem{Lundberg:2017SHAP}
S.~Lundberg and S.-I. Lee, ``A unified approach to interpreting model
  predictions,'' 2017.
\newblock \url{https://arxiv.org/abs/1705.07874}.

\bibitem{Lundberg:2018TreeSHAP}
S.~M. Lundberg, G.~G. Erion, and S.~Lee, ``Consistent individualized feature
  attribution for tree ensembles,'' {\em CoRR} {\bfseries abs/1802.03888}
  (2018) , \href{http://arxiv.org/abs/1802.03888}{{\ttfamily 1802.03888}}.
  \url{http://arxiv.org/abs/1802.03888}.

\end{thebibliography}\endgroup

\end{document}